\documentclass[10pt,a4paper]{article}
\usepackage[utf8]{inputenc}  
\usepackage[hang,small,tight]{subfigure}   
\usepackage{amssymb}   
\usepackage{amsmath}   
\usepackage{amsbsy}   
\usepackage{natbib}   
\usepackage{graphicx}   
\usepackage{longtable}
\usepackage{pdfpages}
\usepackage{multirow}
\usepackage{color}   
\usepackage{indentfirst}   
\usepackage[pdftex,bookmarks=true,colorlinks=true,linkcolor=blue,citecolor=red]{hyperref}   
\usepackage{here}
\usepackage{appendix}
\usepackage{arydshln}  
\usepackage{dsfont}
\usepackage{multirow}
\usepackage{enumitem} 
\usepackage{hyperref}
\usepackage{bold-extra}
\usepackage{listings}
\usepackage{setspace}

\usepackage{algorithm}
\usepackage{algpseudocode}

\newcommand{\bfm}{{\bf m}}

\newcommand{\bfx}{{\bf x}}
\newcommand{\bfy}{{\bf y}}
\newcommand{\bfz}{{\bf z}}

\newcommand{\bfD}{{\bf D}}

\newcommand{\bftau}{\mbox{\boldmath $\tau$}}

\newcommand{\bfbeta}{\mbox{\boldmath $\beta$}}

\newcommand{\bftheta}{\mbox{\boldmath $\theta$}}

\newcommand{\bfkappa}{\mbox{\boldmath $\kappa$}}

\newcommand{\R}{\mathbb{R}}

\begin{document}

\clearpage
\thispagestyle{empty}
\title{Variational  Full Bayes Lasso: \\  Knots Selection in  Regression Splines}
\author{Larissa Alves$^{1}$, Ronaldo Dias$^{2}$, Helio S. Migons$^{3}$}\thanks{The authors,(LA)$^1$ENCE, (RD)$^{2}$IMECC-UNICAMP and (HSM)$^3$IM-UFRJ, contributed equally to the design and implementation of the research, to the analysis of the results and to the writing of the manuscript. {\emph{Corresponding author. E-mail addresses:}}\tt{ migon@im.ufrj.br} }
\date{\today}
\maketitle
\setcounter{page}{1}
\pagenumbering{arabic}

\begin{abstract}
 We develop a fully automatic Bayesian Lasso via variational inference. This is a scalable procedure for approximating the posterior distribution. Special attention is driven to the knot selection in  regression spline. In order to carry through our proposal, a full automatic variational Bayesian Lasso, a Jefferey's prior is proposed for the hyperparameters and a decision theoretical approach is introduced to decide if a knot is selected or not. Extensive simulation studies were developed to ensure the effectiveness of the proposed algorithms. The performance of the algorithms were also tested in some real data sets, including data from the world pandemic Covid-19. Again, the algorithms showed a very good performance in capturing the data structure.
\end{abstract}
 
\section{Introduction}

In the recent literature one finds many alternative proposals for modeling and estimating a smooth function. In this article we focus on variants of smoothing splines, called penalized regression splines (\cite{Montoya14}, \cite{brian:eilers:1996}). This is an attractive approach for modeling the nonlinear smoothing effect of covariates. This work discusses the selection of knots given a fixed maximum number of knots. A roughness penalty is introduced to control the selection of knots and consequently to balance the two conflicting goals, goodness of fit and smoothness.
Our approach will be through a full Bayesian Lasso with variational inference. It is related to the work of \cite{Osborne99} where an efficient algorithm to calculate the classical Lasso estimator was presented.
Our contribution, therefore, includes the application of the mean field variational inference (\cite{Blei17}, \cite{Ormerod10}) for the complete Bayesian lasso penalty (\cite{Park08}
 and \cite{Mallick14})).
 Choosing the ideal number of knots and their position is a difficult problem. We propose a two-step procedure related to the work of \cite{Ruppert02}.
For regularization and model selection, the proposed procedure starts with a fixed maximum number of knots and then uses a full Bayesian lasso, which combines characteristics of shrinkage and variable selection, to obtain the most significant knots to recover the unknown smooth function. The number of knots is chosen based on an approximation of the predictive distribution in a grid of knots values.
 

The original formulation of the Bayesian Lasso is based on a hierarchical representation of the Laplace distribution, as a mixture of scale normal based on exponential (\cite{Park08}) and more recently as a mixture of uniform with exponential (\cite{Mallick14}). 
 
Alternative procedures for selecting the effective number of knots involving least squares and penalized splines regression has been proposed in the recent literature, see (\cite{Spiriti12}, \cite{Montoya14}).

It is well known that MCMC often takes a great deal of computational time and is not  scalable. Therefore, our proposal is to use variational inference (VI) integrated with a decision theoretical approach to knot selection in regression splines. Both are discussed in detail and have shown to be comparatively better than the alternatives presented in the current literature. 

The remainder of the paper is organized as follows. In Section 2 presents a review the Bayesian linear model, the variational inference and the hierarchical formulation of the Laplace distribution. In Section 3, shows the full Bayesian Lasso, including the Jeffrey's prior for the hyperparameters and the Bayes factor criterion for knots selection. Section 4  states the knot selection procedure for regression spline in an almost fully automatic algorithm. Section 5 shows a comparative numerical simulation of the performance of the proposed method and other existing approaches in the literature. A data analysis of real datasets is presented in Section 6.

\section{ A review of the Methodology }

In order to set the notation to be used later,  this section presents a brief summary of Bayesian regression models and variational inference techniques and also establishes the framework for our proposal to select knots in regression spline models to be developed in Section \ref{sec:regsplines}.

We summarize the conjugate Bayesian analysis of a linear model. In addition, we present an introduction to variational inference and the hierarchical representation of the Laplace distribution. For more details see, \cite{Drugowitsch19}, \cite{deni:mall:smit:1998}, \cite{Berry02}, \cite{goepp:hal:2018} and 
\cite{Lang04}.

\subsection{Bayesian estimation in linear models}

Following the notation of \cite{MGL2015} let the linear model  be
$$
\bfy \, | \, \bfbeta, \phi \sim N(X\bfbeta, \phi^{-1} I_n)
$$
where $y$ is n-vector of observed quantities, $X$ is a known $n \times p$ matrix, $\bfbeta$ is a p-vector of parameters and  $\phi$ is the precision associated with each one of the independent observations.
The conjugate prior, a Normal-Gamma, is defined as:
\begin{eqnarray*}
\bfbeta | \phi &\sim& N(\bfm_0, \phi^{-1} C^{-1}_0)\\
\phi &\sim& Ga(a_0, b_0)
\end{eqnarray*}
where $\bfm_0$ and $(\phi \, C_0)^{-1}$ are, respectively, the prior mean  and covariance matrix and $a_0, b_0$ are the parameters of the precision prior distribution.
The posterior distribution is
\begin{eqnarray*}
\bfbeta| \phi, \bfy, X   &\sim& N(\bfm_1, \phi^{-1} C_1^{-1}) \\ \phi|\bfy, X &\sim& Ga(a_1, b_1)\\
\bfm_1 &=& C^{-1}_1 \, ( C_0 \bfm_0  + X^T\bfy )  \, \, \, \, \, \,   \mbox{and} \, \, \,  \, \, \,  C_1 = C_0  + X^TX\\
a_1 & =& a_0 + \frac{n}{2} \, \, \,  \, \, \, \, \mbox{and} \, \, \,  \, \, \,  b_1 = b_0 + \frac{1}{2}[(\bfy - X \bfm_1)^T \bfy + (\bfm_0-\bfm_1)^T \, C_0 \bfm_0]
\end{eqnarray*}

A very useful extension of the above regression model is the Bayesian hierarchical regression models, which will be extensively used latter. It was proposed in the seminal paper of \cite{LindleySmith72} and a dynamic version was introduced in \cite{DaniMigon:93}.

 
\subsection{Variational Inference - main aspects}\label{subsec:VB}
It is well known that Bayesian inference regarding unknown quantities is entirely based on their probabilistic description.
 Therefore, {\it variational inference (VI)}, a method to deal with the approximation of probability densities is very useful for Bayesian inference. In fact, these techniques can be traced back to the field of machine learning (\cite{Jordan99}).
Loosely speaking, they basically exchange {\it sampling}, as in MCMC procedures, for optimization. By choosing a flexible family of approximate densities, an attempt is made to find a member of this family, which minimizes some optimal criterion, for example Kulback-Leibner divergence ($KL$). Variational inference is useful for quickly comparing alternative models and also for dealing with large data sets. \cite{Blei17} pointed out that the accuracy of variational inference has not yet been thoroughly studied and many open questions are still there to be answered.

The basic ideas about variational inference can be easily followed in \cite{Blei17}   and in \cite{Ormerod10}. Many examples are presented in the \cite{Bishop06} book. 
Let $\bfy$ be a vector of $n$ independent identically distributed observations and $\bfz$ a vector including latent variables and  the parameters as well.  The log marginal data distribution, also known as  evidence integral, is denoted by  $p(\bfy)$. Evidence integrals that are often unavailable in closed form require exponential time to be evaluated and present difficulties in making the inference for a model such as this. To avoid calculating the evidence integral, one tries to find a lower bound, which is known as {\it ELBO$(q)$ - Evidence Lower Bound} and will be denoted by ${ \cal L}(q)$. It is easy to verify that:
$$
\log\, p(\bfy) = {\cal L}(q) \, + KL(q| |p), 
$$
where ${\cal L}(q) = \int  q(\bfz) \log \, \frac{p(\bfy,\bfz)}{q(\bfz)} d\bfz$ \, and \,  $KL(q | | p) = - \, \int q(\bfz) \log \, \frac{p(\bfz | \bfy)}{q(\bfz)} d\bfz$, since \,  $p(\bfy) =\frac{p(\bfy,\bfz) /q(\bfz)}{p(\bfz|\bfy)/q(\bfz)}$.

It is clear that $max_{q} \, {\cal L}(q) \simeq min_{q} \, KL(q | | p)$ and also that $KL(q| |p) \ge 0$ with equality if and only if $p(\bfz|\bfy)=q(\bfz)$.
In general, it is difficult to obtain this posterior distribution, therefore, the approach is to choose a family of tractable densities. Let's assume the following: 
$$
q(\bfz) = \prod_{l=1}^m q_l(\bfz_l)
$$ 
where a partition of the $\bfz$ into $m$ disjoint groups is denoted as $\bfz_l$. It is worth pointing out that there is no restriction on the functional forms of the variational densities $q_l(\bfz_l)$.

The central idea is to maximize each factor (blocks of z's) of $q(\bfz)$ in turn.  We keep  $q_{l\ne h}$ fixed and maximize ${\cal L}(q)$. Note that:
\begin{eqnarray}
\label{eq:lq}
{ \cal L}(q) &=& \int \, \prod_{l=1}^m   \,  q_l(\bfz_l)  [\log\, p(\bfy,\bfz) - \,  \log\, q_l(\bfz_l)  ] d\bfz \nonumber\\
&=& \int  \, q_h(\bfz_h) \, [ \int \, \log\,p(\bfy,\bfz) \prod_{l \ne h}   q_l(\bfz_l) \, d\bfz_l] \, d\bfz_h - \int q_h(\bfz_h) \, \log\, q_h(\bfz_h)  \,d\bfz_h + const \nonumber \\ 
&=& \int \, q_h(\bfz_h) \, \log  \, \tilde{p}(\bfy, \bfz_h) \, d\bfz_h - \int \, q_h(\bfz_h) \, \log \, q_h(\bfz_h) \, d\bfz_h + const
\end{eqnarray}
where $ \tilde{p}(\bfy, \bfz_h) = E_{l \ne h} \, \log \, p(\bfy,\bfz) + const$.
The ${ \cal L}(q)$ will be presented for the specific case of Lasso in subsection \ref{subsec:VBLasso}. Note that it depends on the variational parameters.

Worth emphasizing that the problem of approximating the posterior distribution for the parameters of interest was replaced for a maximization problem. The algorithm to solve the optimization problem was introduced by \cite{Bishop06} and denoted by CAVI - coordinate ascent variational inference. The CAVI optimizes one factor of the mean field variational density at a time.

Since (\ref{eq:lq}) is equal to $- KL(\cdot||\cdot)$, maximizing it is equivalent to minimizing  $KL$.
Therefore, the optimal solution  is:
$$
log \, q^*(\bfz_l) =  E_{l \ne h} [ \log \, p(\bfy, \bfz)] + const
$$

As one can see, $q^*(\bfz_l)$ depends on the full conditional distributions, as usually denoted in the MCMC literature (\cite{George:Casella:1992}). Therefore, there is a natural link with Gibbs Sampling but the proposed approach leads to tractable solutions involving only local operations.


 \subsection{ Hierarchical representation of the Laplace distribution}\label{subsec:Hierarchical_laplace}
 

It is well known that the original Lasso formulation (\cite{Tibishirani96}) is related to the Laplace distribution which can be represented as hierarchical mixture of distributions and is relevant to a hierarchical modeling, which in turn is important for the implementation of Gibbs Sampling.

One of these forms of representation is a scale mixture of a Normal distribution with an Exponential distribution (\cite{West87}) and the other is a mixture of a Uniform distribution with a Gamma distribution (\cite{Mallick14}).
 
Specifically, following \cite{Andrews74}, it is easy to verify that the hierarchical representation: ${\beta}|\tau \sim N[0,\tau]$ and $\tau|\lambda \sim \exp\left(\frac{\lambda^2}{2}\right)$ leads, by marginalizing on $\tau$, to the standard Laplace distribution, whose density is:

\begin{eqnarray}\label{eq:Laplace}
\frac{\lambda}{2} \exp(- \lambda \, |\beta|) = \int_0^{\infty} \left[\frac{ \tau^{-1/2}}{\sqrt{2 \pi}} \, \exp\left(-\frac{  \beta^2  }{2 \tau}\right)\right] \, \, \left[ \frac{\lambda^2}{2} \exp\left(-\frac{\lambda^2}{2}\tau\right)\right]
d \tau.
\end{eqnarray}

The above hierarchical representation of the Laplace distribution is important to introduce the full Bayesian Lasso. The penalty term in the classical Lasso can be interpreted as independent Laplace prior distribution over the regression parameters. Moreover, the posterior mode can be seen as the Lasso estimates.

 
 
 
\section{ The full Bayesian Lasso} 
\label{sec:fullbayesianlasso}
Following the hierarchical representation for the Laplace distributions in Subsection \ref{subsec:Hierarchical_laplace}, \cite{Park08} shows a Bayesian formulation of the Lasso regression model. The hierarchical model is defined as: 
\begin{eqnarray*}
\bfy|X, \bfbeta, \phi &\sim& N[X\bfbeta, \phi^{-1} I_n]\\
\bfbeta|\phi, \bftau &\sim& N[0, \phi^{-1} \bfD_{\tau}] \\ \tau_j|\lambda  &\sim& Exp(\lambda) \;\;\;\; \mbox{with} \;\;\; j = 1, \ldots, p
\end{eqnarray*}
where $\bfD_\tau = diag(\tau_1, \ldots, \tau_p)$ and $\tau_j|\lambda$ are conditionally independent for all $j$.
The model can be completed with the hyperparameters of the priors $ \phi \sim Ga (a_0, b_0) $ and $\lambda \sim Ga (g_0, h_0) $. In Subsection \ref{Jeffreysprior} we propose an independent Jeffreys prior for $\phi$ and $\lambda$ to automate the Lasso, and this implies supposing $a_0$, $b_0$, $g_0$ and $h_0$ tending to zero.  

Let $\bftheta = (\bfbeta,\phi,\bftau,\lambda)$ be the vector of the parameters and the latent variables of the model.
The posterior distribution  is obtained as  proportional to the   model  distribution times  the prior distribution for the latent component and the parameters:
 $$
 p(\bftheta | \mathbf{y},X) \propto p( \bfy | X, \bfbeta, \phi) \, \, p(\bfbeta | \phi, \bftau) \,\,  p(\bftau|\lambda) \, \,  p(\phi) \, \,  p(\lambda).
 $$
For instance, the above joint posterior is often intractable. An almost obvious numerical approach, since the breakthrough paper of \cite{Gelfand:Smith:1990}, is to use stochastic simulation.

 
\subsection{Jeffreys prior using Fisher decomposition} \label{Jeffreysprior}

In order to develop an automatic Bayesian Lasso procedure it is worth to introduce non informative priors for the hyperparameters involved.
Following \cite{Fonseca19} and exploring the conditional independence involved in the Lasso model,  the Fisher information decomposition for Lasso follows as:
\begin{eqnarray}{\label{FisherDec}}
I_\bfy(\lambda) = I_{\bftau}(\lambda) - E_{\bfy}\, [ I_{\bfbeta,\bftau} (\lambda| \bfy)],
\end{eqnarray}
where $ I_{\bfbeta,\bftau} (\lambda | \bfy)$ is the information obtained from the  full conditional distribution $ p(\bfbeta, \bftau | \bfy,  \lambda)$. We also are using the conditional independence described by the graph that represents the Bayesian Lasso model.

We will develop, in turn, each of the components in the expression (\ref{FisherDec}). The quantity $I_{\bftau}( \lambda)$ is based on the independent marginal distribution of $\tau_j$, leading directly to $I_{\bftau}(\lambda) = \frac{p}{\lambda^2}$.
 
In order to obtain $I_{\bfbeta, \bftau}(\lambda | \bfy) $, \, we take advantage of the known full conditional distribution of $(\bfbeta, \bftau| \lambda, \bfy) $ (see (\ref{Gibbs})). Since $(\bfbeta | \bftau, \lambda, \bfy) $ does not depend on $\lambda$, then it is easy to obtain $E_{\bfy}\, [I_{\bfbeta,\bftau} (\lambda| \bfy)] = \frac{p}{\lambda}$.

Then substituting in (\ref{FisherDec}),  it follows 
$I_\bfy(\lambda) = \frac{p}{\lambda^2}  +  \frac{p}{\lambda^2}$
and so the prior for $\lambda$ \,  is  \,  $p(\lambda) \propto \lambda^{-1}$.
This result is similar to the one  reported in \cite{Fonseca19}, using the Uniform Gamma mixture.

It is well known that the Jeffrey's prior of $\phi$ is proportional of $\phi^{-1}$.

\subsection{ The MCMC formulation}\label{MCMC}
   
Considering the model and the  prior distribution already specified, we know that the posterior distribution in this case has an unknown form. Therefore, we can use the MCMC to obtain a sample of the posterior distribution through the complete conditional distributions (Gibbs Sampler). Calculations of complete conditionals are as follows.

\begin{eqnarray}\label{Gibbs}
(\bfbeta|\bfy, \bftheta_{-\bfbeta} ) &\sim& N\left((X^TX + \bfD_{\tau}^{-1})^{-1}X^T\bfy , \frac{1}{\phi} (X^TX + \bfD_{\tau}^{-1})^{-1}\right) \nonumber \\
(\tau_j|\bfy, \bftheta_{-\tau_j} ) &\sim&  GIG\left(\frac{1}{2}, 2 \lambda, \bfbeta_j^2 \phi\right)  \nonumber \\
(\phi|\bfy, \bftheta_{-\phi} ) &\sim&  Ga\left(\frac{n}{2}+\frac{p}{2}+a_0 , b_0 + \frac{1}{2}[(\bfy-X\bfbeta)^T(\bfy-X\bfbeta)+\bfbeta^T \bfD_{\tau}^{-1}\bfbeta]\right)  \nonumber \\
(\lambda|\bfy, \bftheta_{-\lambda} ) &\sim&  Ga\left(g_0 + p, h_0 + \sum_{j=1}^p \tau_j\right)
\end{eqnarray}
where $\bftheta_{-}$ stands for the entire vector $\bftheta$ without the parameter followed by symbol "\rule{0.15cm}{0.15mm}", and {\it GIG} denotes  the generalized inverse Gaussian distribution. See appendix. 
   
 \subsection{ The variational approximation applied to Lasso}\label{subsec:VBLasso}
 
In order to obtain a scalable inference procedure, we introduce an alternative methodology.

\begin{figure}[h!]
\begin{center}
\includegraphics[width=15cm]{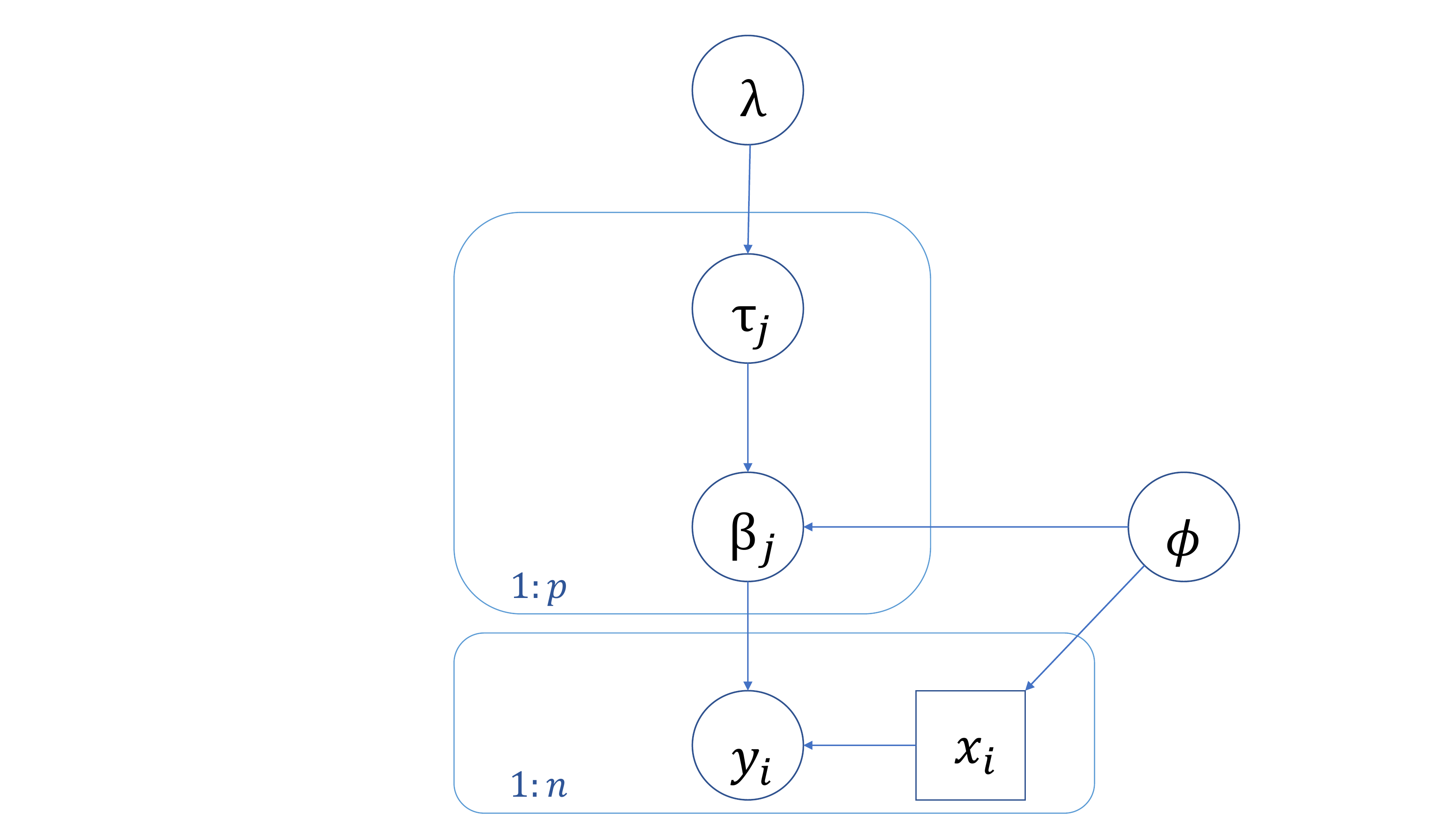}
\end{center}
\caption{Directed acyclic graph}\label{Graph1}\label{DAG}
\end{figure}

To make the notation consistent,  the vector including latent variables and parameters denoted by $\bfz$ in the subsection \ref{subsec:VB} is represented in this section by the vector $\bftheta$.
Let the independent Jeffrey's prior be $p(\phi) \propto \frac{1}{\phi}$ and $p(\lambda) \propto \frac{1}{\lambda} .$
The joint distribution of the observations, latent components and parameters can easily be followed from the Figure \ref{DAG} which in turn summarizes the model.

 It is worth remembering the expression of the {\it mean field} posterior approximation for the latent components and parameters:  
 $$
\log( q(\bftheta)) = \log(q_1(\bfbeta, \phi)) + \log(q_2(\bftau|\lambda)) + \log(q_3(\lambda))
 $$
 

After quoting \cite{Blei17} the optimal $q_l(\bftheta_l)$ is proportional to the exponential of the log of the complete conditional distribution that is calculated in (\ref{Gibbs})
$$q^\ast_l(\bftheta_l) \propto \exp\{E_{-l}[\log p(\bftheta_l|\bftheta_{-l},\bfy)]\}, \;\; l = 1, 2, 3.$$

In the first step,  the variational posterior for $\bfbeta$ and $\phi$,   that maximizes the variational bound ${\cal L}(q)$ while holding $q_2(\bftau|\lambda)$ and $q_3(\lambda)$ fixed, is given by 
\begin{eqnarray*}
\log q_1^\ast(\bfbeta,\phi) &=& \log(p(\bfy|\bfbeta,\phi)) + E_{\tau}[\log(p(\bfbeta,\phi|\bftau))] + const\\
&=& \log N(\bfbeta|m_\beta, \phi^{-1} C_\beta) \times Ga(\phi|a_\phi,b_\phi)
\end{eqnarray*}


It is easy to see that this is a normal-gamma distribution with parameters:

\begin{eqnarray*}
C_\beta^{-1} = E_{\tau} (\bfD_\tau^{-1}) + X^TX, \, \, \,  \, \, \,   &\mbox{and}&  \, \, \,  \, \, \, m_\beta = C_\beta X^T\bfy, \\
a_\phi = a_0 + n/2, \, \, \, \, \,  \,  &\mbox{and}&  \, \, \,  \, \, \, b_\phi = b_0 + \frac{1}{2} (\bfy^T\bfy - m_\beta^T C_\beta^{-1}m_\beta).
\end{eqnarray*}

Next, the variational  distribution of   $\bftau$,    that  maximizes the variational bound ${\cal L}(q)$ while holding $q_3(\lambda)$ fixed is given by 

\begin{eqnarray*}
\log q_2^\ast(\tau_j) &=& E_{\lambda}[\log (p(\tau_j|\lambda))] + E_{\beta,\phi}[\log (p(\beta_j,\phi|\tau_j))] + const\\ 
&=& \log GIG(\tau_j|c_\tau,d_\tau,f_{\tau_j})
\end{eqnarray*}

with GIG being generalized inverse Gaussian distribution, where
$$c_\tau = \frac{1}{2}\;;\; d_\tau = 2 E_\lambda[\lambda] \; ; \; f_{\tau_j} = E_{\beta,\phi}[\phi \beta_j^2].$$ 
Therefore, 
$$\log q_2^\ast(\bftau) = \log \prod_{j=1}^p GIG(\tau_j|c_\tau,d_\tau,f_{\tau_j}).$$
 
Finally, we will identify the variational distribution of $\lambda$:

\begin{eqnarray*}
\log q_3^\ast(\lambda) &=& \log (p(\lambda)) + E_\tau[\log (p(\bftau|\lambda))] + const\\
&=& \log Ga(\lambda|g_\lambda, h_\lambda)
\end{eqnarray*}

which is a gamma distribution with parameters $$g_\lambda = g_0 + p  \; ;\; h_\lambda = h_0 + \sum_{j=1}^p E_\tau(\tau_j).$$

The expected values involved in the definition of the above variational distributions are  computed as follows (see the appendix for details and \cite{Jorgensen82}).

\begin{eqnarray}
E_\tau(\tau_j) &=& \frac{\sqrt{f_{\tau_j}} \kappa_{c_\tau+1}(\sqrt{d_\tau f_{\tau_j}})}{\sqrt{d_\tau} \kappa_{c_\tau}(\sqrt{d_\tau f_{\tau_j}})},\label{eq:Esptau}\\ 
Var_\tau(\tau_j) &=& \frac{f_{\tau_j}}{d_\tau}\left[\frac{K_{c_\tau+2}(\sqrt{d_\tau f_{\tau_j}})}{K_{c_\tau}(\sqrt{d_\tau f_{\tau_j}})} - \left( \frac{K_{c_\tau+1}(\sqrt{d_\tau f_{\tau_j}})}{K_{c_\tau}(\sqrt{d_\tau f_{\tau_j}})}\right)^2 \right],\label{eq:Vartau}\\
E_\tau[\bfD_\tau^{-1}] &=& \mbox{diag} (E_\tau(\tau_1^{-1}), \ldots, E_\tau(\tau_p^{-1})), \;\;\mbox{where} \;\; E_\tau(\tau_j^{-1}) = \frac{\sqrt{d_\tau} \kappa_{c_\tau+1}(\sqrt{d_\tau f_{\tau_j}})}{\sqrt{f_{\tau_j}} \kappa_{c_\tau}(\sqrt{d_\tau f_{\tau_j}})} - \frac{2 c_\tau}{f_{\tau_j}},\nonumber\\ 
E_{\beta,\phi}[\phi \beta_j^2] &=& m_{\beta_j}^2 a_\phi/b_\phi + (C_\beta)_{jj},\nonumber\\
E_\lambda(\lambda) &=& \frac{g_\lambda}{h_\lambda}\nonumber
\end{eqnarray}
where $\kappa_p(\cdot)$ is the Bessel modified function of the second kind.



The evidence lower bound (ELBO) for this model consists of:
\begin{eqnarray*}
{\cal{L}} (q) &=& E_{\beta,\phi}(\log p(\bfy|X,\bfbeta,\phi)) + E_{\beta,\phi,\tau}(\log p(\bfbeta,\phi|\bftau)) + E_{\tau,\lambda}(\log p(\bftau|\lambda)) +\\ 
&& + E_{\lambda}(\log p(\lambda)) - E_{\beta,\phi}(\log q_1(\bfbeta,\phi)) - E_{\tau,\lambda}(\log q_2(\bftau|\lambda)) - E_{\lambda}(\log q_3(\lambda))
\end{eqnarray*}

Each of the above terms are evaluated as function of the variational parameters, as follows:

\begin{eqnarray*}
E_{\beta,\phi}(\log p(\bfy|X,\bfbeta,\phi)) &=& \frac{n}{2}(\psi(a_\phi) - \log b_\phi - \log 2\pi) +\\
&& - \frac{1}{2} \left[\frac{a_\phi}{b_\phi} (\bfy - X m_\beta)^T(\bfy-X m_\beta) + tr(X^TX C_\beta)\right]\\ 
E_{\beta,\phi,\tau}(\log p(\bfbeta,\phi|\bftau)) &=& \frac{p}{2} (\psi(a_\phi) - \log b_\phi - \log 2\pi) + (a_0 - 1) (\psi(a_\phi) - \log b_\phi) - b_0 \frac{a_\phi}{b_\phi} +\\
&& + \frac{1}{2} \sum_{j=1}^p E_\tau(\log \tau_j) -\frac{1}{2} \sum_{j=1}^p \left[m_{\beta_j} \frac{a_\phi}{b_\phi} + (C_\beta)_{jj}\right] E_\tau\left(\frac{1}{\tau_j}\right)\\
E_{\tau,\lambda}(\log p(\bftau|\lambda)) &=& p (\psi(g_\lambda) - \log h_\lambda) - \frac{g_\lambda}{h_\lambda} \sum_{j=1}^p E_\tau(\tau_j)\\
E_{\lambda}(\log p(\lambda)) &=& g_0 \log h_0 - \log \Gamma(g_0) + (g_0 - 1) (\psi(g_\lambda) - \log h_\lambda) - h_0 \frac{g_\lambda}{h_\lambda}\\
E_{\beta,\phi}(\log q_1(\bfbeta,\phi)) &=& \frac{p}{2} (\psi(a_\phi) - \log b_\phi - \log 2\pi) - \frac{1}{2} \log |C_\beta| + a_\phi \log b_\phi - \log \Gamma(a_\phi) +\\
&& + (a_\phi - 1) (\psi(a_\phi) - \log b_\phi) - a_\phi\\
E_{\tau,\lambda}(\log q_2(\bftau|\lambda)) &=& \sum_{j=1}^p \left[ \frac{c_\tau}{2} \log \frac{d_\tau}{f_{\tau_j}} - \log 2 - \log K_{c_\tau}(\sqrt{d_\tau f_{\tau_j}}) + (c_\tau -1) E_\tau(\log \tau_j) \right. +\\ 
&& - \left[ \frac{d_\tau}{2} E_\tau(\tau_j) - \frac{f_{\tau_j}}{2} E_\tau \left(\frac{1}{\tau_j}\right) \right]\\
E_{\lambda}(\log q_3(\lambda)) &=& -\log \Gamma(g_\lambda) + (g_\lambda-1) \psi(g_\lambda) + \log h_\lambda - g_\lambda\\ 
\end{eqnarray*}

The second order Taylor expansion for $\log \tau_j$  at $E(\tau_j)$ is used to obtain the approximation for its expected value:
$E(\log \tau_j) \approx \log E(\tau_j) - \frac{Var(\tau_j)}{2 E^2(\tau_j)}$ where the mean and the variance of $\tau_j$ are in equations (\ref{eq:Esptau}) and (\ref{eq:Vartau}).


Note that the variational bound depends on the quantities $m_\beta$, $C_\beta$, $b_\phi$, $d_\tau$, $f_{\tau_j}$ e $h_\lambda$. The algorithm updates these quantities in each iteration. The ELBO is maximized and hence ${\cal{L}} (q)$ reaches a plateau with stabilization of those quantities. The algorithm consists of the following steps:

\begin{algorithm}
\caption{Variational Inference}\label{algo:VI}
\footnotesize{
\begin{algorithmic}[0]
\\\hrulefill
\item\text{\textit{Step} 1. Initialize the variational hyperparameters: $m_\beta$, $C_\beta$, $a_\phi$, $b_\phi$, $g_\lambda$, $h_\lambda$, $c_\tau$, $d_\tau$, $f_{\tau_j}$.}
\item\text{\textit{Step} 2.} 
\While{ELBO does not reach convergence} 
\For{$l = 1, 2, 3$} 
\item Compute $q^\ast_l(\bftheta_l) \propto \exp\{E_{-l}[\log p(\bftheta_l|\bftheta_{-l},\bfy)]\}$ 
\item \text{\textit{Step} 3. Update the variational hyperparameters based on the expected values.}
\State \text{Calculate ELBO}
\EndFor
\EndWhile
\\\hrulefill
\end{algorithmic}
}
\end{algorithm}

Convergence can be achieved by analyzing changes to ELBO in consecutive iterations or by analyzing the quantities on which it depends.

We end this section by showing the predictive distribution. Let $y^o$ e $y^p$ be the observed  and the predicted vectors, respectively. Finally, let $p(\beta,\phi|y^o)$ be its variational component. Then, after some algebraic calculations, we have a Student's t-distribution (St) as follows:
\begin{eqnarray*}
p(y^p|y^o,X^p) &=& \int \int p(y^p|\bfbeta,\phi) p(\bfbeta, \phi| y^o) d\bfbeta d\phi \approx \int \int p(y^p|\bfbeta, \phi) q_1(\bfbeta,\phi) d\bfbeta d\phi\\
&=& St\left(y^p|X^Tm_\beta,(1+X^T C_\beta X)\frac{b_\phi}{(a_\phi-1)},2a_\phi\right),
\end{eqnarray*}
where $q_1(\bfbeta,\phi)$ is the variational approximation of the posterior distribution, a normal-gamma distribution. 

\subsection{Variable selection}

We will discuss, from a Bayesian point of view, three alternative procedures for selecting knots (variables) in penalized regression splines (linear regression).
In this work, we propose a new decision criterion based on the Bayes factor. This proposed criterion is fully described below in \ref{BayesFactorDecisionCriteria}


\subsubsection{Bayes Factor decision criteria}
\label{BayesFactorDecisionCriteria}
In general, the selection of predictors/knots in a penalized regression/penalized regression splines () can be seen as a decision problem. Consider the general case where it is necessary to decide for one of the following models  ${\cal M}_0: \theta \in \Theta_0$  or  ${\cal M}_1: \theta \in \Theta_1$ based on  some observations  ($D$). An optimal decision will be based on the posterior probabilities, $ p({\cal M}_0 | D) $ and $ p({\cal M}_1 | D) $ and, also on the cost  of the wrong decisions.
Denote by $a$ the cost associate for the choice of the model ${\cal M} _0$ when, in fact, the true model is ${\cal M} _1$ and let $b$ be the cost of choosing  model ${\cal M} _1$ when the true model is ${\cal M} _0$. Therefore, if $ b \, \, p({\cal M} _0 | D) > a \, \, p({\cal M} _1 | D) $ then $ {\cal M} _0 $ should be chosen as the most plausible model for $ \theta $. By Bayes' theorem, the posterior odds are given by the product of the prior odds times the Bayes factor, 
$FB({\cal M}_0,{\cal M}_1) = {p(D|{\cal M}_0)}/{p(D|{\cal M}_1)}$,  where  $p(D|{\cal M}_i) = \int_{\Theta_i} p(D|\theta,{\cal M}_i) p(\theta|{\cal M}_i) d\theta, \;\;\; i=0, 1$. Hereafter, we will assume that the prior odds is equal one.

Particularly, we consider two alternatives: ${\cal M} _0: \beta_j = 0$ and ${\cal M} _1: \beta_j = \delta$, where $\delta \neq 0$ is a constant to be defined. Under ${\cal M} _0$, let us assume that $\beta_j|D \sim N (0, s_j^2) $ and under ${\cal M}_1$ we have $\beta_j| D \sim N(\delta,s_j^2)$, where $s_j^2=var(\beta_j|D)$. Hence, it is straightforward to get
$log(FB({\cal M}_0,{\cal M}_1) )=  log( \exp\{-\frac{1}{2} \beta_j^2\} / \exp\{-\frac{1}{2}(\beta_j-\delta)^2\} )=  \frac{1}{2} \delta^2 - \beta_j \delta.$

Assuming that at least a moderate evidence against  ${\cal M}_1$ corresponds to $FB({\cal M}_0,{\cal M}_1) \ge 3$  and   $\beta$ is considered to be significantly distant from the ${\cal M}_0$,   if and only if it  is  greater or equal to  the third quartile of the standard normal distribution, that is $\beta_{0.75} = 0.67$,  then a quadratic equation must be solved whose  root is $ \delta = 2.3 $. Thus, our proposal to select knots in spline regression is described in the following algorithm:
\pagebreak 
\begin{algorithm}
\caption{Bayes Factor decision criteria}\label{alg:BFKnotSelection}
\footnotesize{
\begin{algorithmic}[0]
\\ \hrulefill
\item \text{\textit{Step}  1. Take
 $\beta_{j}^{*}= m_j/s_j$,  
a standardized point estimate of $\beta_j$, \; $m_j=E[\beta_j | D]$
and $s_j^2=var(\beta_j|D)$.}
\item \text{\textit{Step} 2.  Compute the   Bayes factor at  $\widehat{\beta}_j^\star$: $BF({\cal M}_0, {\cal M}_1) = \frac{\exp\{-\frac{1}{2} (\widehat{\beta}_j^\star)^2\}}{\exp\{-\frac{1}{2}(\widehat{\beta}_j^\star-\delta)^2\}},$ \; $\delta = 2.3$.}
\item \text{\textit{Step} 3. Compute $\pi^\star = BF({\cal M}_0, {\cal M}_1) /(1+BF({\cal M}_0, {\cal M}_1) )$.}
\item \text{\textit{Step} 4. Define BF evidence}
\item \text{Choose two positive numbers $a$ and $b$, (with $a = 1$ and $b = 3$,  corresponding to a Bayes factor equal to 3)}
\If {$\pi{^\star}<\frac{a}{a+b}$} 
\State ${\cal M}_0$ is rejected and $j^{th}$  predictor is excluded from the model
 \Else{
 \State the $j^{th}$ predictor  is  not excluded from the model.}
\EndIf
\\\hrulefill
\end{algorithmic}
}
\end{algorithm}

\subsubsection{Other criteria}

One procedure, due  to (\cite {Li10}), is based on   a  $ 50 \% $ credible interval. That is, if the credible interval of a given coefficient contains zero, the explanatory variable associated with it must be removed from the model. A second criterion, named scaled neighborhood (\cite{Li10}) corresponds to evaluate the posterior probability of $[-s_j, s_j]$, where $s_j^2= var(\beta_j|y)$ and decide the exclusion of a predictor if this probability exceeds a certain threshold, for instance \cite{Li10} suggests $ {1}/{2} $ as this limit.

\section{Knots Selection in Regression Spline}\label{sec:regsplines}

\begin{figure}[h!]
\begin{center}
{\includegraphics[scale=0.3]{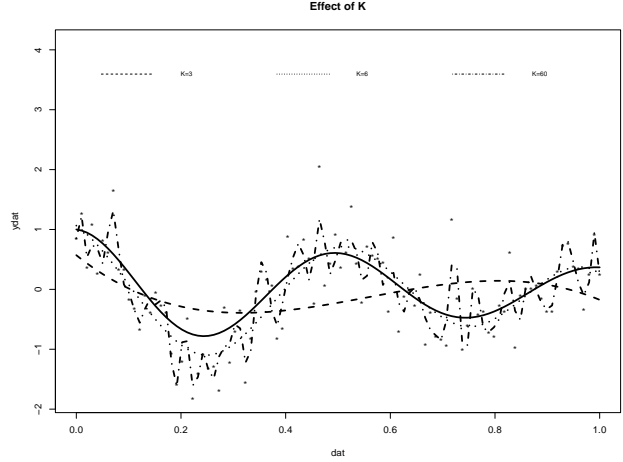}}
\end{center}
\caption{Large values of K causes over-fitting.}\label{fig:K-effect}
\end{figure}

We start with the standard setup for nonparametric regression
models. Let's suppose we have a collection of observations $(y_i,x_i)$ for
$i=1,\ldots, n$ such
that
\begin{eqnarray}\label{reg_spline1}
 y_i = f(x_i) + \epsilon_i ,
 \end{eqnarray}
where $f(x_i) = E[y_i|x_i]$ are the values obtained by a unknown
smooth function $f$ that takes values on the interval $[a,b] \subset
\R $ and  $\epsilon_i$ is a sequence of random variables 
that are uncorrelated with mean zero and unknown precision $\phi$. A possible
approach to estimate $f$ is to assume that the regression curve $f$
can be well approximated 
by a spline function. See details in \cite{dias:1999a}. That this, given a
sequence of knots $\bfkappa = (\kappa_1, \ldots, \kappa_K)$ so that
$\kappa_1 < \kappa_2 \ldots < \kappa_{K-1} <  \kappa_K $,  a spline regression
model can be written as:
\begin{equation}\label{reg_spline2}
    f(x, \bfbeta) = \beta_0 + \sum_{j=1}^{p} \beta_j x^{j} +  \sum_{k=1}^{K} \beta_{p+k}
(x-\kappa_k)^p_{+} ,
\end{equation}
 where $p$ is the degree of the polynomial
spline and $\bfbeta$ is the vector of coefficients of dimension $K+p+1$. The functions
$(u)_{+} $ are the well known truncated power basis,  $(u)^p_{+}
= \max(0,u^p)$. Note that, for a fixed $K$, the vector of 
knots $\bfkappa$ and the set of basis functions
$\{1,x,x^2,\ldots, x^p, (x-\kappa_1)^p_{+}, \ldots
(x-\kappa_K)^p_{+} \}$, an estimate of $f$, say
$\hat{f}$, can be obtained by estimating the vector $\bfbeta$. Such
that,
$$ \hat{f} = f(x,\widehat{\bfbeta}) = \hat{\beta}_0+ \sum_{j=1}^{p}
\hat{\beta}_j x^{j} +  \sum_{k=1}^{K} \hat{\beta}_{p+k}(x-\kappa_k)^p_{+}
.$$ 
It's well known (\cite{dias:1998a}, \cite{dias:game:2002}, \cite{dias:garcia:2007}, \cite{koop:ston:1991}) when $K$ increases the
bias gets smaller causing over-fitting but at the same time
substantially increases the variance. On the other hand, if $K$ goes to
zero then variance might drastically be reduced causing under-fitting
and  considerably increases bias. Thus, $K$ acts as the smoothing
parameter in regression spline fit and hence it balances the trade-off
between over-fitting and under-fitting. A good procedure should not only
provide an ideal number of knots (or basis functions) but also quantify the uncertainty of
adding or removing them.
Figure ~\ref{fig:K-effect} shows the
effect of different values of $K$ for a spline regression model.

There are other basis functions that can represent a regression
function such as B-splines, wavelets radial basis etc. For all, it is still
necessary to balance between under-fitting and over-fitting. Even in the
case of smoothing splines, the regularization parameter needs to be obtained. Specifically, in this work we are dealing with the following optimization problem:
Find $\widehat{\bfbeta}(\lambda)$ the minimizer of 
\begin{equation}\label{eq:plsregre}
    \sum_{i=1}^{n} (y_i - f(x_i,\bfbeta))^2 + \lambda \sum_{k=1}^{K} |\beta_{p+k}|,
\end{equation}
where $\lambda$ is the smoothing parameter. For large values of $\lambda$ the solution of this optimization problem tends to the polynomial regression fit, that is over-fitting. Note that the penalty term involves only the coefficients associated to the knots sequence $\kappa_1 < \kappa_2 \ldots <  \kappa_K $. Consequently, selecting knots is equivalent to selecting the coefficients that  contribute most  to the fitting. Under the Bayesian point of view, this work presents a novel and scalable procedure for selecting knots. 

\subsection{The variational inference for knots selection}


Following the idea of the Lasso procedure for variable selection, under the Bayesian point of view, the selection of knots in the spline regression models can be made by assuming an independent Laplace prior distribution for the coefficients associated with the knots. We will  denote $\bfbeta = \left({\bfbeta^{(1)}}^T,{\bfbeta^{(2)}}^T\right)^T$ where $\bfbeta^{(1)} = (\beta_0, \beta_1, \ldots, \beta_p)^T$ is the polynomial coefficients with dimension $p+1$ and $\bfbeta^{(2)} = (\beta_{p+1}, \ldots, \beta_{p+K})^T$ of dimension $K$ is the penalised coefficients. The hierarchical structure presented in the subsection \ref{subsec:Hierarchical_laplace} is maintained and in this way we complete the model defined by the equations (\ref{reg_spline1}) and (\ref{reg_spline2}):

\begin{eqnarray}\label{prior_spline}
{\bfbeta^{(1)}}^T &\sim& N(m_0, C_0)\nonumber\\
{\bfbeta^{(2)}}^T|\phi, \bftau &\sim& N(0, \phi^{-1} D_{\tau}) \nonumber\\
\tau_j|\lambda  &\sim& Exp(\lambda), \;\;\; j=1, \ldots, K \nonumber \\ 
\phi &\sim& Ga(a_0, b_0)\nonumber \\
\lambda &\sim& Ga(g_0, h_0)
\end{eqnarray}

The Bayesian inference procedure in this case must be carried out with caution since the vector of coefficients contains the coefficients of the polynomial, which will not be penalized, and the coefficients of the basis functions  to which we assume  a prior Lasso distribution for knots selection. The design matrix is then partitioned  $X = (X_1,X_2)$ with $X_1$ of dimension $(n\times p+1)$ and $X_2$ $(n\times K)$. Then, the i-th row of the matrix  $X$ is given by:
$$X_i = \{ \underbrace{1,x_i,x_i^2,\ldots, x_i^p}_{X_{1i}}, \underbrace{(x_i-\kappa_1)^p_{+}, \ldots (x_i-\kappa_K)^p_{+} }_{X_{2i}} \}.$$
and $X\bfbeta = X_1 \bfbeta^{(1)} + X_2 \bfbeta^{(2)}.$


In this context, we have a prior distribution of the vector $\bftheta = (\beta^{(1)}, \beta^{(2)}, \phi, \tau, \lambda)$ and the posterior distribution is given by:
$$ p(\bftheta|y) \propto p( y | X, \bfbeta, \phi) \,\, p(\bfbeta^{(1)}) \,\, p(\bfbeta^{(2)} | \phi, \bftau) \,\,  p(\bftau|\lambda) \, \,  p(\phi) \, \, p(\lambda).$$
Slight adaptations need to be made in the variational inference method and the main one refers to the fact that we now have 4 partitions of the parametric vector giving rise to the following variational densities:
$$\log(q(\bftheta)) = \log(q_1(\beta^{(2)},\phi)) + \log(q_2(\tau)) + \log(q_3(\lambda)) + \log(q_4(\beta^{(1)}))$$
where $\log q_4(\beta^{(1)}) = \log N(\beta^{(1)}|m_{\beta_1}, C_{\beta_1})$. The calculations for this and other variational densities are similar to those developed for the regression model in subsection \ref{subsec:VBLasso} and can be found in the appendix.


\subsection{Algorithm for automatic knot selection}
An alternative approach to determine the maximum number of knots $K$ is to consider it as an unknown parameter and estimate it. However, as in this work, $ K $ is not a parameter of direct interest since the selection of knots is carried out through the Lasso scheme. Despite this, in order to have a good fit, it is important to properly define the number of knots and their positions.
In section \ref{section:app} presents some exercises involving selecting knots and  shows the importance of the correct specification of the value of $K$. Thus, an algorithm will be proposed for the automatic choice of the number of knots $K$ that is based both on the VB algorithm for estimating the Lasso model and on the criterion for selecting variables (knots).


The algorithm starts by proposing a grid of possible values of $ K $ the maximum number knots. Naturally, the grid of values is an issue to be discussed. Note that it is not necessary to propose a grid that covers all the natural numbers, since computational time can be excessively high. Moreover, given a maximum number of knots, the most significant knots will be selected. For instance, start the grid with the maximum number of knots $K=20$. By using Lasso and the selection criteria, it is possible to have  6  among these 20 knots as the most significant ones. On the other hand, simulations show that starting with a very large maximum number of knots may cause problems in the selection, since the penalty criteria acts  more severely when the number of knots is extremely big for the size of the data set. See numerical simulations in Section \ref{section:app}. Therefore, our proposal is to provide a grid that increases 10 units at a time, so that the knots are placed in the quantiles or evenly spaced in the explanatory variable domain. This spacing allows us to position knots  in different locations before and after selecting significant knots. In the simulated exercises presented in Section \ref{section:app}, the grid starts with $K=10$ knots.

In summary, ELBO is taken as a stopping criterion and the objective is to maximize it. The algorithm consists of: for fixed grid values, start with the lowest value. The model is adjusted via VB together with the selection criteria and then calculates the ELBO. Move to the next grid value and repeat the procedure. As long as the ELBO increases with the grid values, the algorithm continues. The detailed algorithm is given below:

\begin{algorithm}\label{Alg:MAxNos}
\caption{Maximum Number of Knots }
\footnotesize{
\begin{algorithmic}
\\\hrulefill

\State \text{\textit{Step} 1. $j \gets 1$. Initialize $K_j=10$.}
\State \text{\textit{Step} 2.  Fit  model via VB algorithm.}
\State \text{\textit{Step} 3. Compute ELBO.}
\State \text{\textit{Step} 4. Apply BF to select the most significant knots.}
\State \text{\textit{Step} 5. $K_{j+1}=K_j+10$ and repeat steps 2, 3 and 4.}

\If{ELBO($K_{j+1}$) $\geq$ ELBO($K_j$)} 
\State \text{ Set $K_j \gets K_{j+1}$. Repeat steps 2 to 5.}
\EndIf
\If{ELBO($K_{j+1}$) $<$ ELBO($K_j$)}
\State \text{ Stop and deliver ELBO and the most significant knots}
\EndIf
\\\hrulefill
\end{algorithmic}
}
\end{algorithm}



\section{Simulations Studies} \label{section:app}


This section proposes 5 exercises with artificial data. The first three based on Lasso for linear regression models and the last two focused on the use of Lasso to select knots in spline regression.


The inference procedure assumes, for all exercises, the following  prior distributions: $\phi \sim Ga(0.1,0.1)$, $\lambda \sim Ga(0.1,0.1)$ e $\bfbeta_1 \sim N(1,100)$ (in case of regression splines).
For MCMC, 15,000 iterations were necessary to achieve convergence. The first 5,000 iterations were discarded as the burn in process and one observation was taken for every ten observations to remove autocorrelation, ending with a sample size of 10000.
These quantities were obtained by using the criterion found in \cite{RafteryLewis97}, that provides the number of iterations needed to guarantee the convergence in the Gibbs Sampler. The VB algorithm is repeated until the changes in $ m_ \beta $, $ C_ \beta $, $ b_ \phi $, $ d_ \tau $, $ f _ {\tau_j} $ and $ h_ \lambda $ between two consecutive iterations are less than $ 0.01 \% $.
When applied, the classic procedure was implemented using the R software glmnet package, which in turn applies 5-fold cross-validation to estimate the penalty parameter $\lambda$.

\subsection{Variable Selection for Linear Regression}

The goal of these exercises applied to the linear regression models is twofold. Firstly to compare the estimation methods VB and MCMC (eventually we also use the classic Lasso in the comparison). Secondly, we wish to compare the CI, SN and BF selection criteria. Moreover, different sparsity scenarios, variations in the sample size, different correlations between explanatory variables and different values for the accuracy of the model are considered.


Specifically, exercise 1 aims to estimate Lasso hyperparameters via VB and MCMC. Only one data set is simulated from which the real values of all parameters and hyperparameters of the model are known. VB presents results similar to MCMC and computational time 14 times shorter.
Exercise 2 is based on a simulation study with 100 replicates that presents a lesser sparsity structure. Variations in the sample size and in the correlation among the explanatory variables are considered. Again, VB and MCMC present similar and superior results to the classic Lasso. When the CI, SN and BF selection criteria are compared,  BF gives the best results, with high proportions of exclusion for coefficients that are zero and low exclusion proportions for coefficients that are different from zero.
Exercise 3 is designed for scenarios with 100 replicates and with greater sparsity when compared to exercise 2. This exercise takes into account cases where $n<p$ and different values for the model's precision. The results are similar of those obtained in simulation 2

\subsubsection{Exercise 1: MCMC vs. VB}
The purpose of simulation 1 is to compare the MCMC and VB methods to curve fitting and computational time. For this study we considered $ n = 100 $, $ p = 10 $ and each column of the matrix $X$ was generated from a distribution $ N ({\bf {0}}, I_n) $. For the parameters, were taken $ \phi = 0.4 $, $ \lambda = 5 $ and $ \tau_j | \lambda \sim Exp (\lambda)$, \; $ \forall j $. The regression coefficients and observations were generated considering the Lasso regression model.

\begin{figure}[h!]
\begin{center}
\begin{tabular}{cc}
{\includegraphics[scale=0.3]{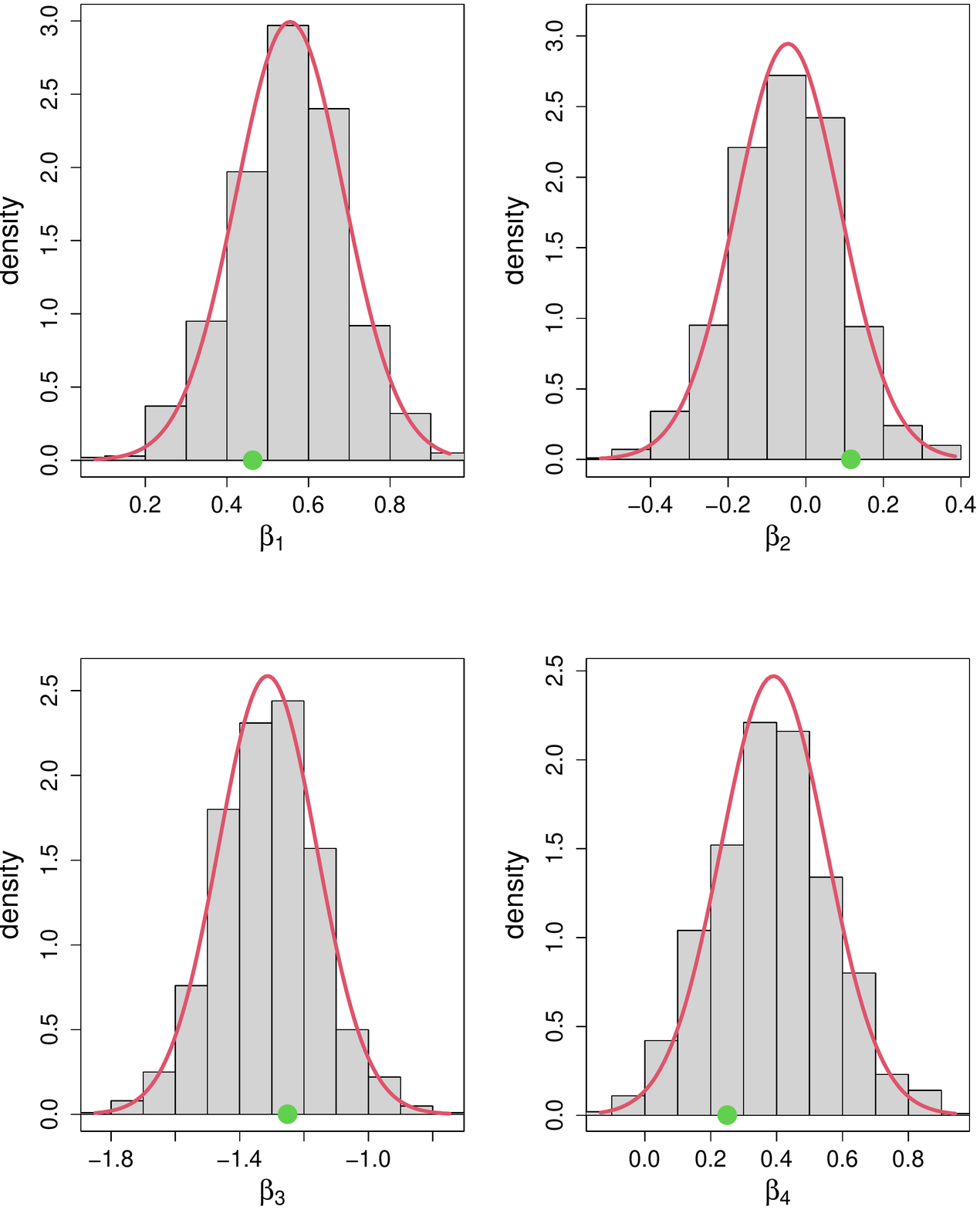}}&
{\includegraphics[scale=0.3]{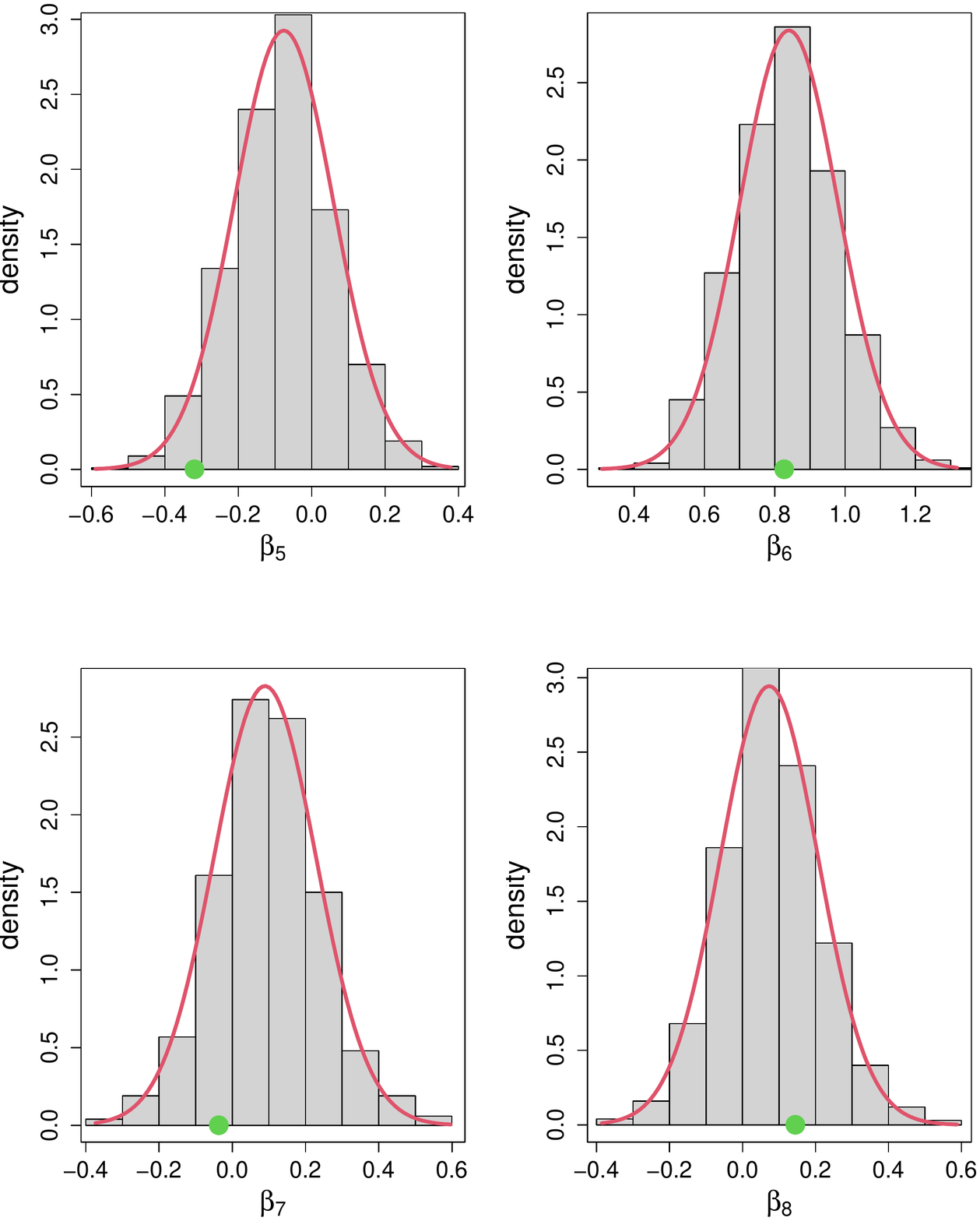}}\\
{\includegraphics[scale=0.3]{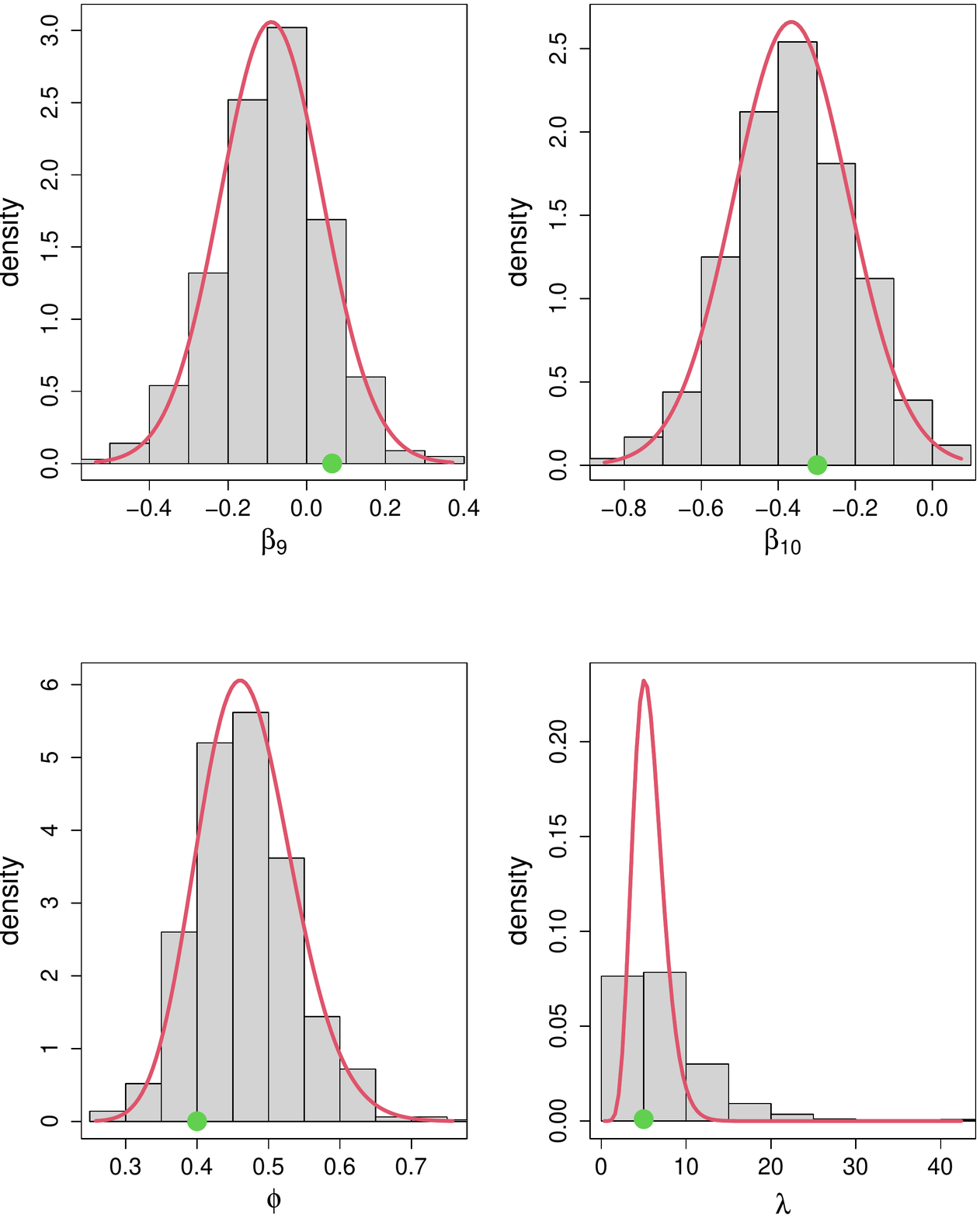}}&
{\includegraphics[scale=0.3]{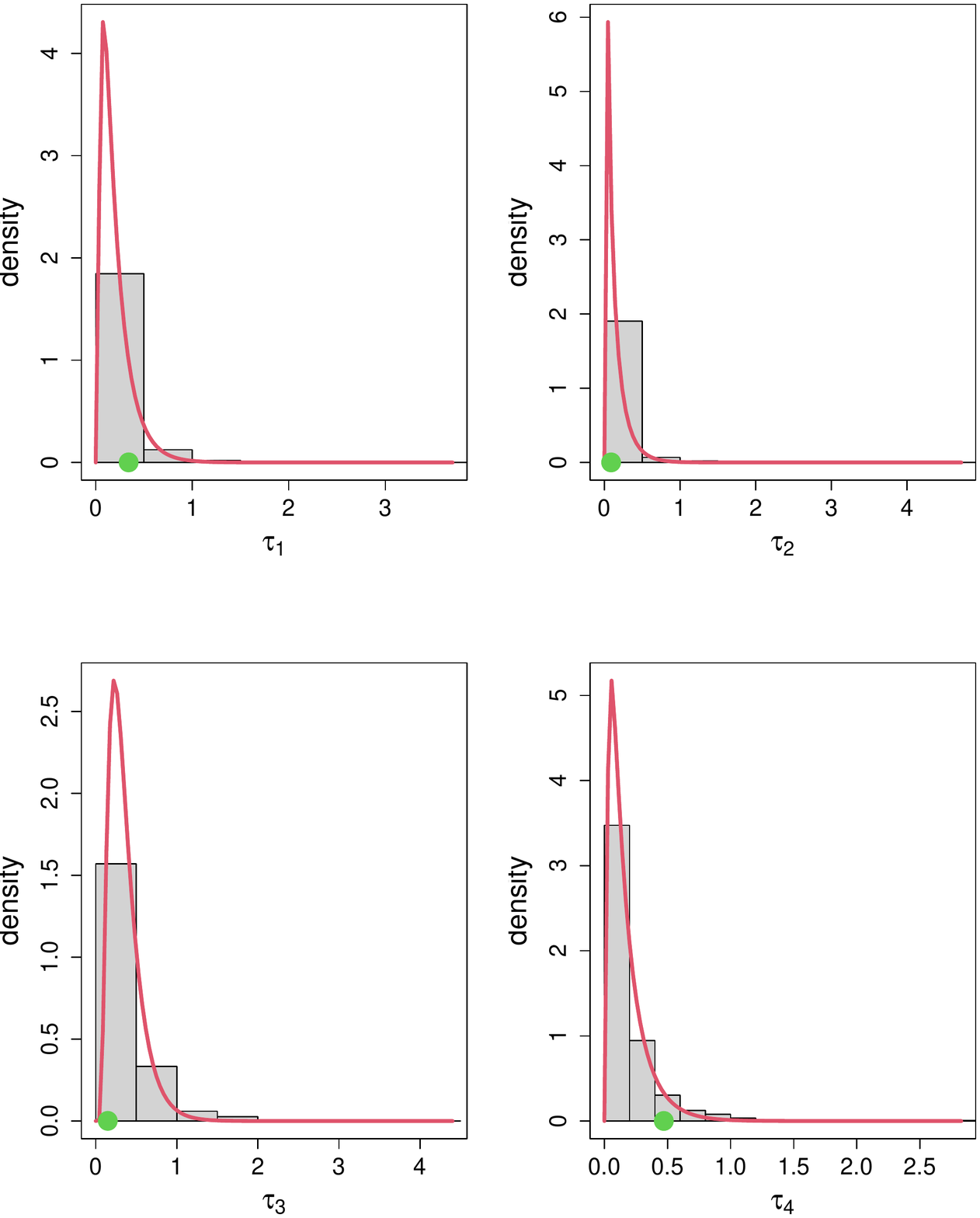}}\\
\end{tabular}
\end{center}\vspace{-0.5cm}
\caption{Comparison  MCMC (histogram) versus  VB (solid line). The dot marks the actual value of the parameter used to generate the data.}.\label{fig1:MCMCvsVB}
\end{figure}

Table \ref{table:postsumm} shows us a posterior summary of the model parameters. There, one can see the mean and the standard deviation of the approximate posterior obtained by using VB. Also, the posterior mean, the posterior standard deviation  via MCMC and the true value of the parameters. Note that the point estimates obtained by the VB are close to those obtained by the MCMC. In addition, for both methods, the results are close to the real values with small standard deviation. This same conclusion can be seen in Figure \ref{fig1:MCMCvsVB}.
In fact, Figure ~\ref{fig1:MCMCvsVB} exhibits a graphical comparison between MCMC and VB. The histogram represents the sample of the posterior distribution obtained via MCMC and the curve in red the approximate posterior density obtained by the VB. The green dot indicates the true value of the parameters. Note that the curves approximated by the VB are close to the histograms and both centered on the actual values. The remaining parameters $\tau_j$ show similar results.
\pagebreak 

\begin{table}[h!]
\caption{Posterior summary.}\label{table:postsumm} 
\begin{center}
{\footnotesize
\begin{tabular}{c|c|c|c|c|c}
  \hline
 Parameters & Real & Mean VB & Sd VB & Mean MCMC & Sd MCMC \\
\hline
$\beta_1$    & 0.463      & 0.557  & 0.132 & 0.558  & 0.139 \\
$\beta_2$    & 0.116      & -0.046 & 0.136 & -0.048 & 0.138 \\ 
$\beta_3$    & -1.251     & -1.316 & 0.153 & -1.315 & 0.160 \\
$\beta_4$    & 0.250      & 0.396  & 0.161 & 0.383  & 0.171 \\
$\beta_5$    & -0.319     & -0.078 & 0.137 & -0.079 & 0.142 \\
$\beta_6$    & 0.826      & 0.844  & 0.140 & 0.845  & 0.148 \\
$\beta_7$    & -0.036     & 0.091  & 0.142 & 0.096  & 0.149 \\
$\beta_8$    & 0.144      & 0.074  & 0.136 & 0.081  & 0.143 \\
$\beta_9$    & 0.064      & -0.090 & 0.131 & -0.081 & 0.134 \\
$\beta_{10}$ & -0.298     & -0.370 & 0.150 & -0.369 & 0.163 \\
$\phi$       & 0.4        & 0.473  & 0.066 & 0.473  & 0.070 \\
$\tau_1$     & 0.340      & 0.233  & 0.188 & 0.265  & 0.326 \\
$\tau_2$     & 0.088      & 0.137  & 0.159 & 0.162  & 0.250 \\
$\tau_3$     & 0.148      & 0.401  & 0.231 & 0.453  & 0.398 \\
$\tau_4$     & 0.470      & 0.201  & 0.179 & 0.234  & 0.359 \\
$\tau_5$     & 0.048      & 0.140  & 0.160 & 0.152  & 0.209 \\
$\tau_6$     & 0.162      & 0.296  & 0.205 & 0.345  & 0.333 \\
$\tau_7$     & 0.120      & 0.142  & 0.161 & 0.164  & 0.266 \\
$\tau_8$     & 0.069      & 0.139  & 0.160 & 0.173  & 0.365 \\
$\tau_9$     & 0.027      & 0.140  & 0.161 & 0.148  & 0.244 \\
$\tau_{10}$  & 0.275      & 0.194  & 0.177 & 0.212  & 0.278 \\
$\lambda$    & 5          & 4.745  & 1.493 & 5.600  & 3.776 \\
  \hline
\end{tabular}}
\end{center}
\end{table}



Since MCMC and VB present similar results, it is worth to point out the main difference between these estimation methods, which is computational time. For exercise 1, the computational time of the VB was 0.72 seconds while that of the MCMC was 10.15 seconds. In the following  exercises these computational times become even more discrepant as we will be dealing with simulations with replicates.
  
\subsubsection{Exercise 2: High correlation}


In this exercise a simulation was developed based on 100 replicates $p = 8$, $\bfbeta = (3, 1.5, 0, 0, 2, 0, 0, 0)^T$ and the design matrix is generated from a multivariate normal distribution with zero mean, variance 1 and two different correlation structures between $x_i$ e $x_j$: 0 e $0.7^{|i-j|}$, $\forall i$ e $j$. Let's consider $\phi = 1/9$ and 3 nested scenarios varying the sample size with $\{n_T,n_V\} = \{20,10\}, \{100,50\}$ e $\{200,100\}$, where $n_T$ e $n_V$ denote the size of the training set and the size of the validation set, respectively. Therefore, we have a total of 6 different scenarios. Note that the explanatory variables are standardized to have mean 0 and variance 1

A Tabela \ref{Cenarios_Sim2} summarize the results of exercise 2.

\begin{table}[h!]
\caption{Simulation 2 with 100 replicates, $p = 8$ explanatory variables and the vector of coefficients $\bfbeta = (3, 1.5, 0, 0, 2, 0, 0, 0)^T$.}\label{Cenarios_Sim2} 
\begin{center}
{\footnotesize
\begin{tabular}{cccc}
  \hline
 Simulation & $n_T$ & $n_V$ & $cov(X_i,X_j)$ \\
\hline
S2.1     & 20  & 10   & 0 \\
S2.2     & 100 & 50   & $-$ \\ 
S2.3     & 200 & 100  & $-$ \\
S2.4     & 20  & 10   & $0.7^{|i-j|}$ \\
S2.5     & 100 & 50   & $-$ \\
S2.6     & 200 & 100  & $-$ \\
\hline
\end{tabular}}
\end{center}
\end{table} 

The comparison of the MCMC and VB methods is our main objective in this simulation, however, frequentist  Lasso is also considered through the glmnet package of the R software.
For the frequentist Lasso, a 5-fold cross-validation is used to select the parameter $ \lambda $. In addition, different variable selection criteria will be compared as described in \ref{section:app}: credible interval (CI), scaled neighborhood (SN) and Bayes factor (BF).


In order to compare Lasso's predictive power from the different estimation techniques, MCMC, VB and  frequentist Lasso, the mean absolute error (MAE) was calculated for each replicate of the validation set using the following expression:


\begin{eqnarray}
MAE = \frac{1}{n_V} \sum_{i=1}^{n_V} |y_i^P - y_i^V|
\end{eqnarray}\label{EAM}
where $y_i^P$ are the predicted values in the validation set, obtained from the fitted model after the selection of the coefficients. $y_i^V$ are the observed values in the validation set and $n_V$ is the size of the validation set.
Note that  MCMC generates  a sample of the predictive distribution from each iteration of the method. Then, $y_i^P$ is obtained as follows:
$$p(y_i^P|{\bf{y}}) = \frac{1}{AM} \sum_{j=1}^{AM} p(y_i^P|\bftheta^{(j)}),$$
where $AM$ is MCMC number of iterations  and $\bftheta$ is the vector of coefficients.

\begin{figure}[h!]
\begin{center}
\begin{tabular}{ccc}
{\includegraphics[scale=0.23]{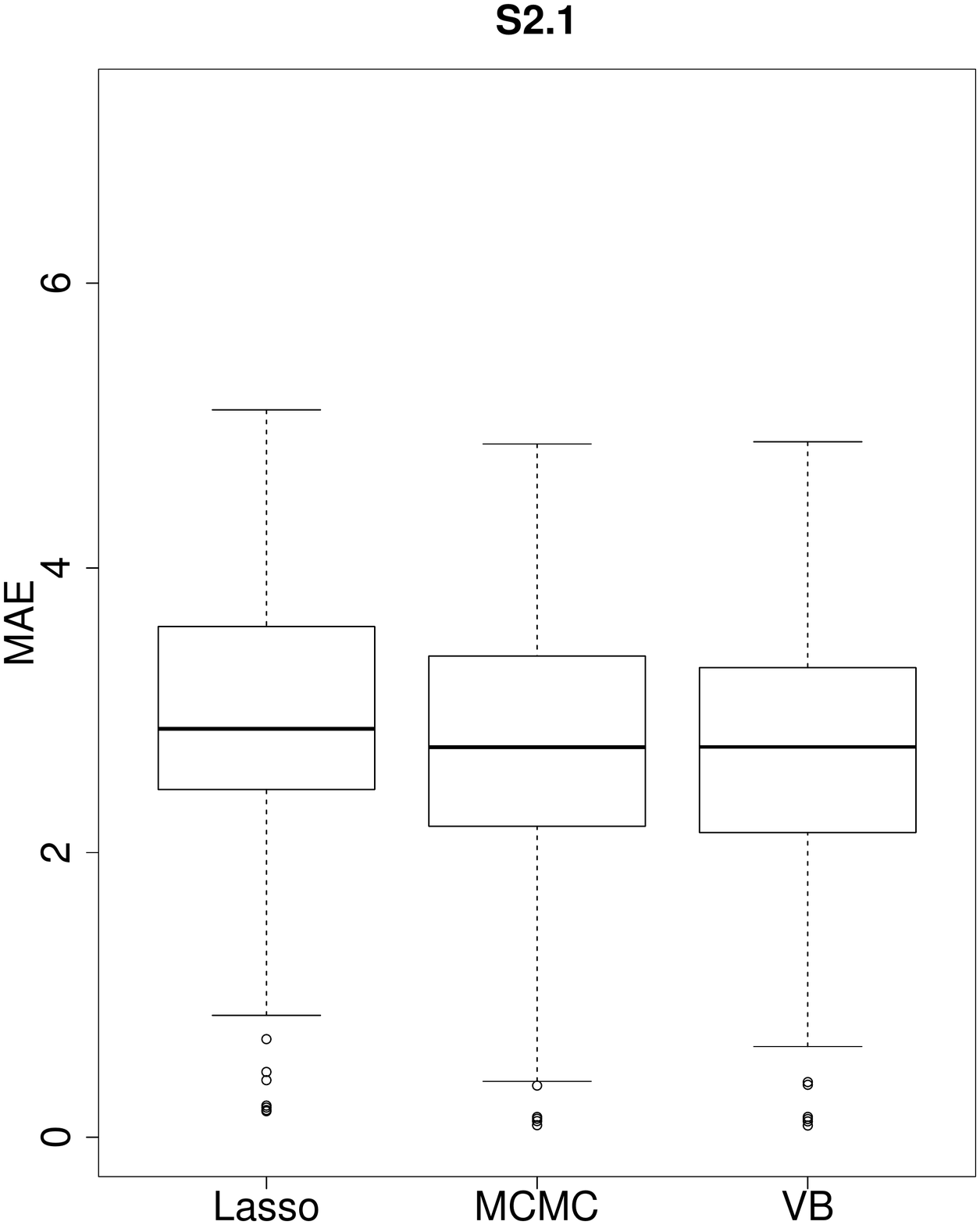}}&
{\includegraphics[scale=0.23]{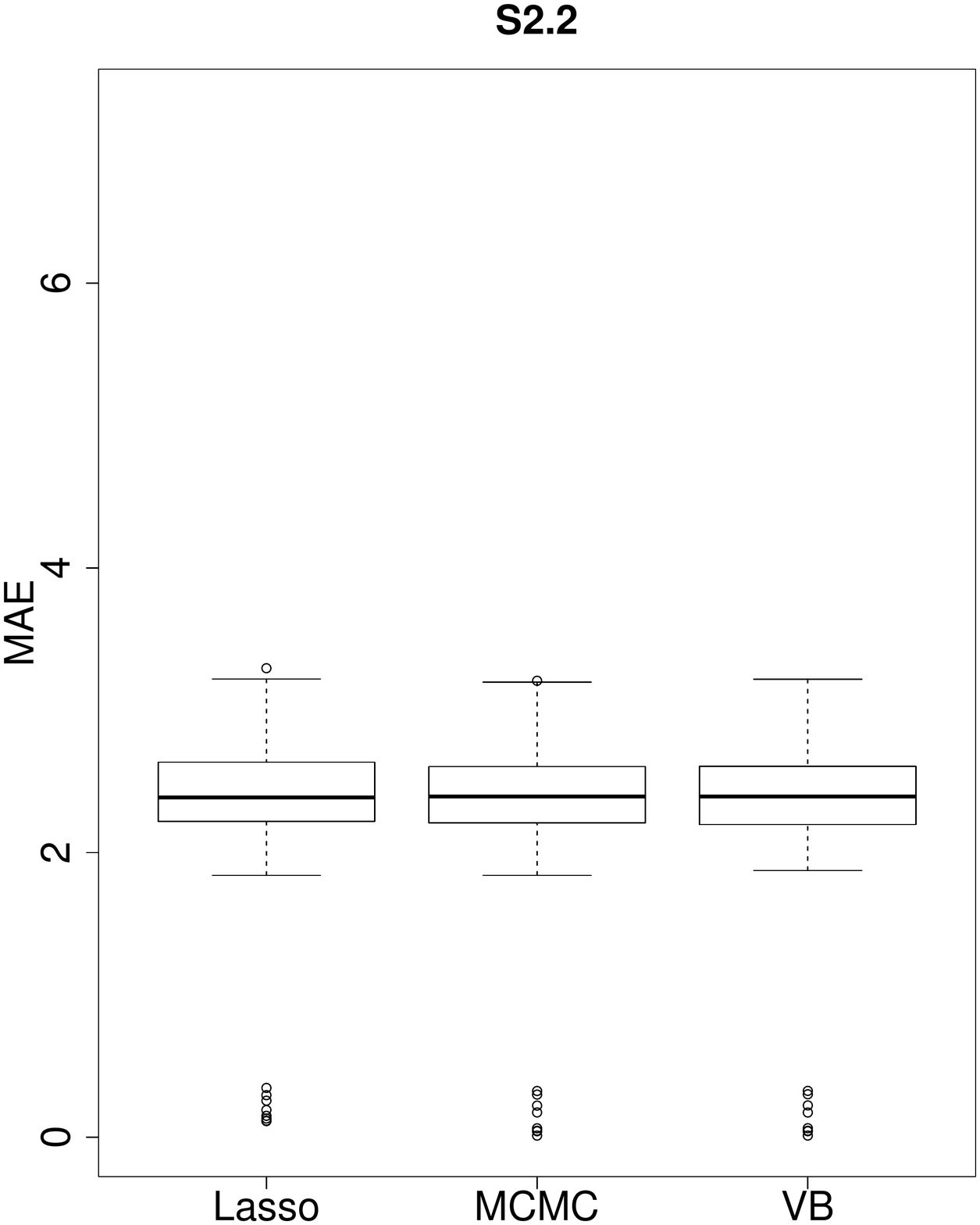}}&
{\includegraphics[scale=0.23]{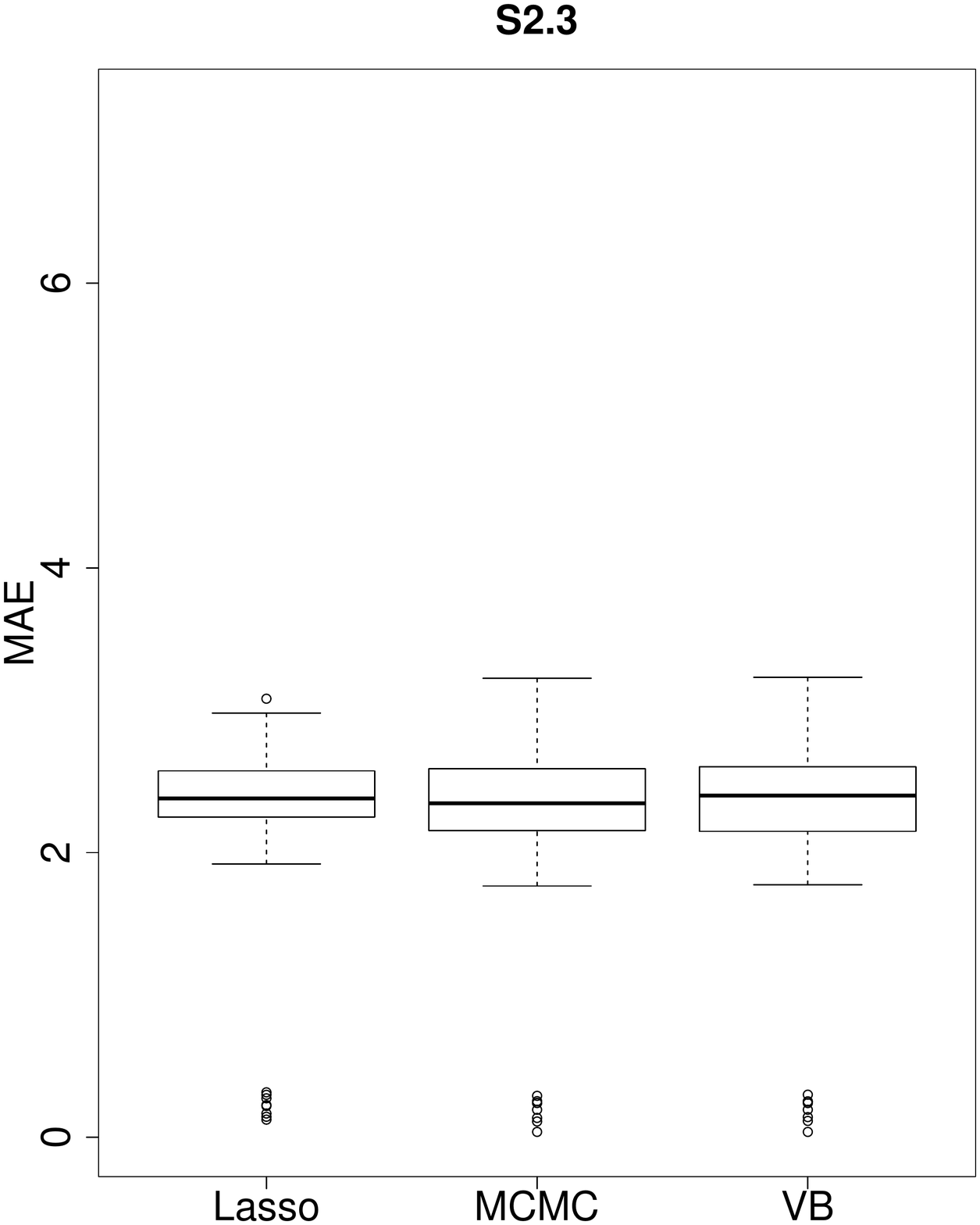}}\\
{\includegraphics[scale=0.23]{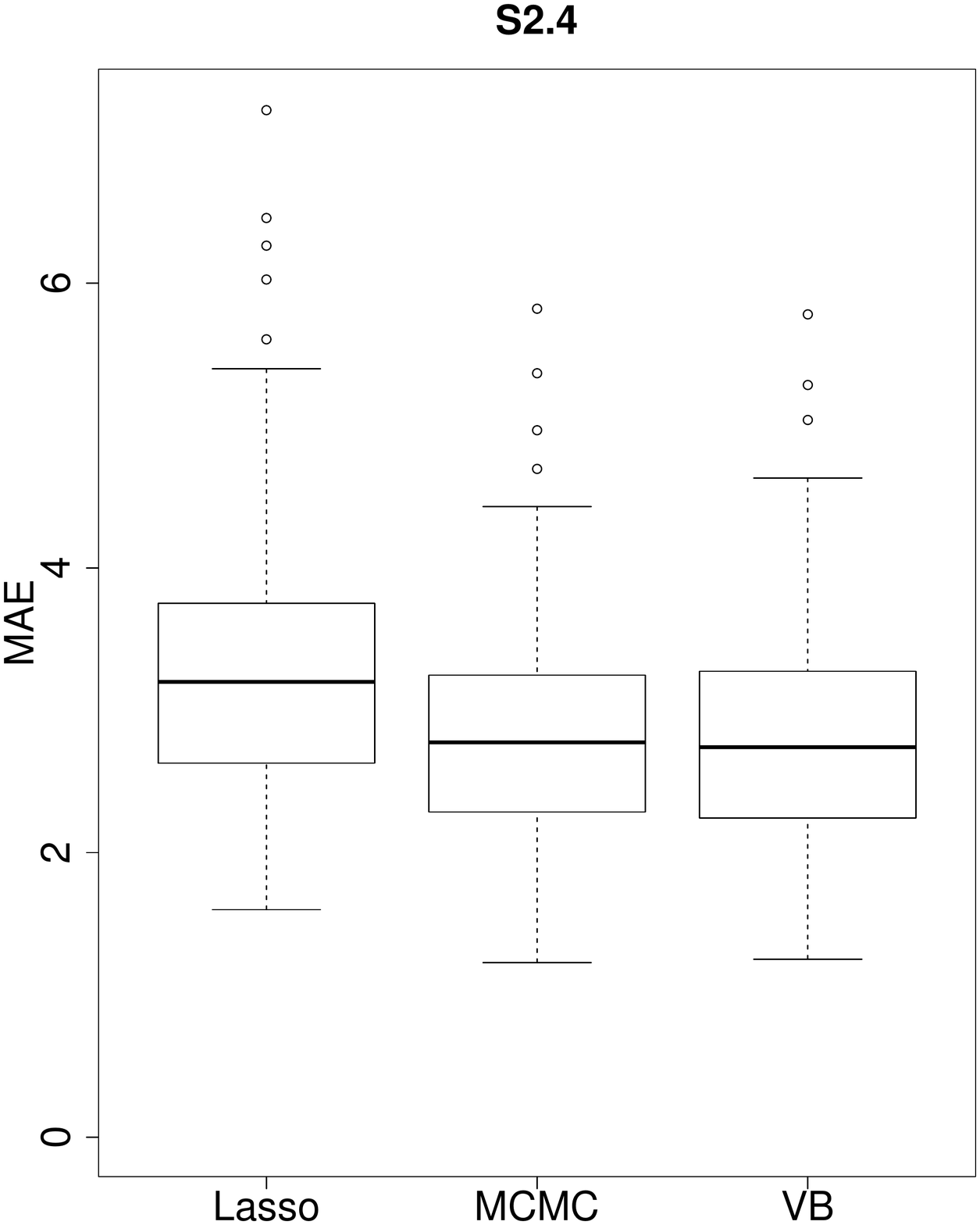}}&
{\includegraphics[scale=0.23]{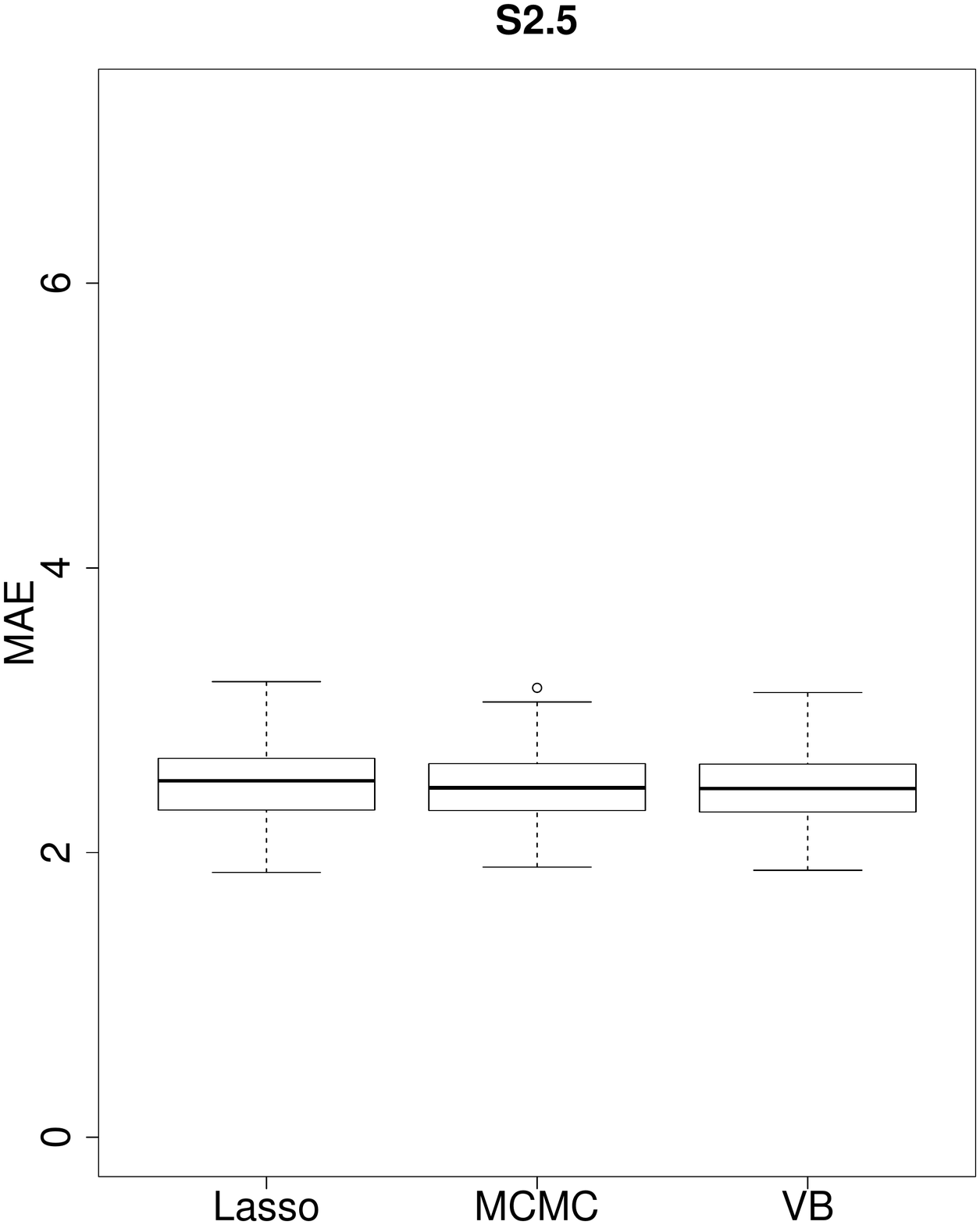}}&
{\includegraphics[scale=0.23]{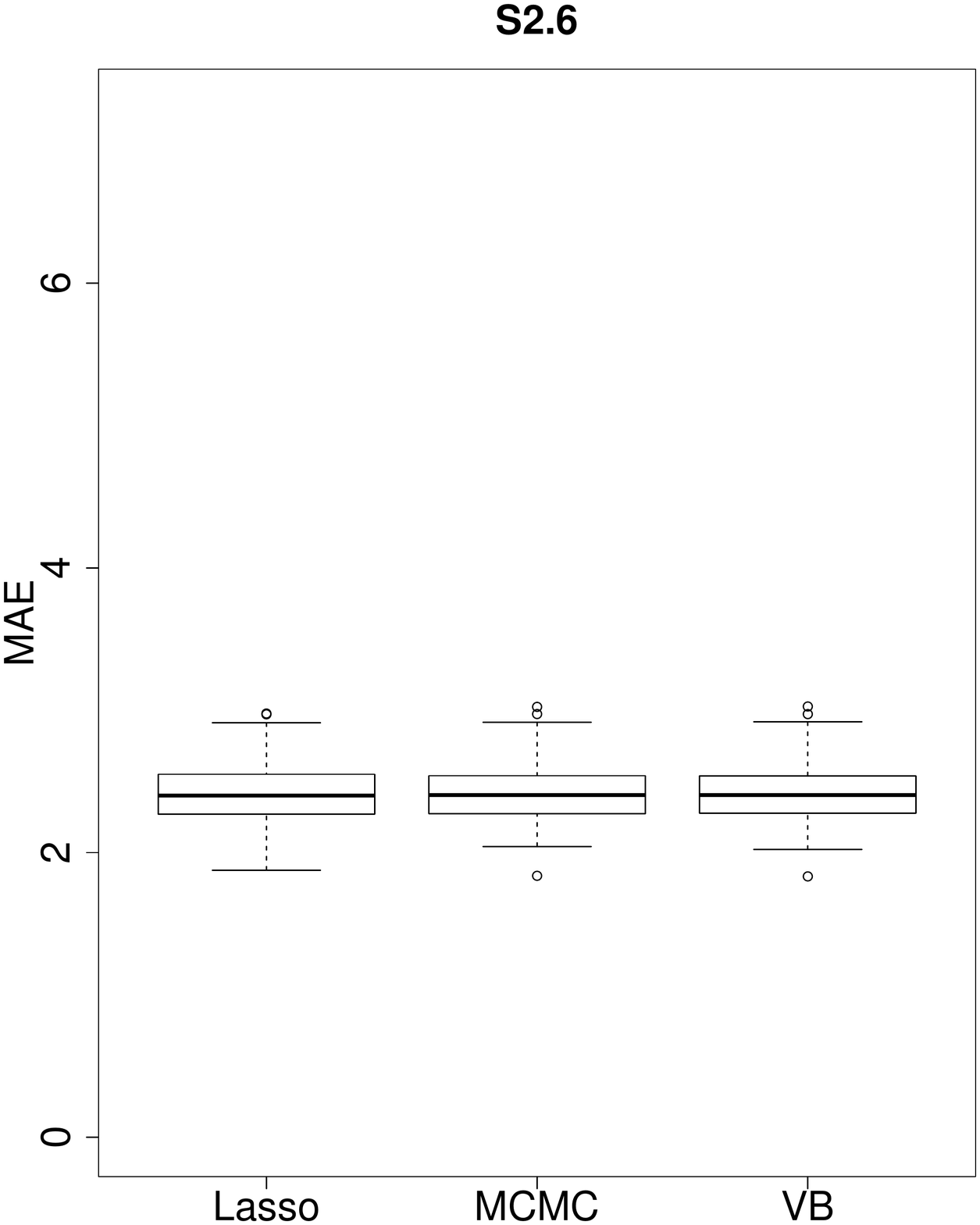}}\\
\end{tabular}
\end{center}\vspace{-0.5cm}
\caption{ Mean absolute error (MAE) using (\ref{EAM}) for the 6 scenarios of the simulation 2.  Estimation methods: MCMC, VB and frequentist Lasso.}\label{fig:EAM}
\end{figure}

Figure \ref{fig:EAM} shows the box-plots of the mean absolute errors for each of the six proposed scenarios. As the sample size increases, we observe a smaller difference between the three estimation methods. When the sample is small, similar results are obtained between MCMC and VB. These have the median MAE and the lowest dispersion when compared to the frequentist Lasso. Next, we will detail the performance of the selection criteria for each $ \beta_j $.



The Table \ref{tab:selecao} shows the frequency of times that the predictor $ x_j $, $ j = 1, \ldots, 8 $ was excluded in the 100 replicates, considering the three variables selection methods and all six scenarios built in Simulation 2. We present the proportions only for the VB because so far its results are similar to those of the MCMC. Note that for this simulation exercise the BF presents the best results in all scenarios, with a greater proportion of exclusion when the actual values of $ \beta_j $ are zero and a small proportion when the $ \beta $ 's are different from zero. In addition, it is noted that as the sample size increases, the three criteria tend to correctly choose  coefficients that are zero and the coefficients that are different from zero.
From exercises 1 and 2, one may notice that the approximations of the VB are as good as the results obtained by the MCMC. Nevertheless, the gain in computational time provided by VB is far superior than MCMC. In addition, we saw that BF is a variable selection criterion that presents superior results when compared with CI and SN.
In the following subsection we show the performance of the VB estimation method and the BF selection criterion for a more complex numerical experiment with greater sparsity.

\begin{table}[h!]
\caption{Comparison of the three methods on variable selection accuracy using VB for the six scenarios (the frequency of exclusions for the predictor $x_j$, $j = 1, \ldots, 8$) with $\bfbeta = (3, 1.5, 0, 0, 2, 0, 0, 0)^T$.}\label{tab:selecao} 
\begin{center}
{\footnotesize
\begin{tabular}{p{2.5cm}|c|cccccccc}
\hline
\centerline{Simulation} & Method & $\beta_1$ & $\beta_2$ & $\beta_3$ & $\beta_4$ & $\beta_5$ & $\beta_6$ & $\beta_7$ & $\beta_8$ \\
\hline\hline \multirow{3}{*}{\centerline{S2.1}} &
VB + CI & 0.01 & 0.09 & 0.64  & 0.57 & 0.13 & 0.66 & 0.65 & 0.62\\
& VB + SN & 0.02 & 0.15 & 0.73  & 0.71 & 0.20 & 0.77 & 0.75 & 0.71\\ 
& VB + BF & 0.00 & 0.09 & 0.88 & 0.70 & 0.13 & 0.82 & 0.75 & 0.92 \\
\hline \hline \multirow{3}{*}{\centerline{S2.2}} &
VB + CI & 0.00 & 0.00 & 0.47 & 0.51 & 0.00 & 0.57 & 0.67 & 0.56 \\
& VB + SN & 0.00 & 0.00 & 0.70 & 0.62 & 0.00 & 0.71 & 0.78 & 0.72 \\ 
& VB + BF & 0.00 & 0.00 & 0.73 & 0.78 & 0.00 & 0.73 & 0.84 & 0.72 \\
\hline\hline \multirow{3}{*}{\centerline{S2.3}} &
VB + CI & 0.00 & 0.00 & 0.50 & 0.47 & 0.00 & 0.67 & 0.60 & 0.58 \\
& VB + SN & 0.00 & 0.00 & 0.71 & 0.62 & 0.00 & 0.71 & 0.73 & 0.76 \\ 
& VB + BF & 0.00 & 0.00 & 0.72 & 0.73 & 0.00 & 0.77 & 0.80 & 0.77 \\
\hline\hline \multirow{3}{*}{\centerline{S2.4}} &
VB + CI & 0.02 & 0.09 & 0.53 & 0.53 & 0.19 & 0.71 & 0.60 & 0.64 \\
& VB + SN & 0.06 & 0.11 & 0.70 & 0.62 & 0.24 & 0.75 & 0.73 & 0.74 \\ 
& VB + BF & 0.02 & 0.09 & 0.65 & 0.80 & 0.18 & 0.85 & 0.81 & 0.88 \\
\hline\hline \multirow{3}{*}{\centerline{S2.5}} &
VB + CI & 0.00  & 0.00 & 0.57 & 0.49 & 0.00 & 0.60 & 0.51 & 0.57 \\
& VB + SN & 0.00  & 0.00 & 0.73 & 0.68 & 0.00 & 0.75 & 0.70 & 0.75 \\ 
& VB + BF & 0.00  & 0.00 & 0.78 & 0.77 & 0.00 & 0.76 & 0.80 & 0.82 \\
\hline\hline \multirow{3}{*}{\centerline{S2.6}} &
VB + CI & 0.00 & 0.00 & 0.55 & 0.47 & 0.00 & 0.60 & 0.46 & 0.57 \\
& VB + SN & 0.00 & 0.00 & 0.70 & 0.66 & 0.00 & 0.77 & 0.64 & 0.75 \\ 
& VB + BF & 0.00 & 0.00 & 0.77 & 0.75 & 0.00 & 0.79 & 0.72 & 0.77 \\
\hline
\end{tabular}}
\end{center}
\end{table} 

\subsubsection{Exercise 3: High sparsity with small n and large p}

\textcolor{blue}{
}

In this exercise we consider a situation with sparsity given by $p = 40$ e $\beta = ({\bf{0^T}},{\bf{3^T}},{\bf{0^T}},{\bf{3^T}})^T$, where ${\bf{0}}$ e ${\bf{3}}$ are vectors of dimension 10 and each of their entries are 0 and 3 respectively. The design matrix $X$ is generated from a multivariate normal distribution with mean zero, variance 1 and the correlation between the columns  $x_i$ e $x_j$ is equal to 0.5, $\forall i \neq j$. We analyze 4 different scenarios by varying the sample size and the precision parameter $\phi$. The simulated data were analyzed as follows, $\{n_T,n_V\} = \{20,10\}$ e $\{200,100\}$ where $n_T$ e $n_V$ are the size of the training set and the size of the validation set respectively. In addition, we set the precision parameter as  $\phi = 1/9$ and $\phi = 1/225$. For each scenario we consider 100 replicates. Table \ref{Cenarios_Sim3}  summarizes all the scenarios considered in this simulation exercise 3. It is worth mentioning that in scenarios S3.1 and S3.3 we have $n<p$

\begin{table}[h!]
\caption{Scenarios in Simulation 3}\label{Cenarios_Sim3} 
\begin{center}
{\footnotesize
\begin{tabular}{cccc}
  \hline
 Simulation & $n_T$ & $n_V$ & $\phi$ \\
\hline
S3.1   & 20  & 10  & $1/9$ \\
S3.2   & 200 & 100 & - \\ 
S3.3   & 20  & 10  & $1/225$ \\
S3.4   & 200 & 100 & - \\
\hline
\end{tabular}}
\end{center}
\end{table} 



\begin{figure}[h!]
\begin{center}
\begin{tabular}{cc}
{\includegraphics[scale=0.25]{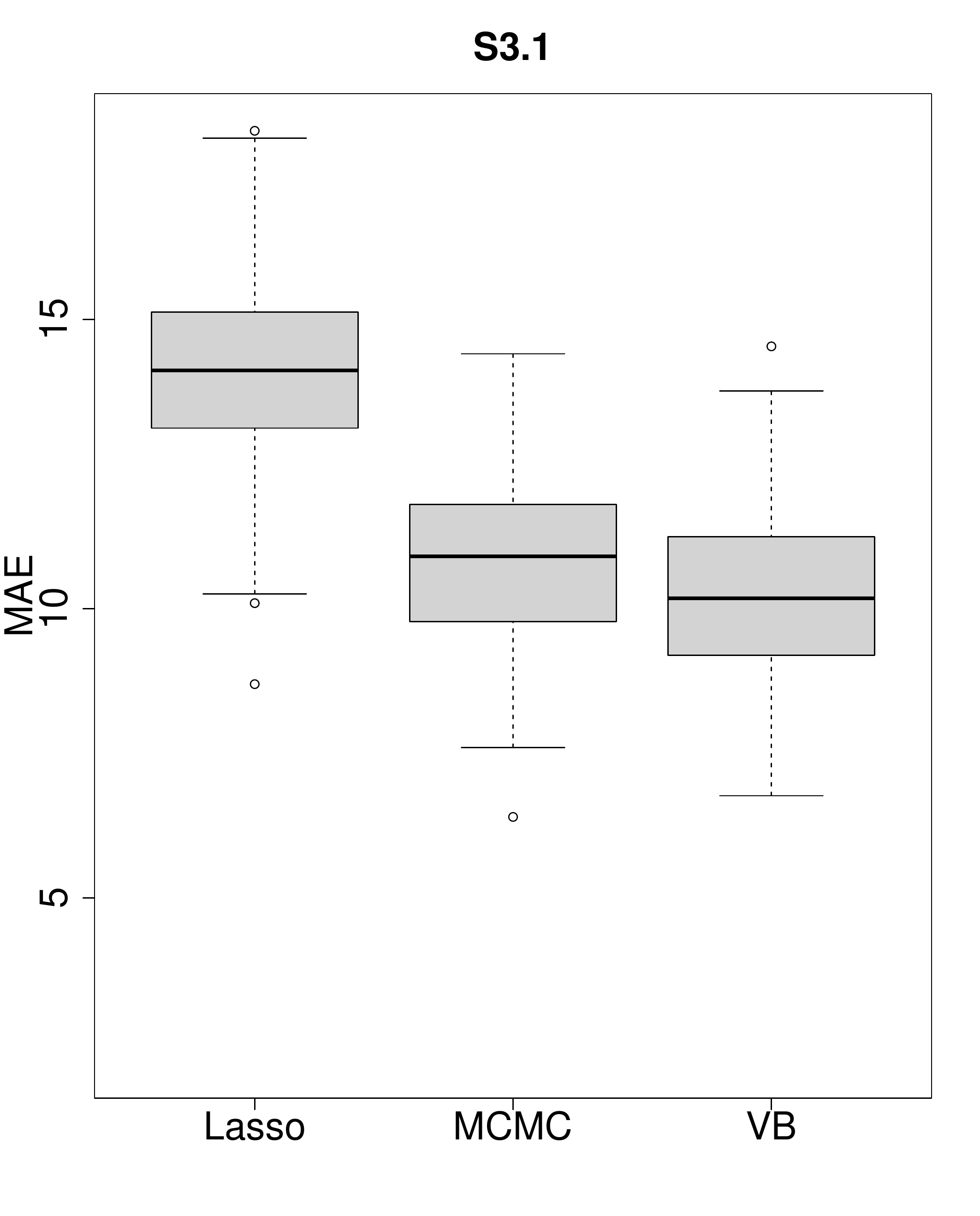}}&
{\includegraphics[scale=0.25]{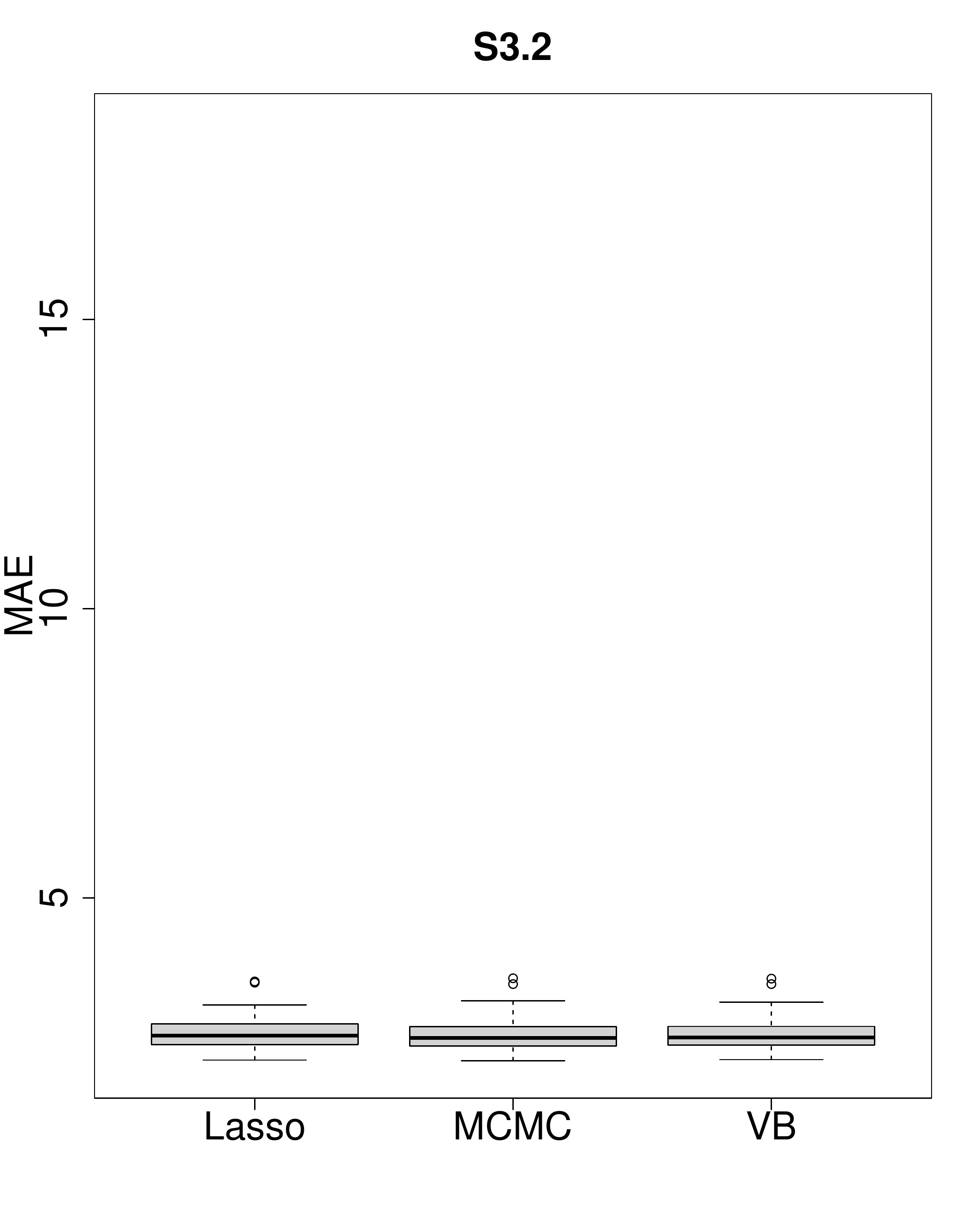}}\\
{\includegraphics[scale=0.25]{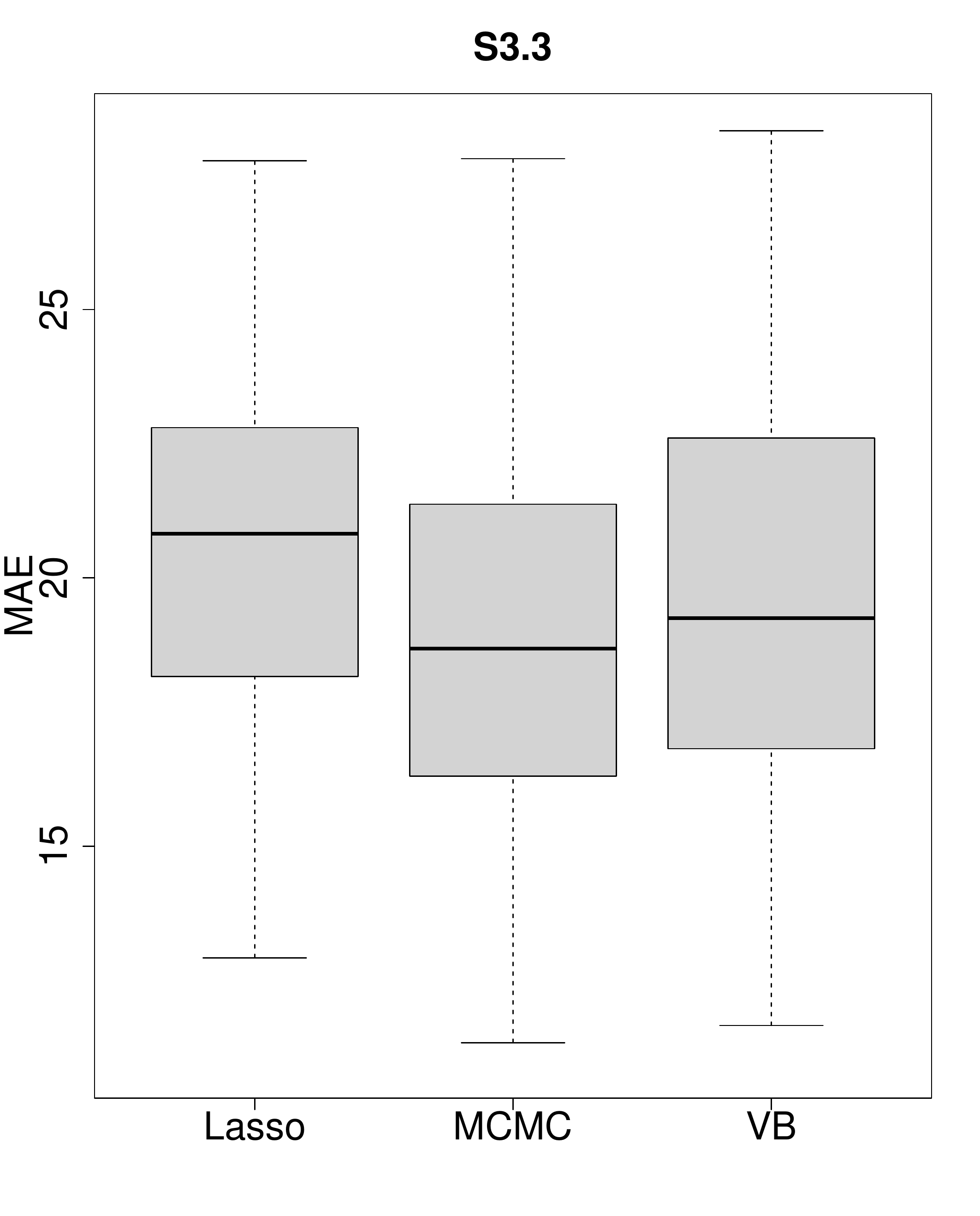}}&
{\includegraphics[scale=0.25]{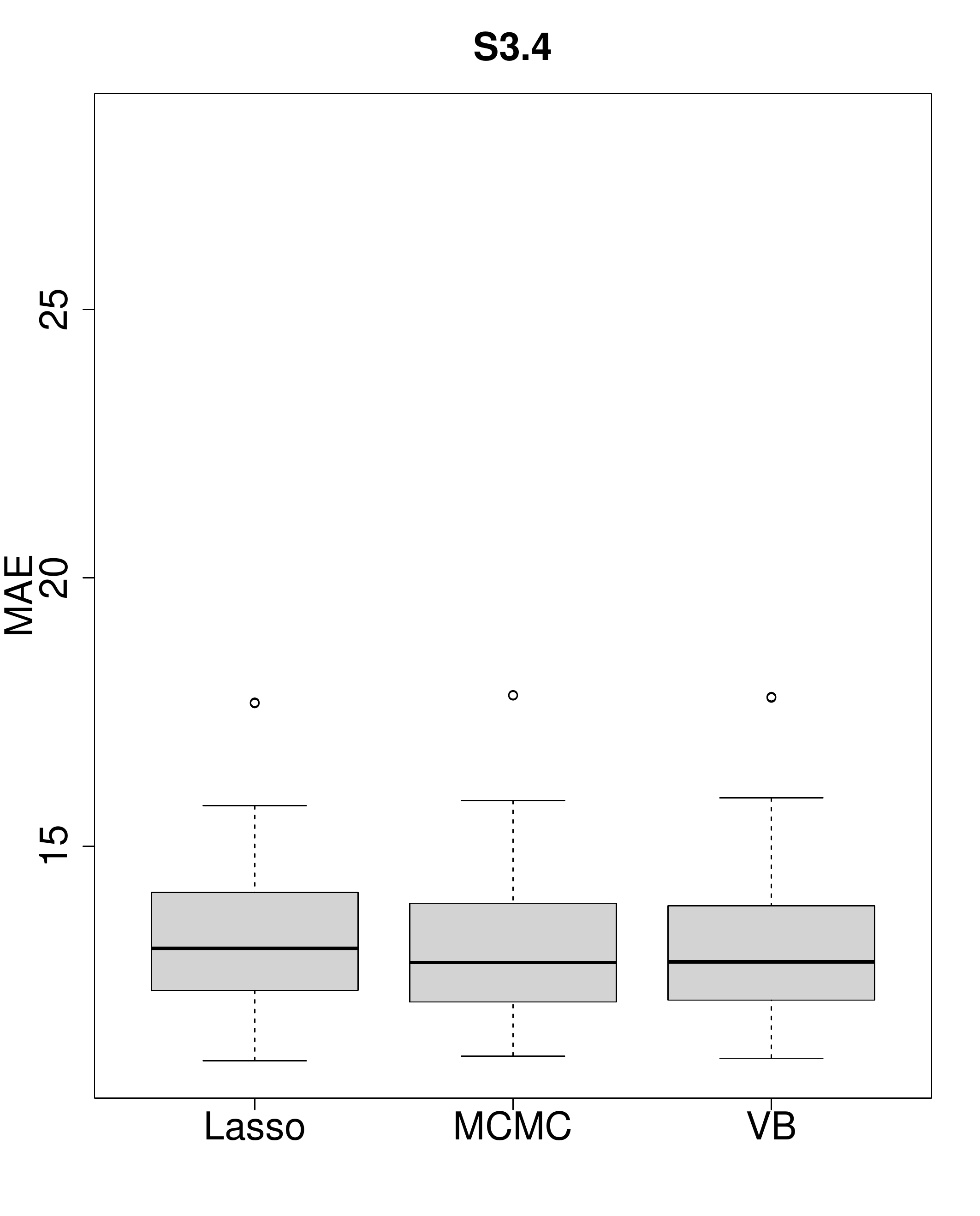}}\\
\end{tabular}
\end{center}\vspace{-0.5cm}
\caption{Mean absolute error (MAE) obtained by using  (\ref{EAM}) for the 4 scenarios in exercise 3, comparing the estimation methods MCMC, VB and Lasso.} \label{fig:EAMS3}
\end{figure}

\textcolor{blue}{
}
Similarly to exercise 2, the MAE was calculated for each replicate as a predictive measure. Figure \ref{fig:EAMS3} shows the box-plots of each scenario for MCMC, VB and Lasso. One may see that the MCMC and VB present similar and superior results to the Lasso when the sample size is small. As the sample increases the results become similar in the 3 approaches.

\begin{figure}[h!]
\begin{center}
\begin{tabular}{cc}
{\includegraphics[scale=0.24]{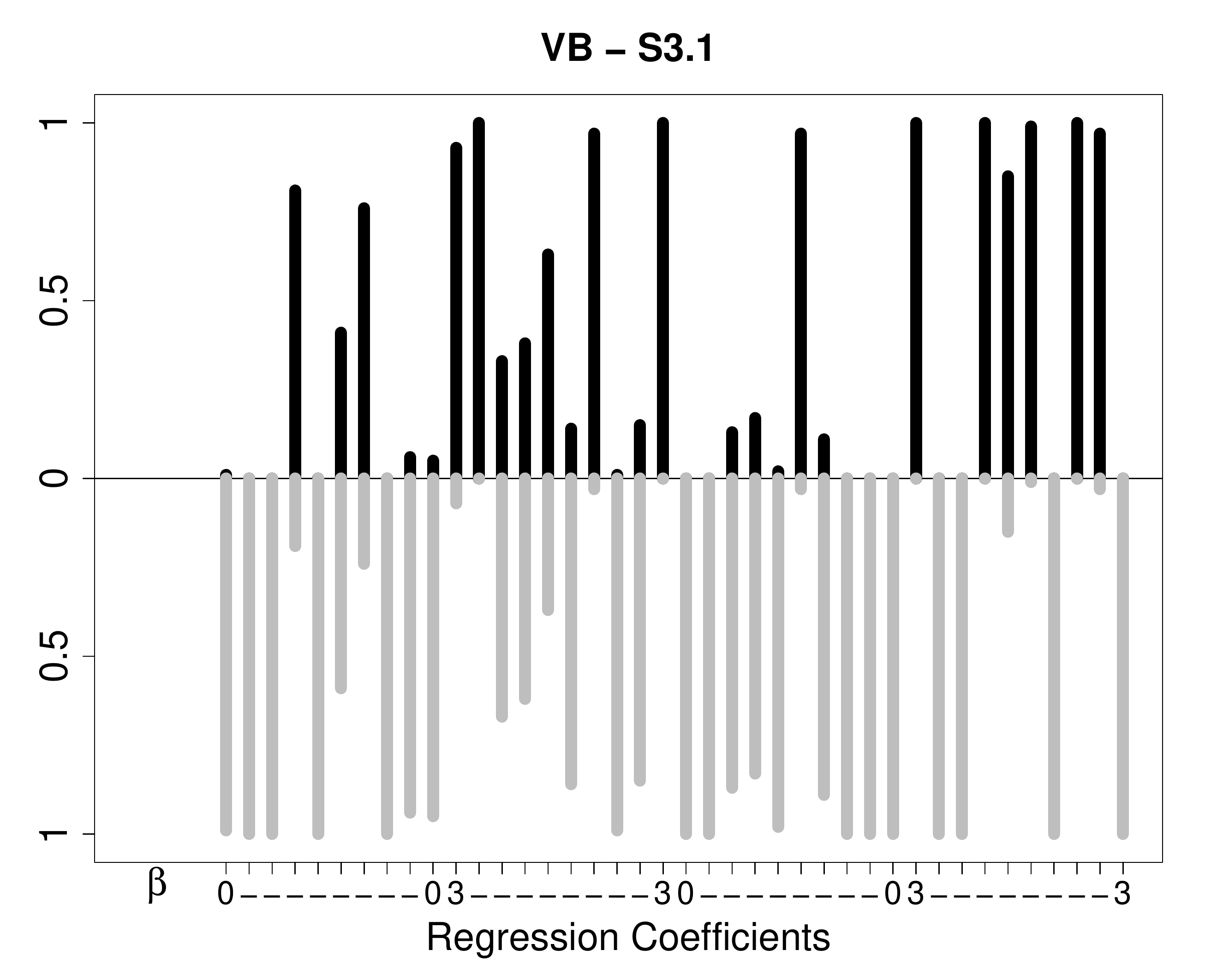}}&
{\includegraphics[scale=0.24]{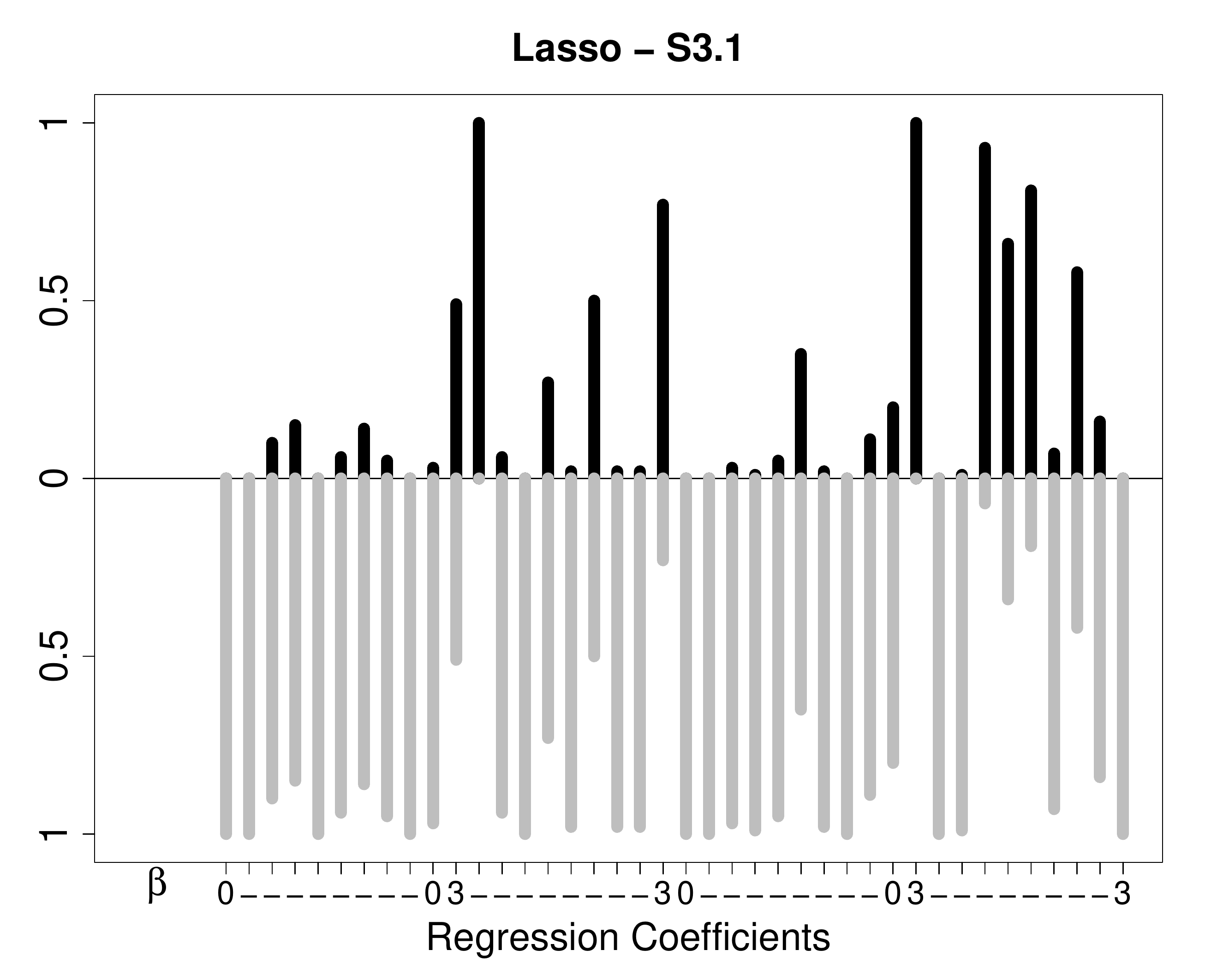}}\\
{\includegraphics[scale=0.24]{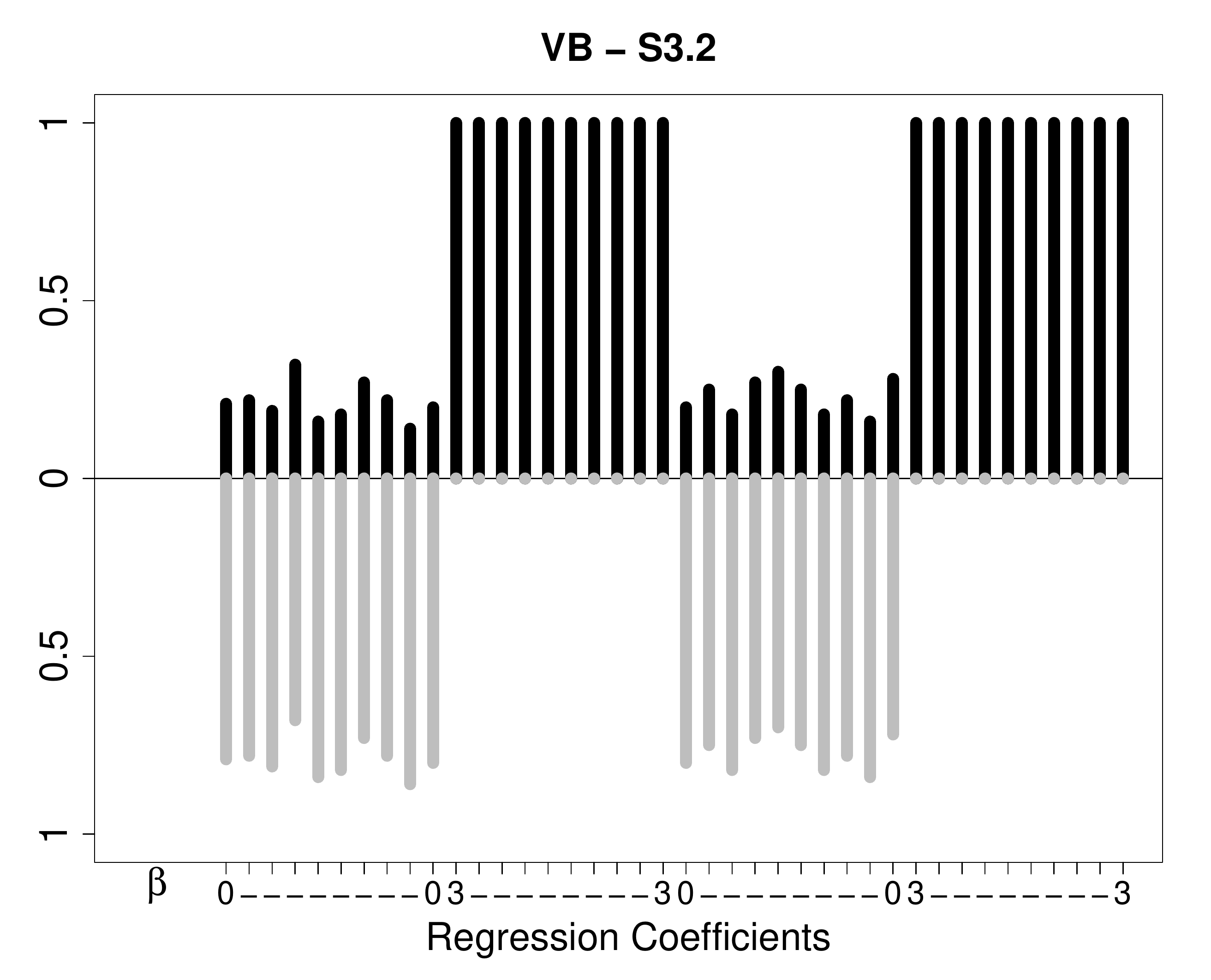}}&
{\includegraphics[scale=0.24]{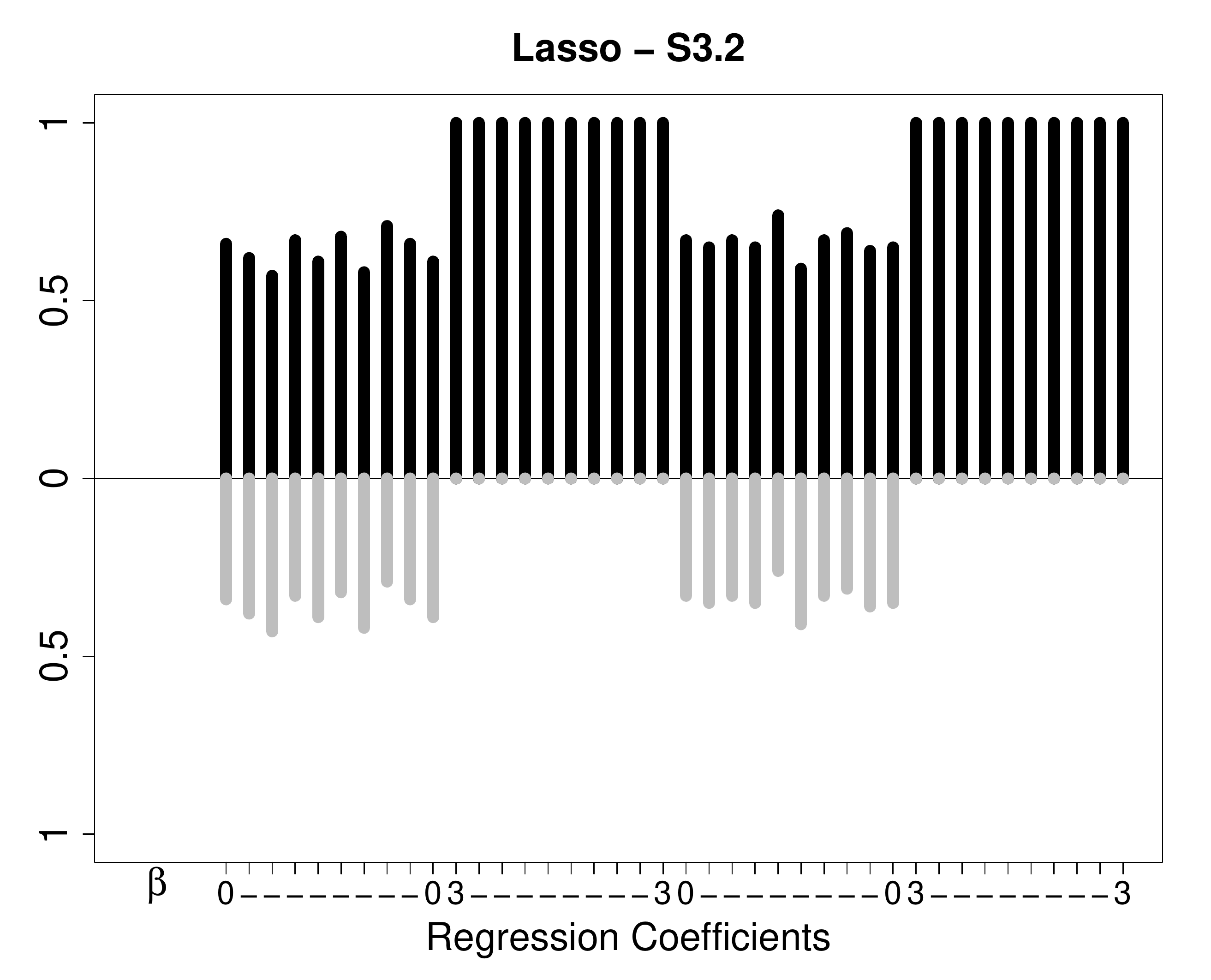}}\\
{\includegraphics[scale=0.24]{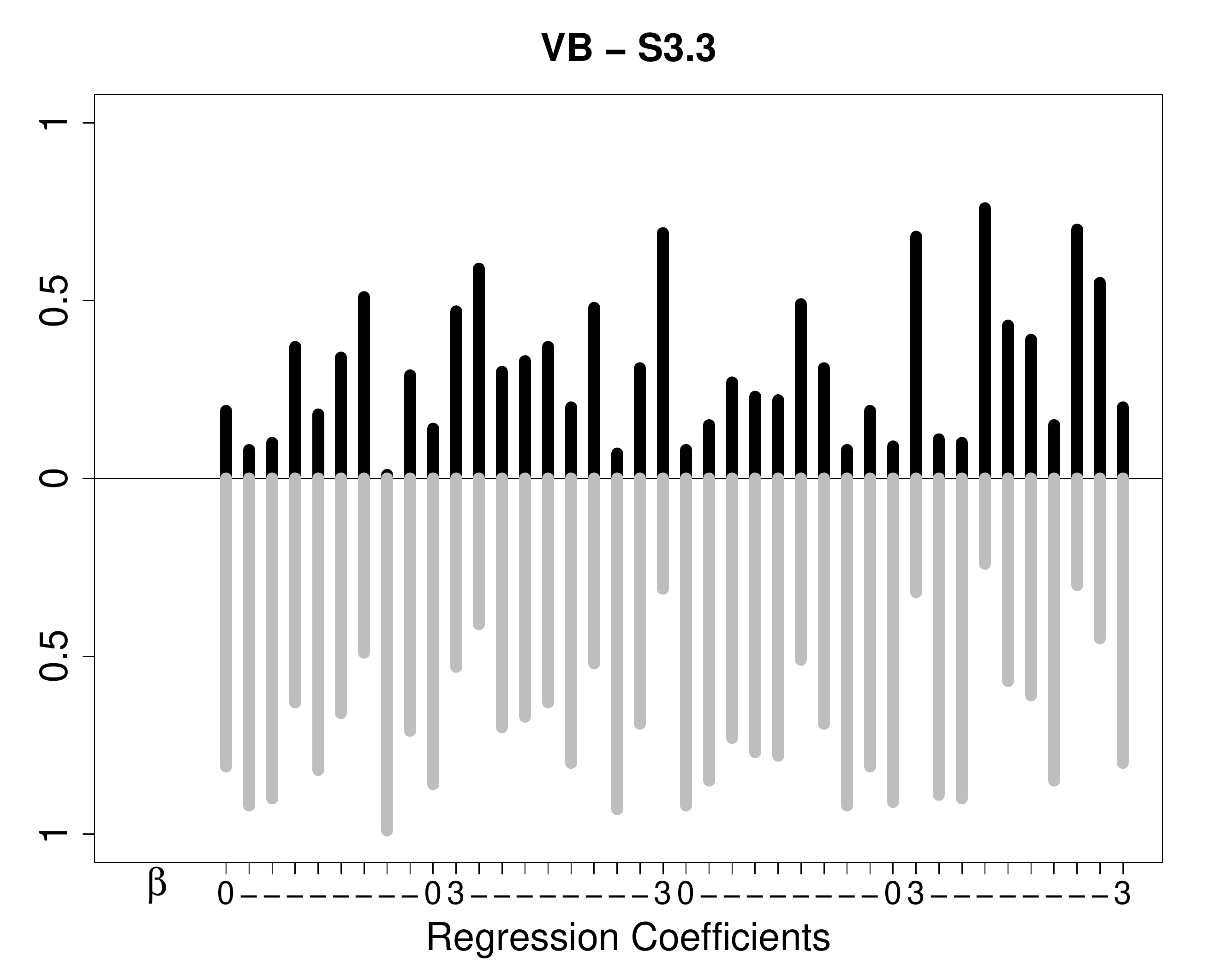}}&
{\includegraphics[scale=0.24]{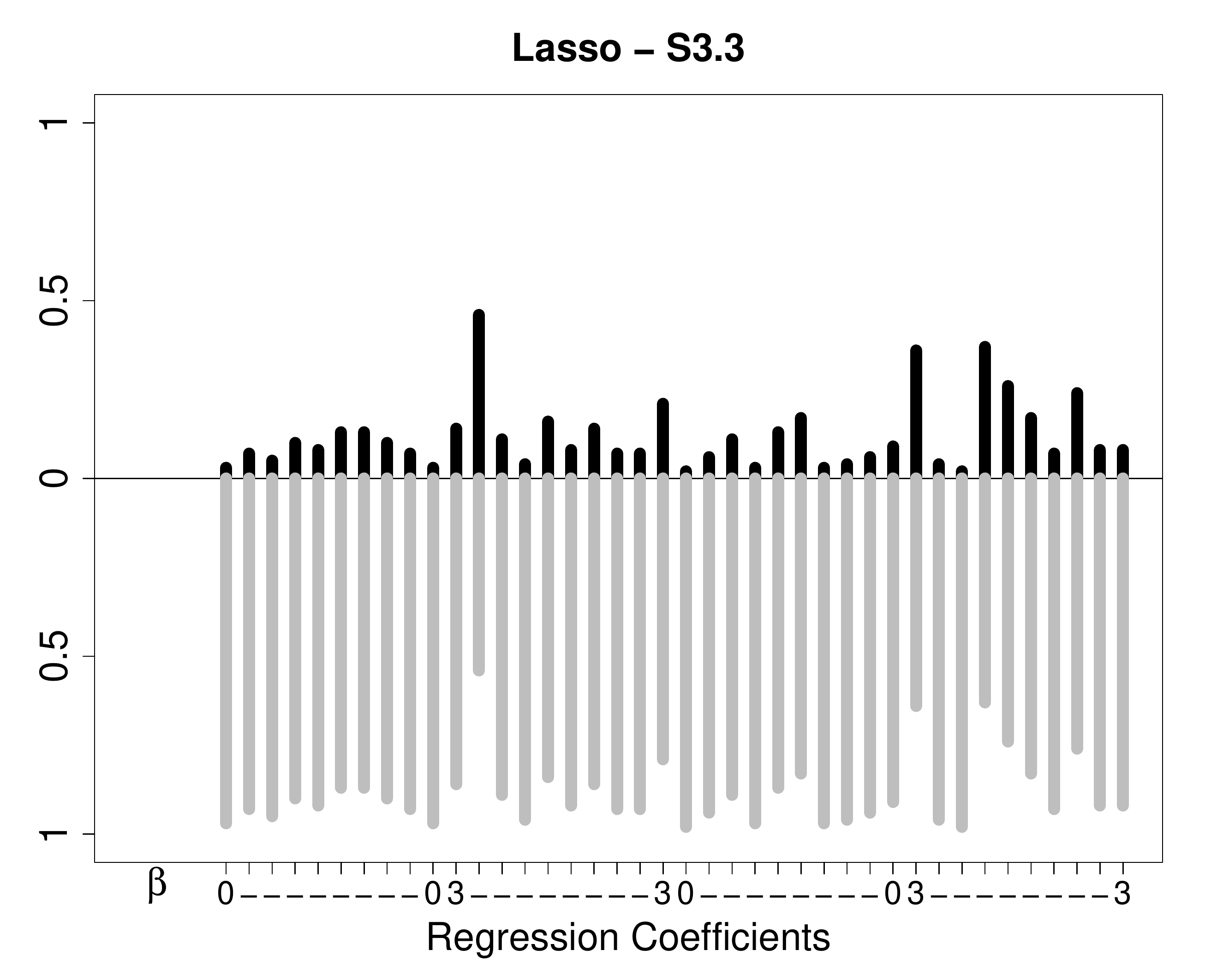}}\\
{\includegraphics[scale=0.24]{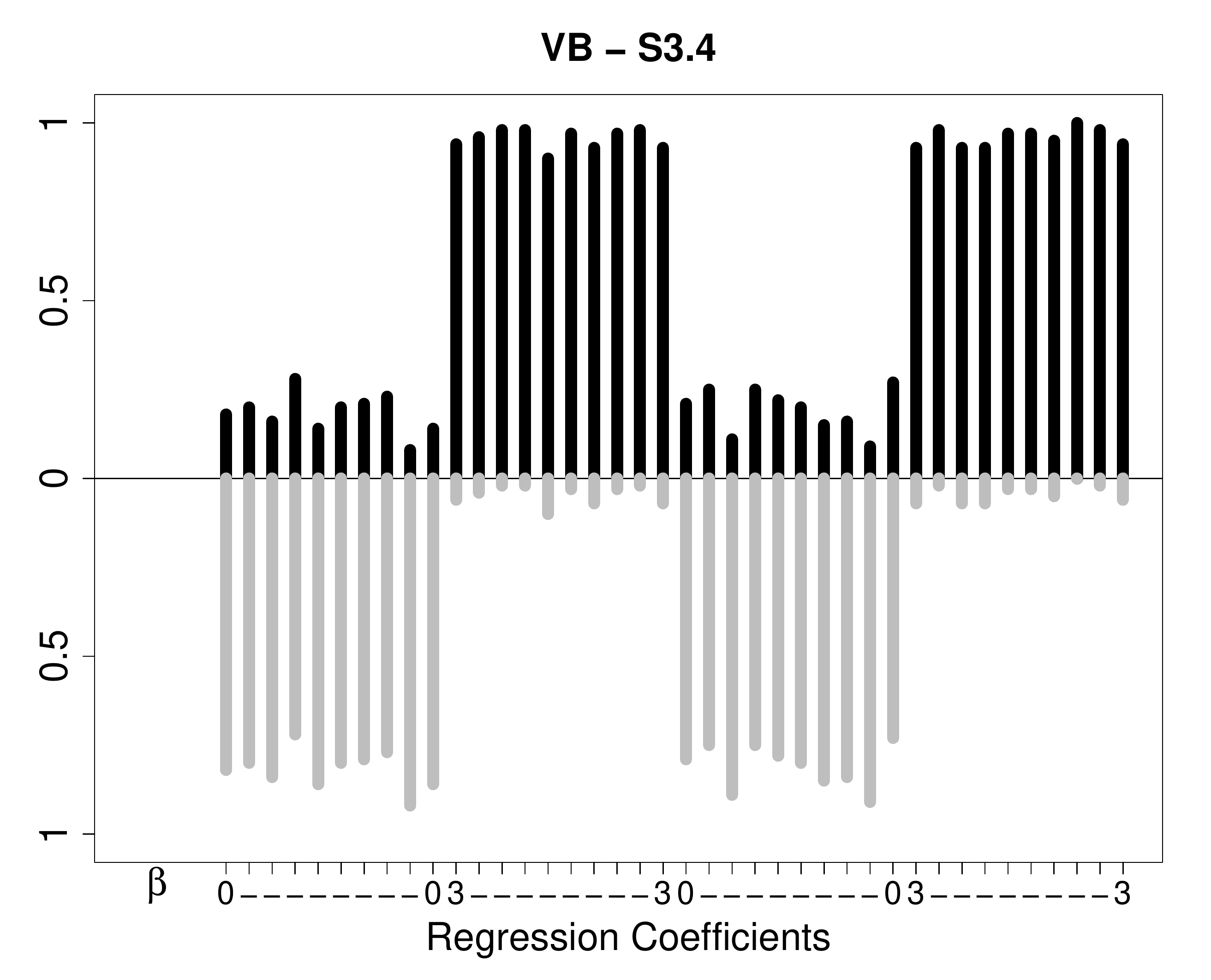}}&
{\includegraphics[scale=0.24]{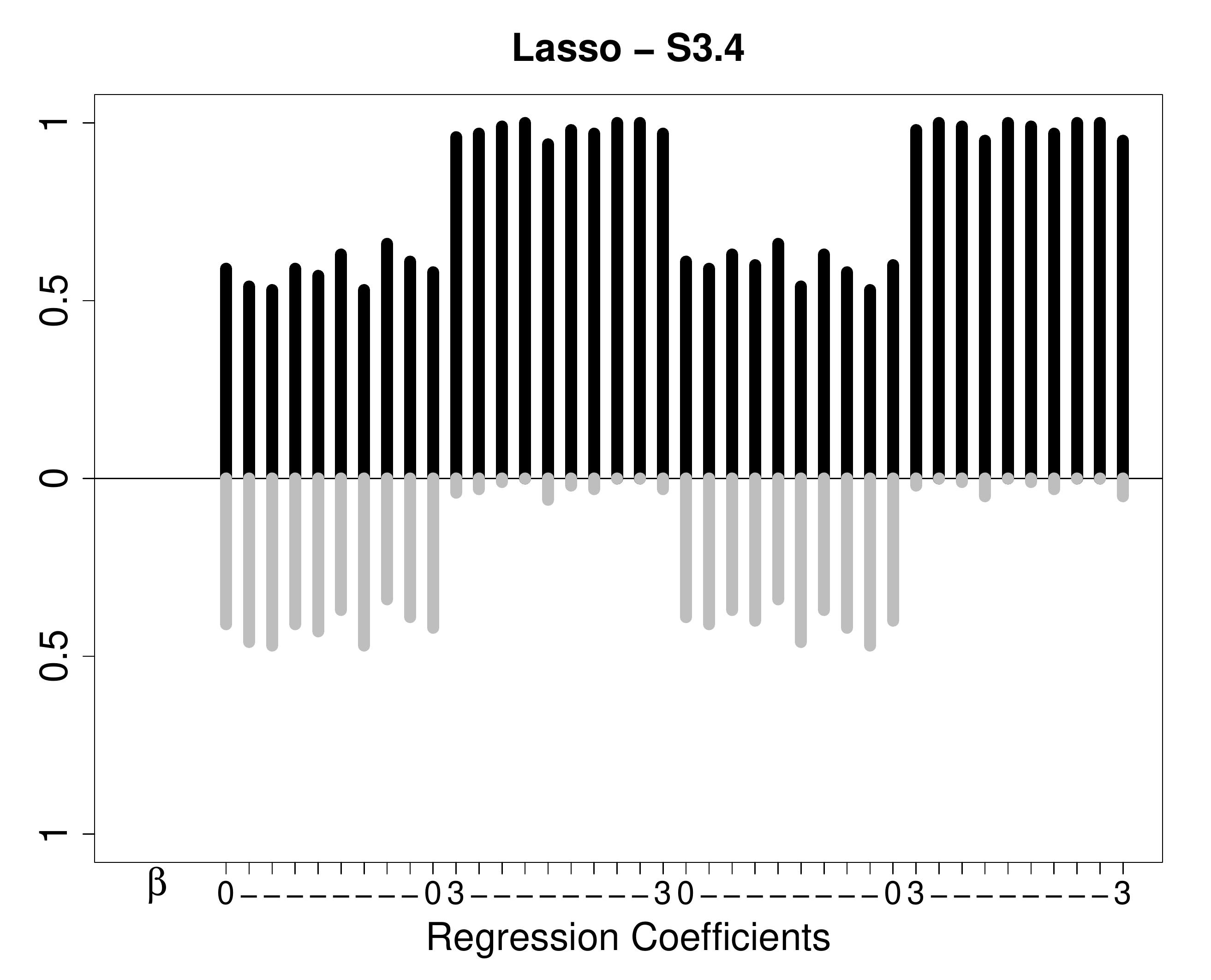}}\\
\end{tabular}
\end{center}\vspace{-0.5cm}
\caption{Proportion of selected  (black) and excluded coefficients (gray) for the 4 scenarios in exercise 3 with the estimation methods  VB (left column ) and Lasso (right column).} \label{fig:selecaobarra}
\end{figure}

\textcolor{blue}{
}
Figure \ref{fig:selecaobarra} shows the proportions of exclusions (gray bars) and selections (black bars) for each of the 40 coefficients in the 100 replicates, when comparing the estimation methods, VB and Lasso. MCMC was omitted for presenting results similar to VB. In the Bayesian context, the selection criterion used in all scenarios was the BF. It is expected that the black bars will be larger when the true coefficients are different from zero and that the gray bars will be large when the true coefficients are equal to zero. The proportions of the errors are represented by the black bars when the coefficients are zero (type I error) and by the gray bars when the coefficients are different from zero (type II error
\textcolor{blue}{
}
Thus, it can be seen for $n<p$, both VB and Lasso do not have a good selection and exclusion performance, with a slight advantage of VB. On the other hand, as the sample increases, the VB presents good results, better than those presented by Lasso. Also note that when $n>p$ both VB and Lasso have the same type II error. However, for all coefficients, the type I error is considerably less in VB than in MCMC


\subsection{Knots selection}

\textcolor{blue}{
}

As the VB presents results similar to the MCMC, but with considerably less computational time, therefore, in the two exercises applied to the spline regression models, only the VB is used. The goal is to define the maximum number of knots from a grid of values and, in turn, to select the most significant knots and their positions. Exercises 4 and 5  consist of a simulation study of the penalized spline regression model defined in equations (\ref{reg_spline1}), (\ref{reg_spline2}) and (\ref{prior_spline}). What differs in the two simulation studies consists of the number of bumps in the smooth function $f$. In exercise 4, shows an example with 1 bump, while exercise 5 analyzes  2 bumps. 

In both exercises we have 100 replicates  with $ n = 100 $, variance equals to $0.3$ ( $ \phi = 1/0.3 $) and $x_i$  taking  equally spaced values in the interval $[0,1]$. In addition, Cubic Splines ($p=3$) are used, the interior knots of the truncated power basis are positioned in the quantiles of the variable $x_i$. The maximum number of knots varies in the grid $K = 10, 20, 30, 40 $ and $ 50 $. 

The ELBO  is used to indicate the maximum number of knots and for the  both exercises we have an optimal initial guess of $K=30$ knots. The BF and CI selection criteria indicate that around 8 of these 30 initial knots have a higher frequency of being selected as the most significant ones.

\subsubsection{Exercise 4: Single structure/One bump}



In exercise 4 we use the smooth function $f$ called "Bump" given by $f(x) = x + 2 \exp \{- (16 * (x-0.5)) ^ 2 \} $.

\begin{figure}[h!]
\begin{center}
\begin{tabular}{ccc}
{\includegraphics[scale=0.50]{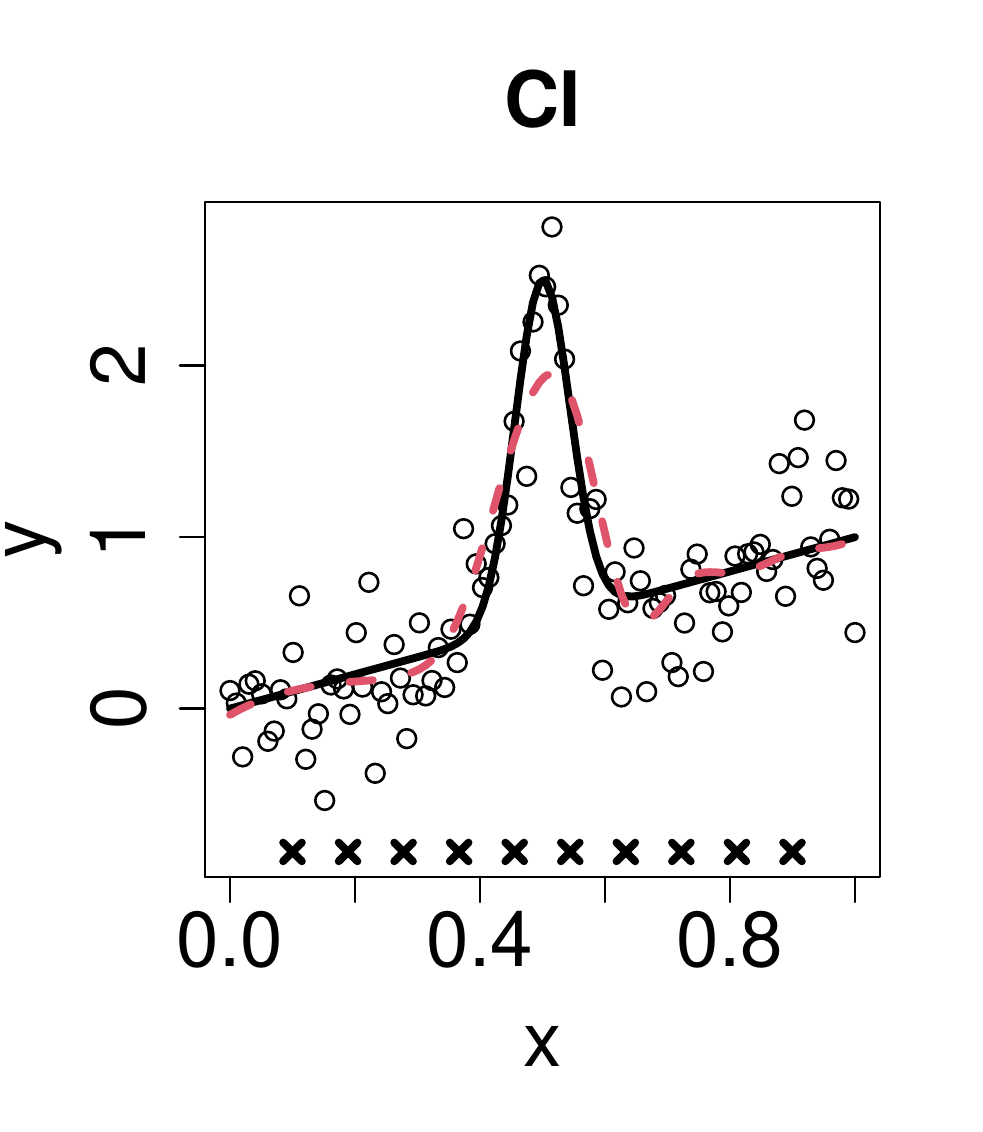}}&
{\includegraphics[scale=0.50]{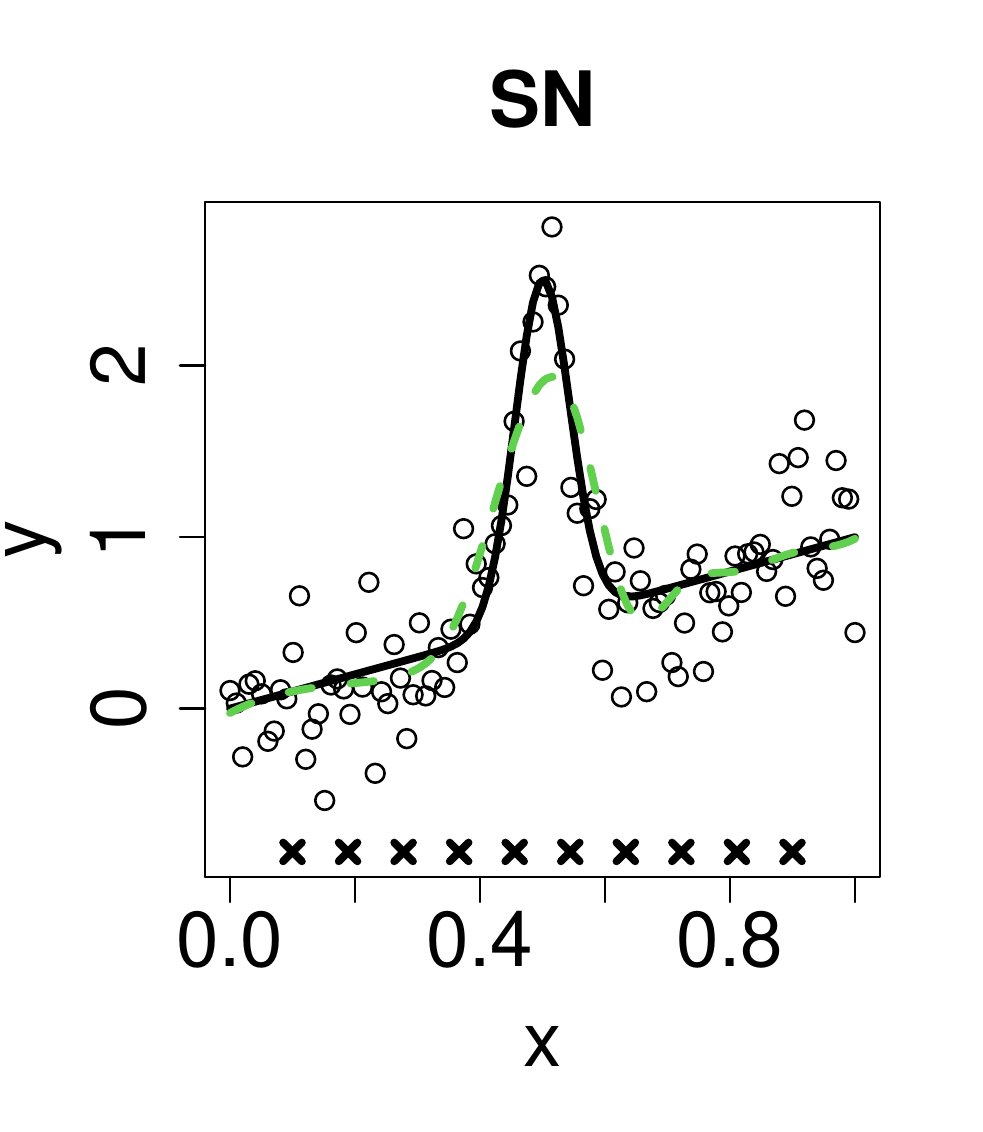}}&
{\includegraphics[scale=0.50]{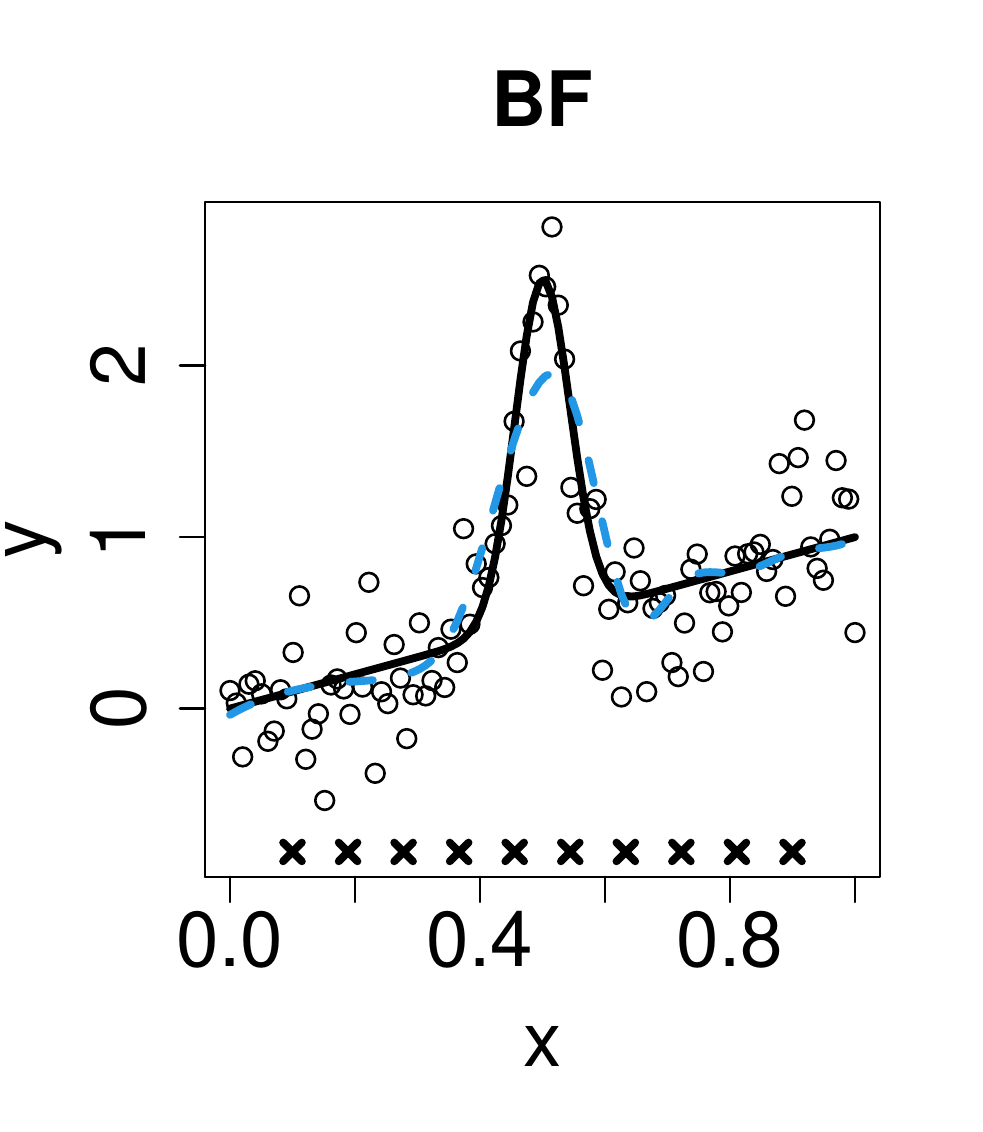}}\\
{\includegraphics[scale=0.50]{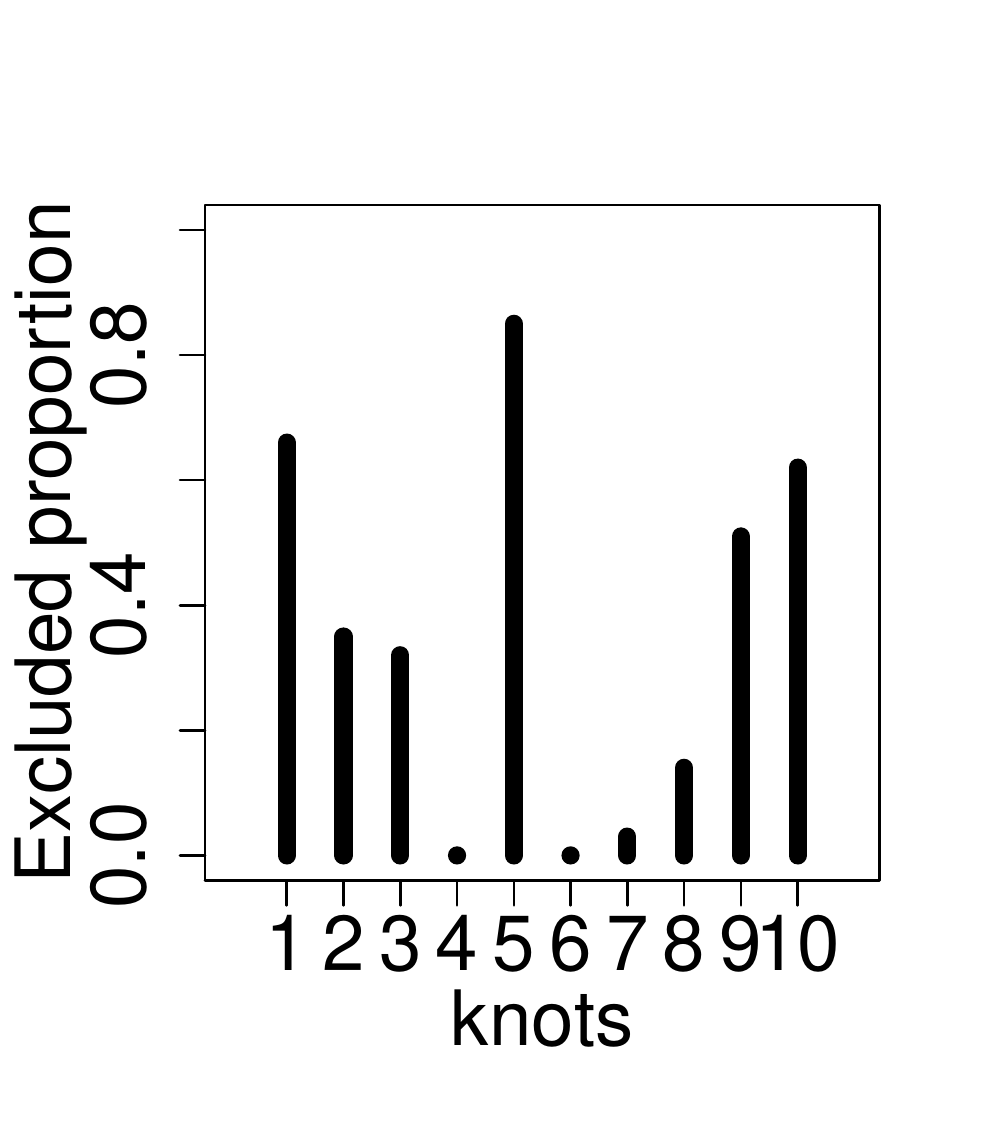}}&
{\includegraphics[scale=0.50]{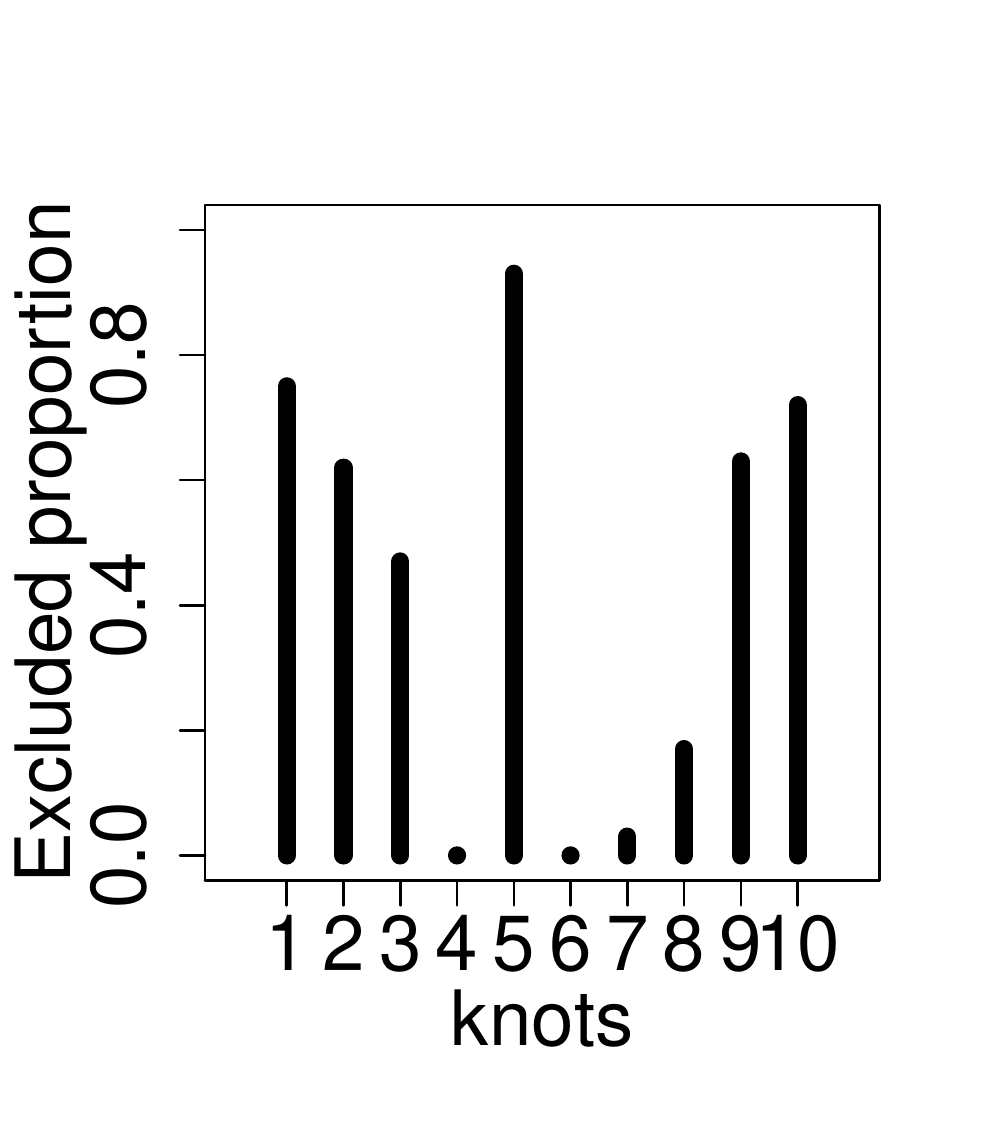}}&
{\includegraphics[scale=0.50]{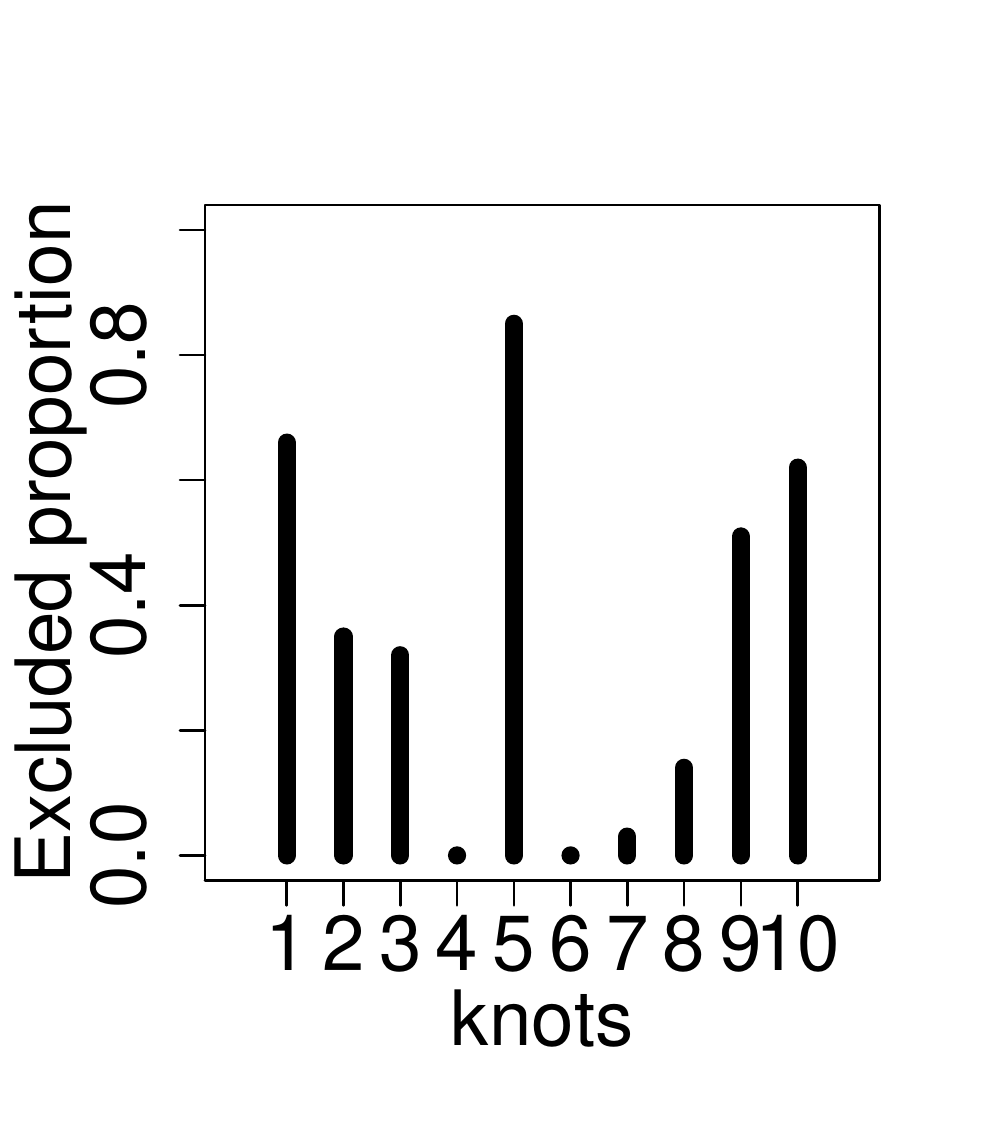}}\\
\end{tabular}
\end{center}\vspace{-0.5cm}
\caption{Proportion of excluded knots and the average of the fittings (dashed line), K=10.}\label{fig:ajuste_1bump_k10}
\end{figure}

\begin{figure}[h!]
\begin{center}
\begin{tabular}{ccc}
{\includegraphics[scale=0.50]{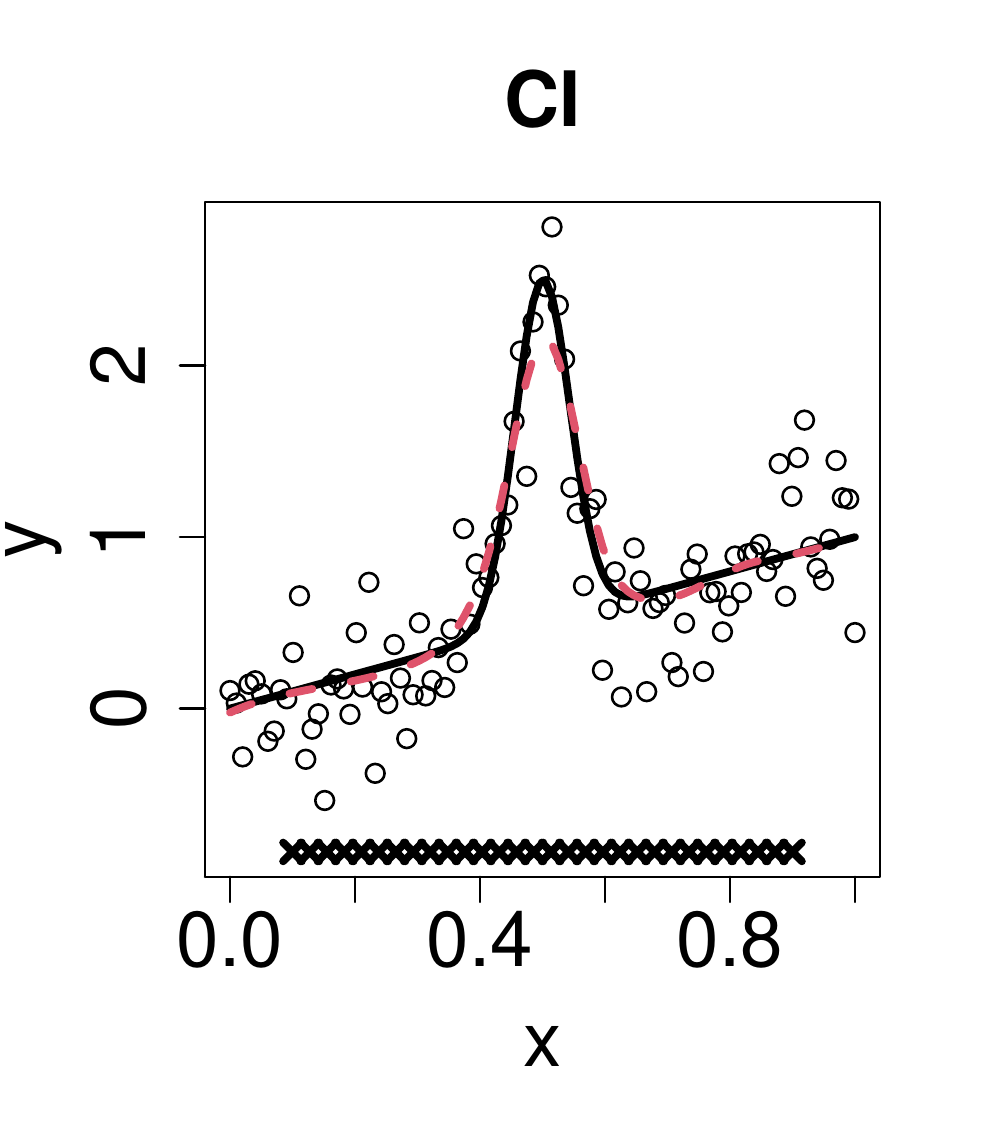}}&
{\includegraphics[scale=0.50]{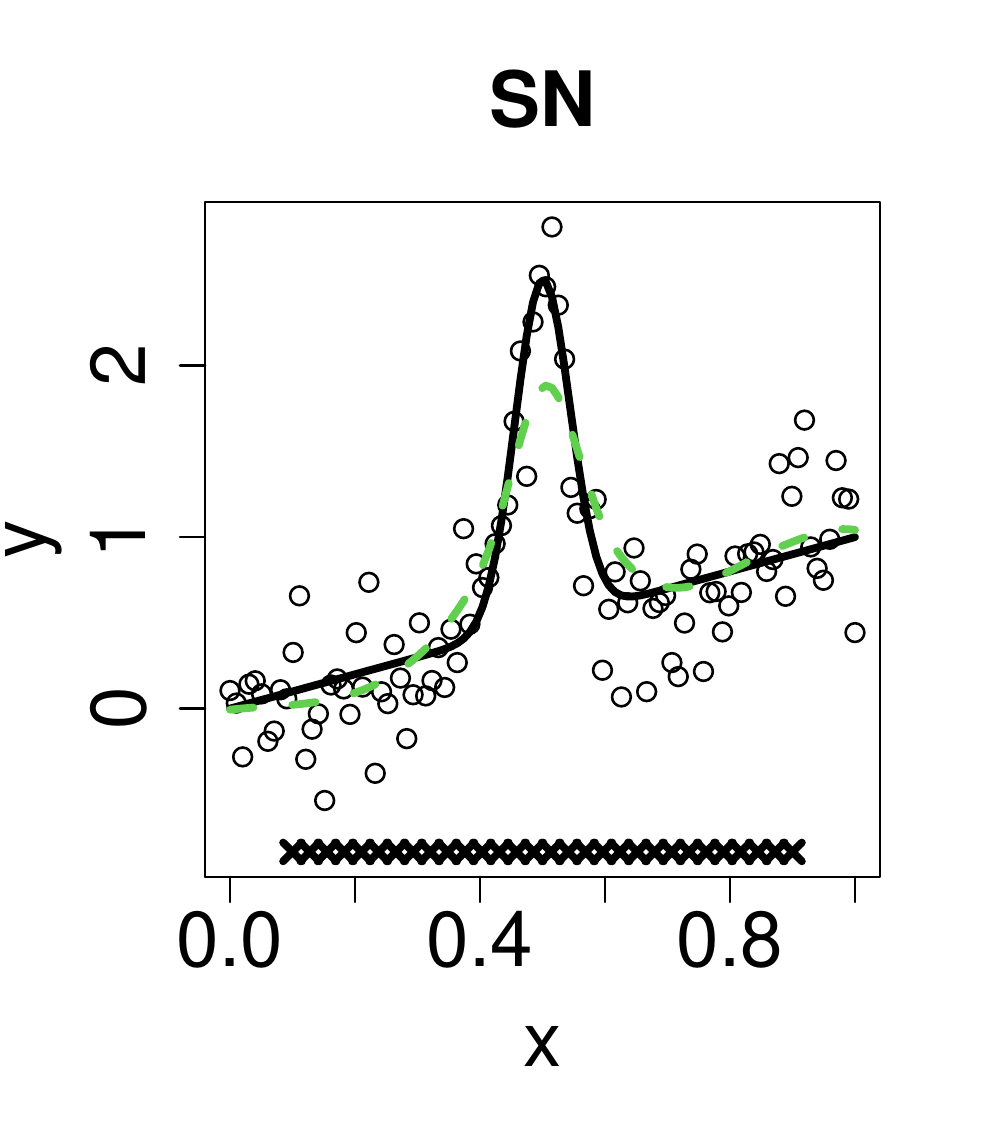}}&
{\includegraphics[scale=0.50]{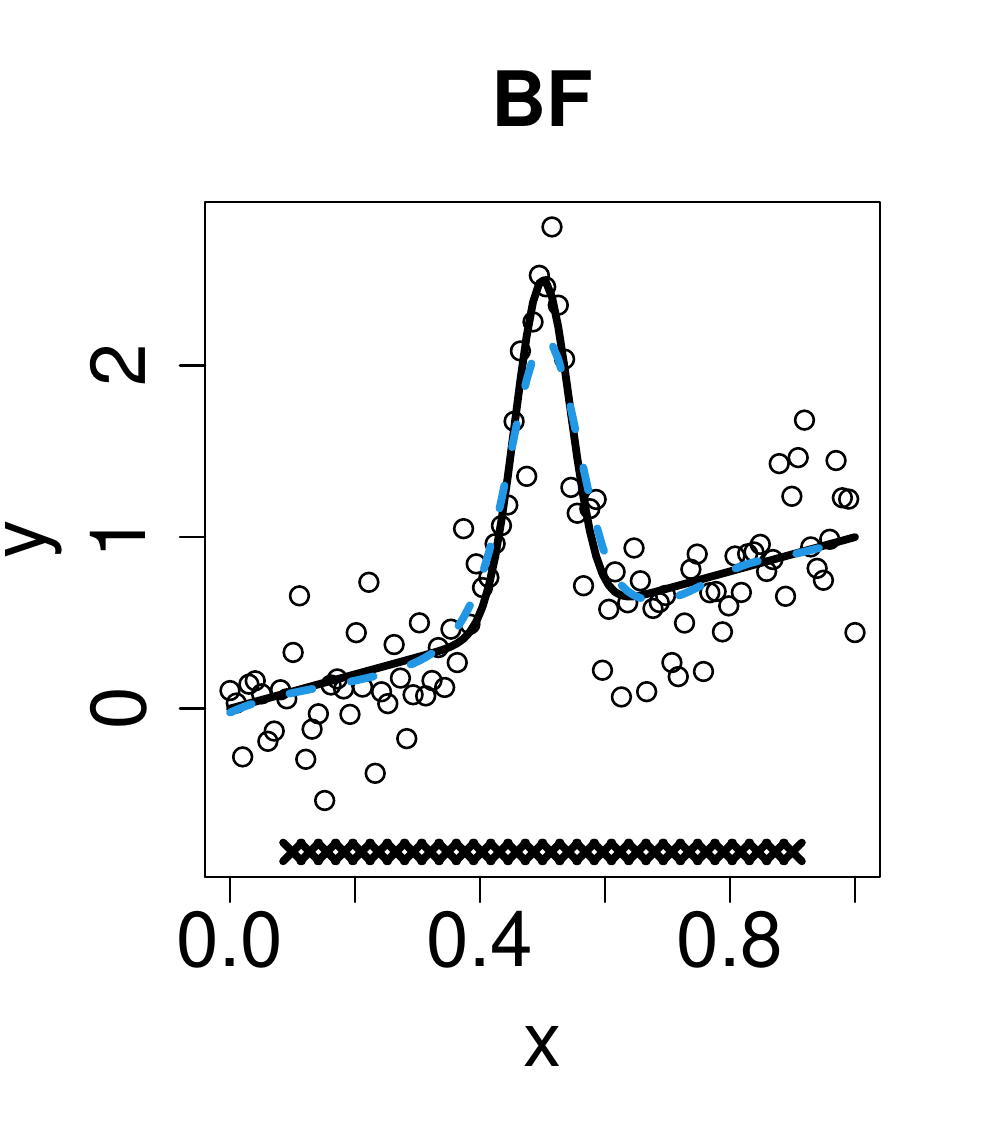}}\\
{\includegraphics[scale=0.50]{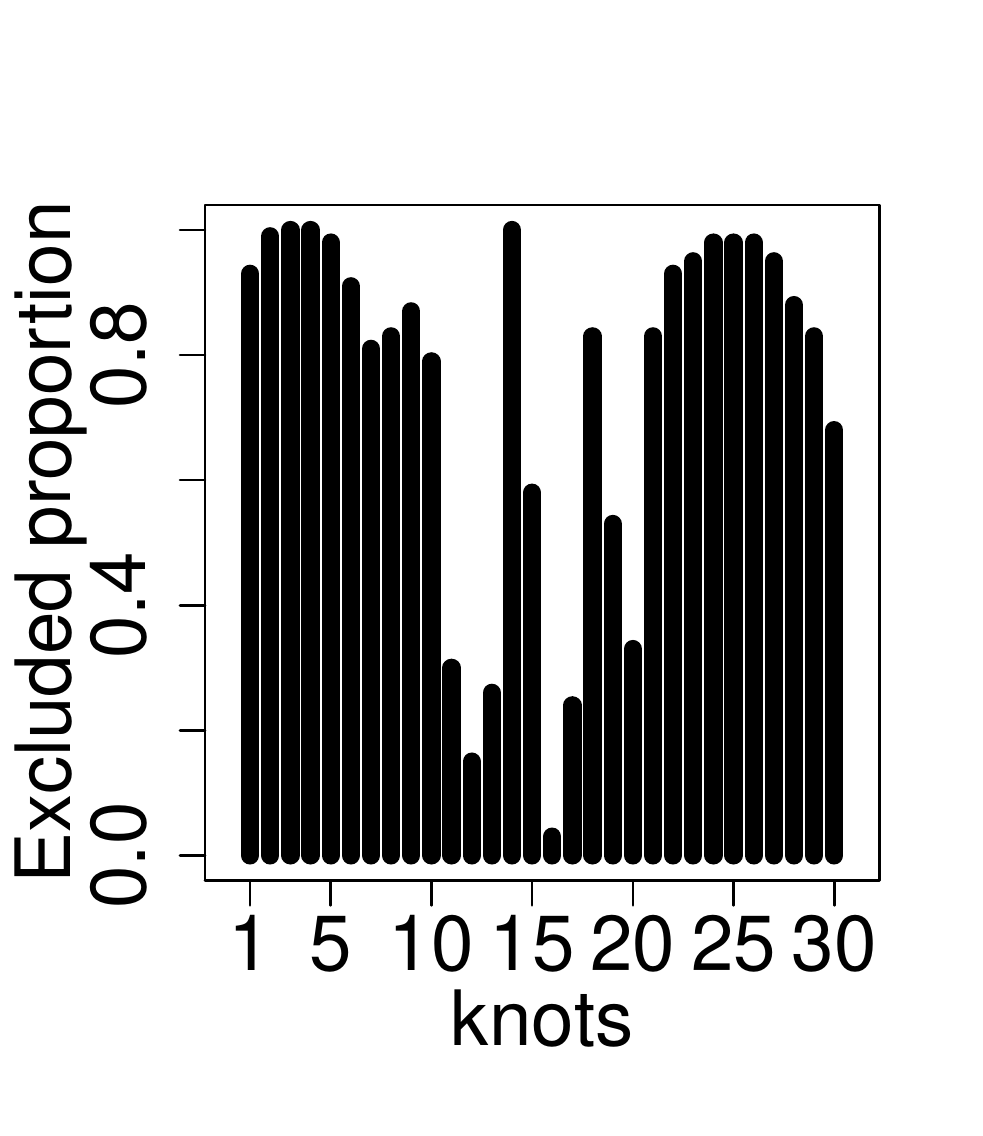}}&
{\includegraphics[scale=0.50]{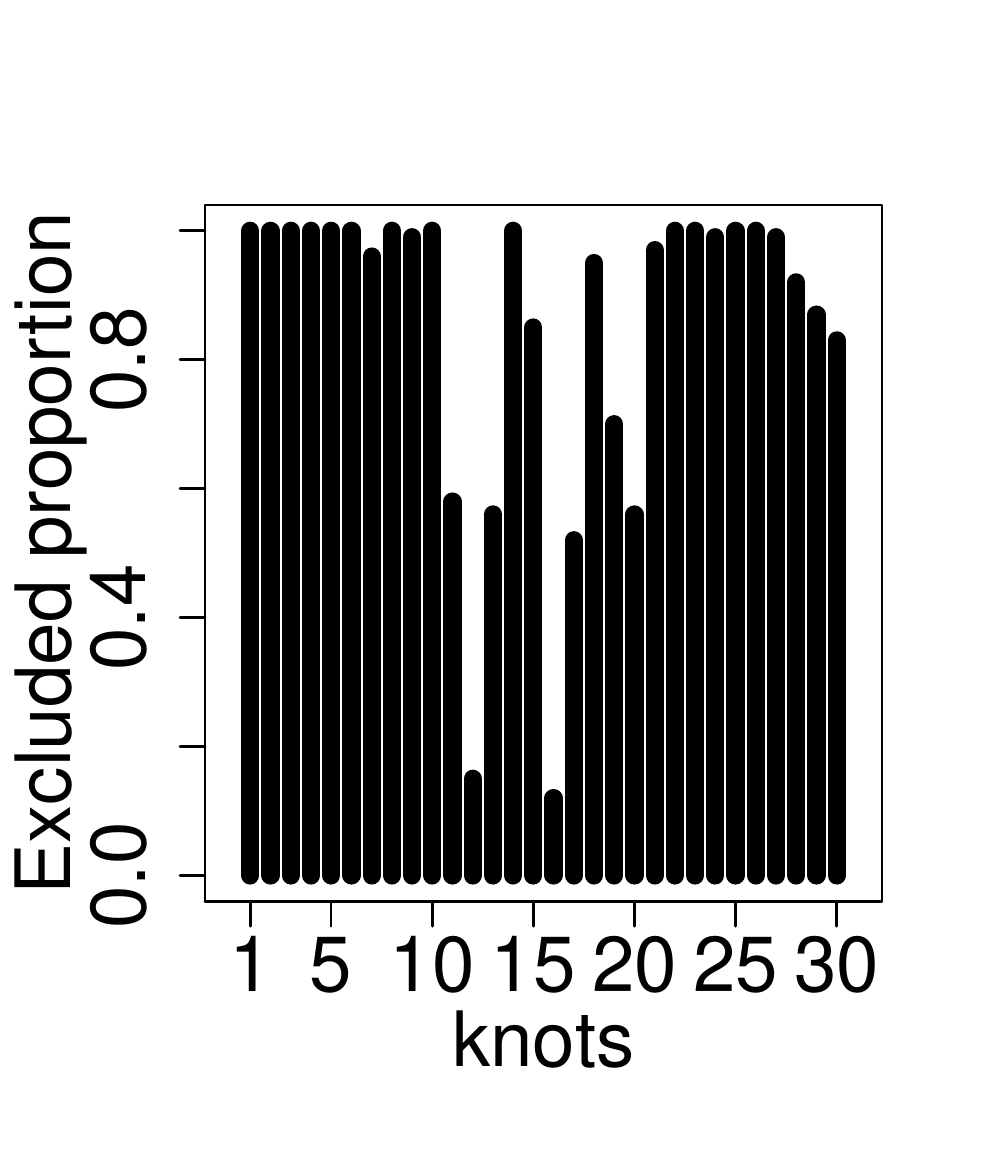}}&
{\includegraphics[scale=0.50]{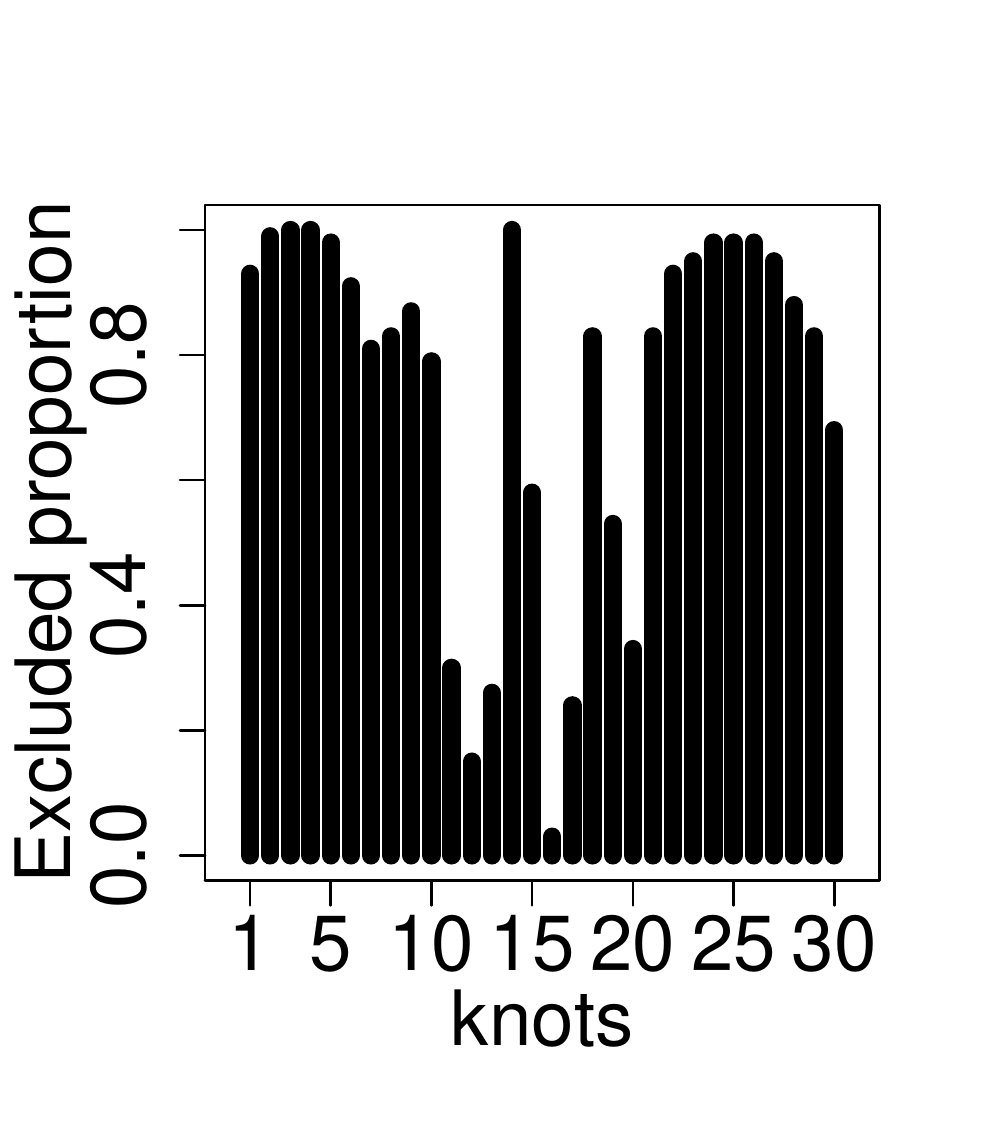}}\\
\end{tabular}
\end{center}\vspace{-0.5cm}
\caption{Proportion of excluded knots and the average of the fittings (dashed line), K=30.}\label{fig:ajuste_1bump_k30}
\end{figure}

\begin{figure}[h!]
\begin{center}
\begin{tabular}{ccc}
{\includegraphics[scale=0.50]{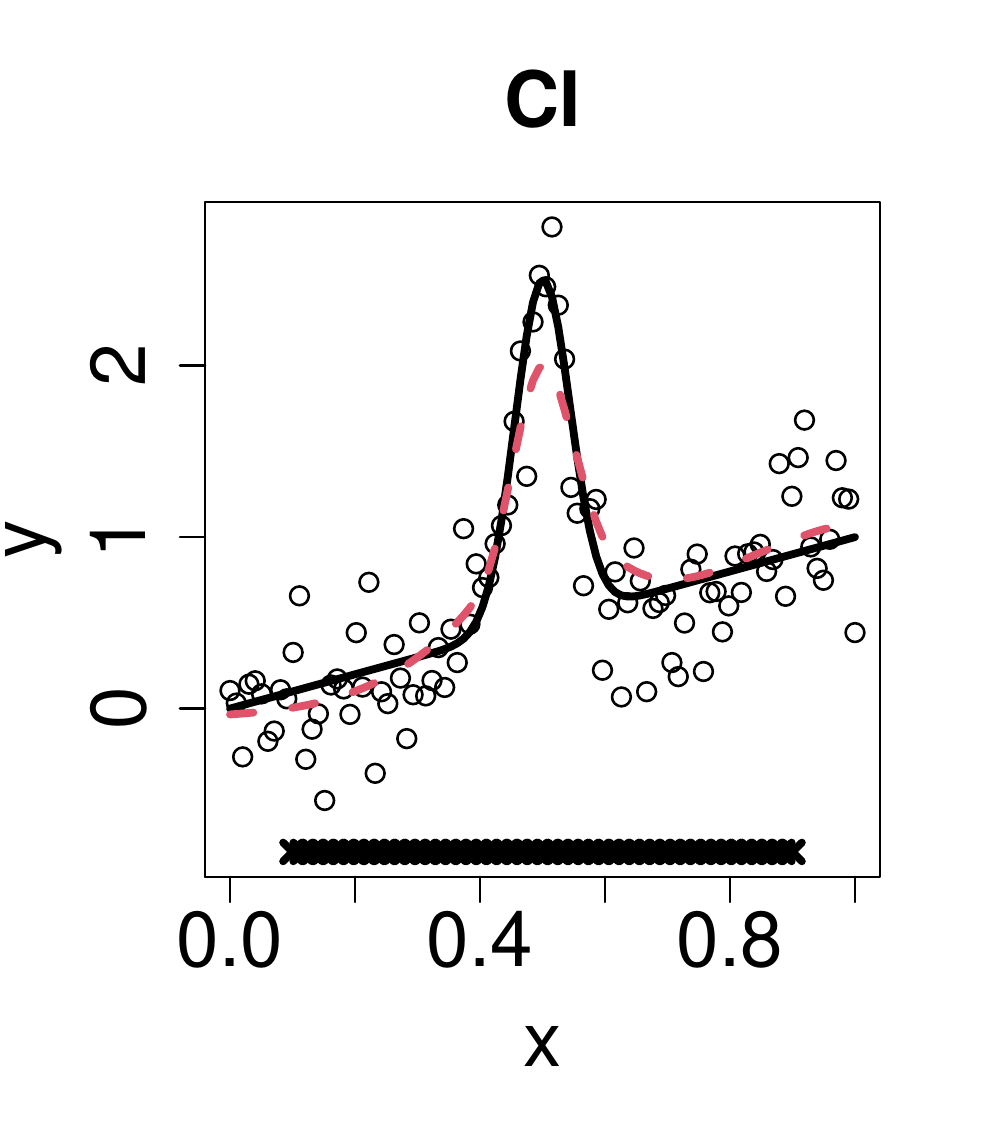}}&
{\includegraphics[scale=0.50]{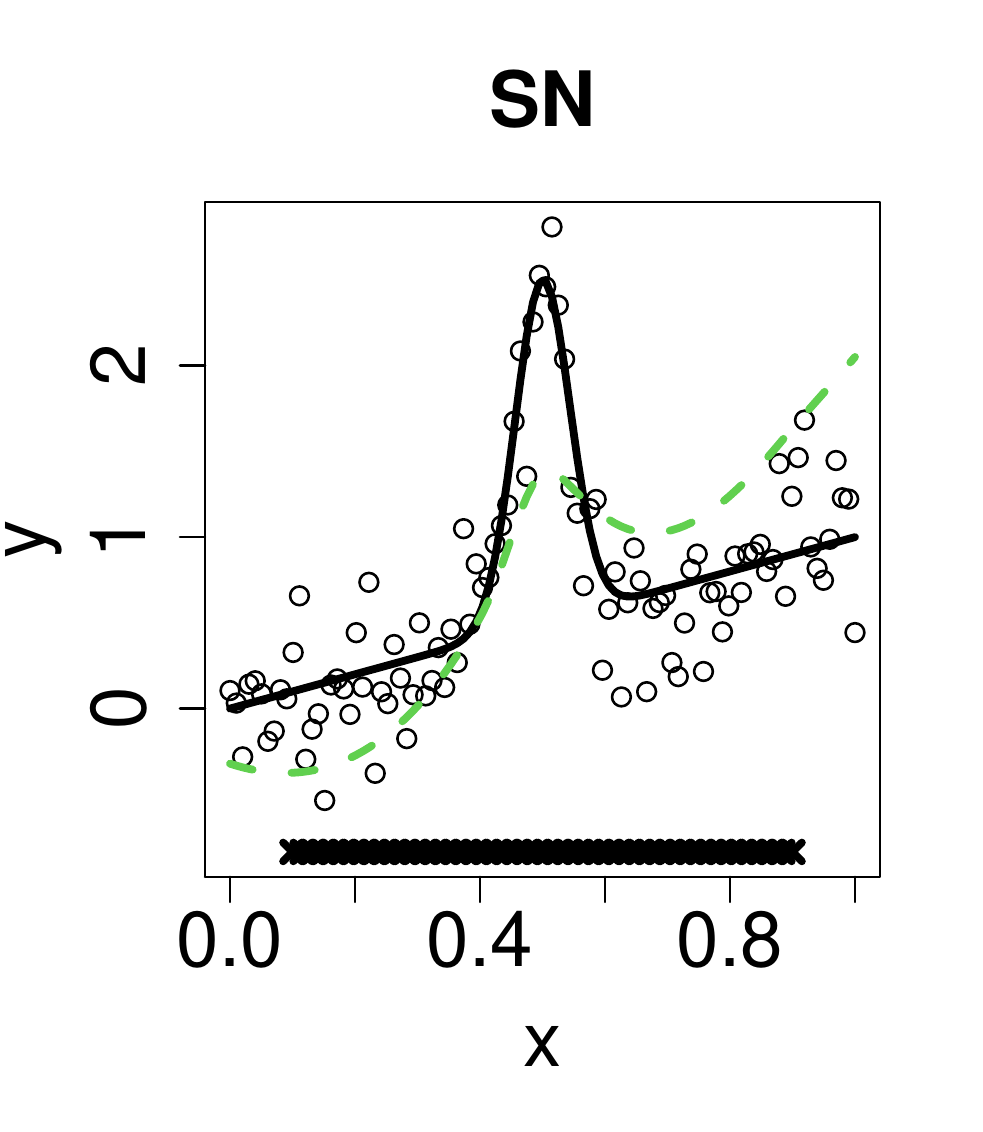}}&
{\includegraphics[scale=0.50]{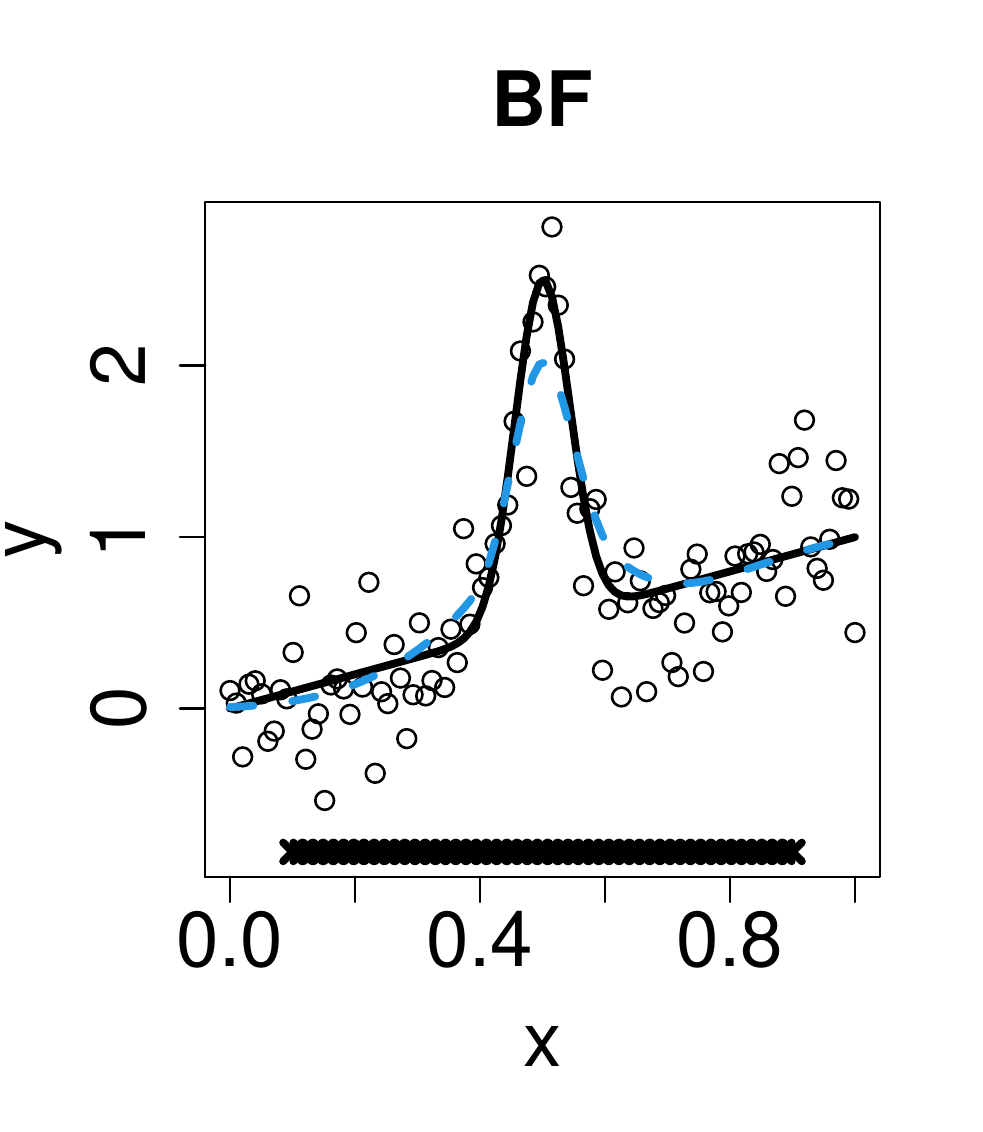}}\\
{\includegraphics[scale=0.50]{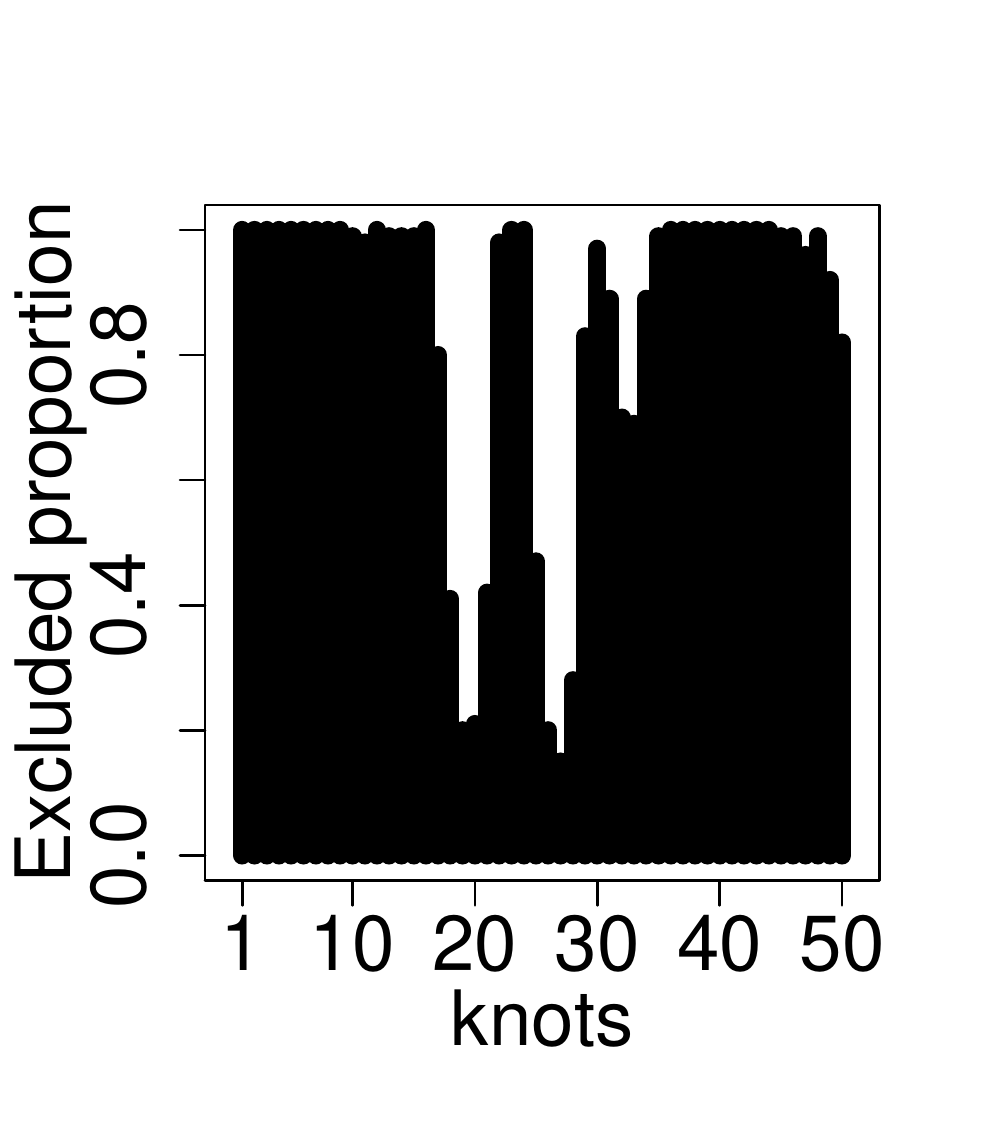}}&
{\includegraphics[scale=0.50]{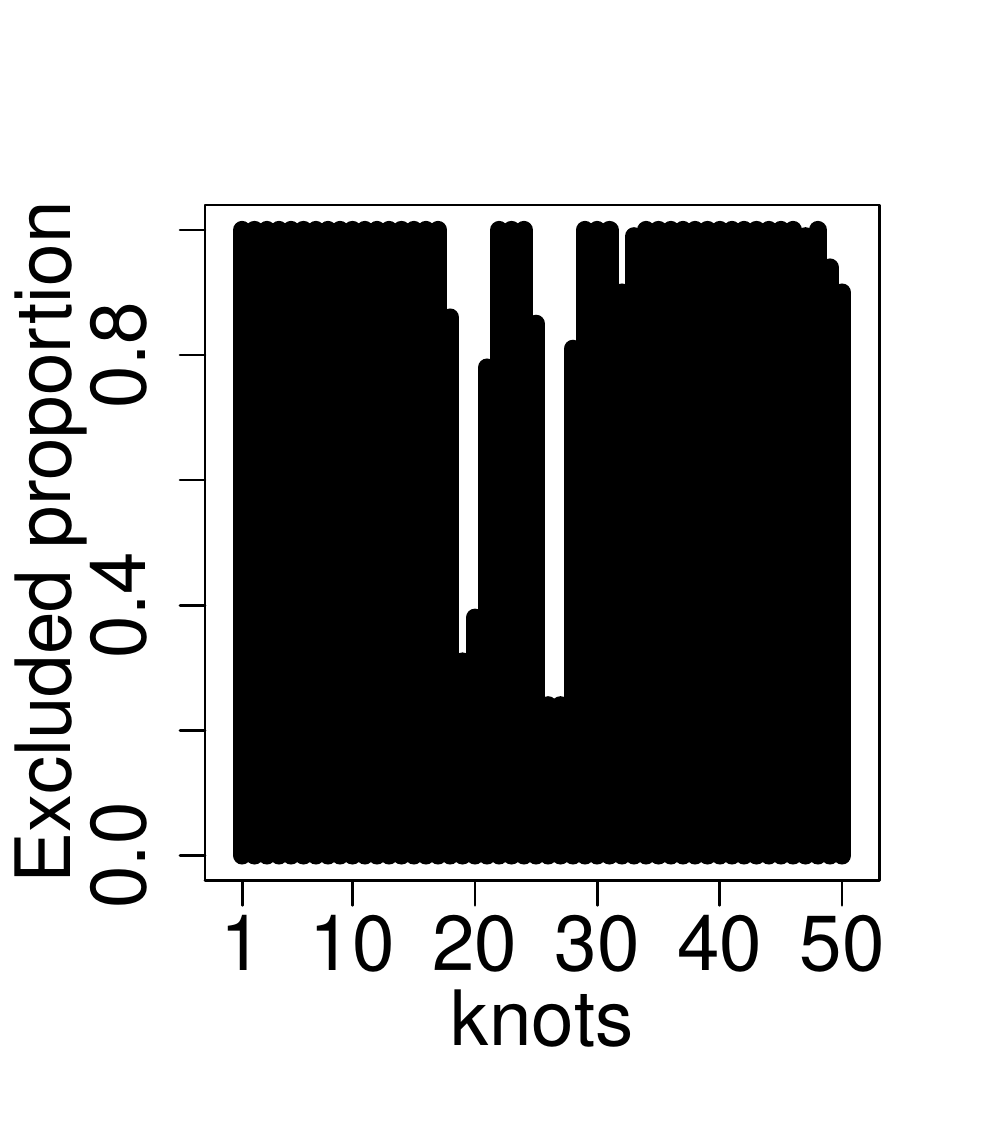}}&
{\includegraphics[scale=0.50]{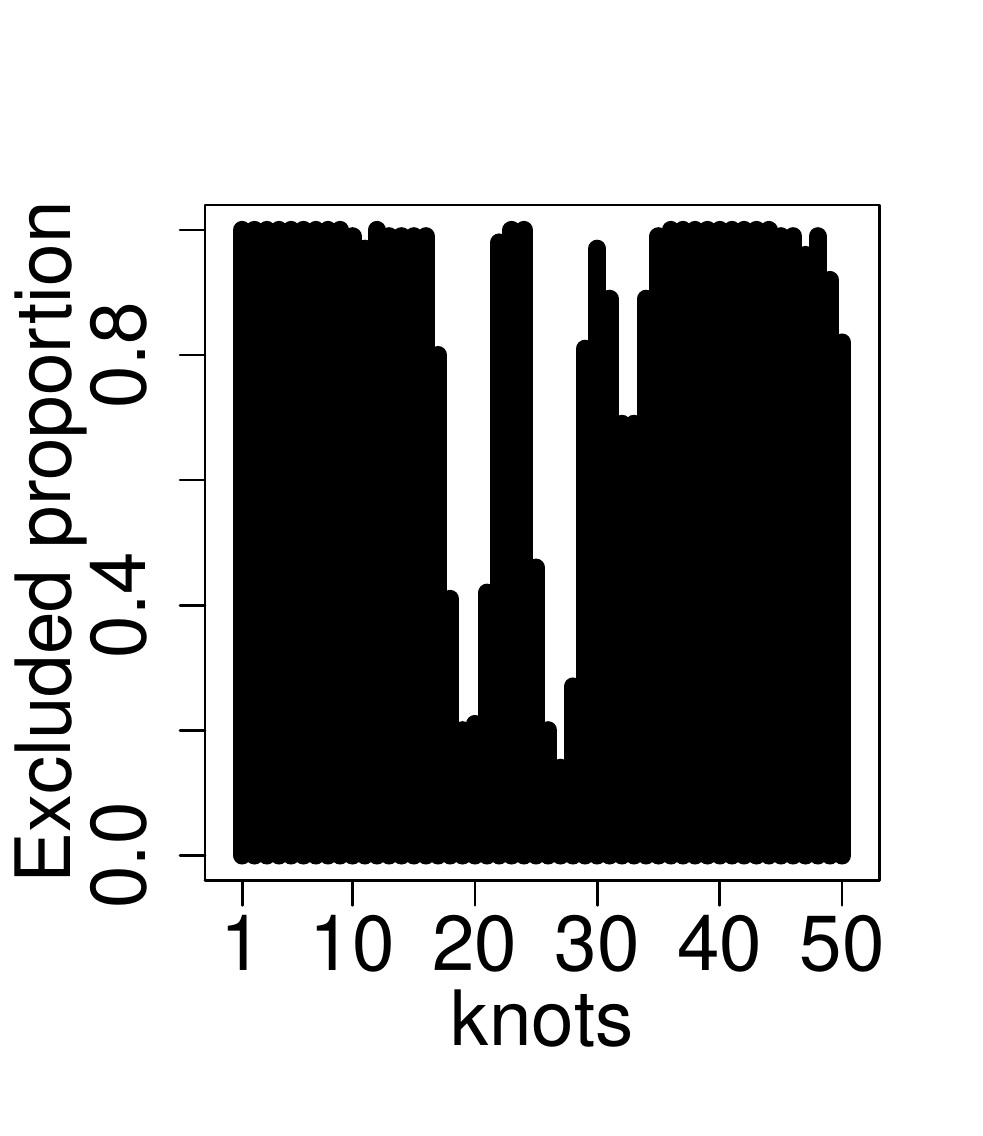}}\\
\end{tabular}
\end{center}\vspace{-0.5cm}
\caption{Proportion of excluded knots and the average of the fittings (dashed line), K=50.}\label{fig:ajuste_1bump_k50}
\end{figure}

Figures  \ref{fig:ajuste_1bump_k10}, \ref{fig:ajuste_1bump_k30} and  \ref{fig:ajuste_1bump_k50} show some results for $K = 10, 30 $ and $ 50 $ knots, respectively. In the first line of graphics there are the plots of the data generated for one of the replicates (dots), the true curve (solid line) and the average of the fittings of the 100 replicates for the three selection criteria (dashed lines). The second line shows the proportion of excluded knots, also for the three selection criteria CI, SN and BF.

Observe that the positions of the knots most selected as significant are in the rise and fall of the bump. In addition, it should also be noted that as $ K $ increases, the three selection criteria tend to be more rigorous in the penalty, hence, excluding more knots. This occurs more severely in the SN criterion, which presents an average of fittings  worse when $ K = 50 $. The results of the CI and BF criteria are similar in all cases. The results for exercises where the maximum number of knots are $ 20 $ and $ 40 $ have been omitted as they are similar, for $ K = 10 $ and $ K = 50 $, respectively

\begin{figure}[h!]
\begin{center}
\begin{tabular}{ccc}
{\includegraphics[scale=0.50]{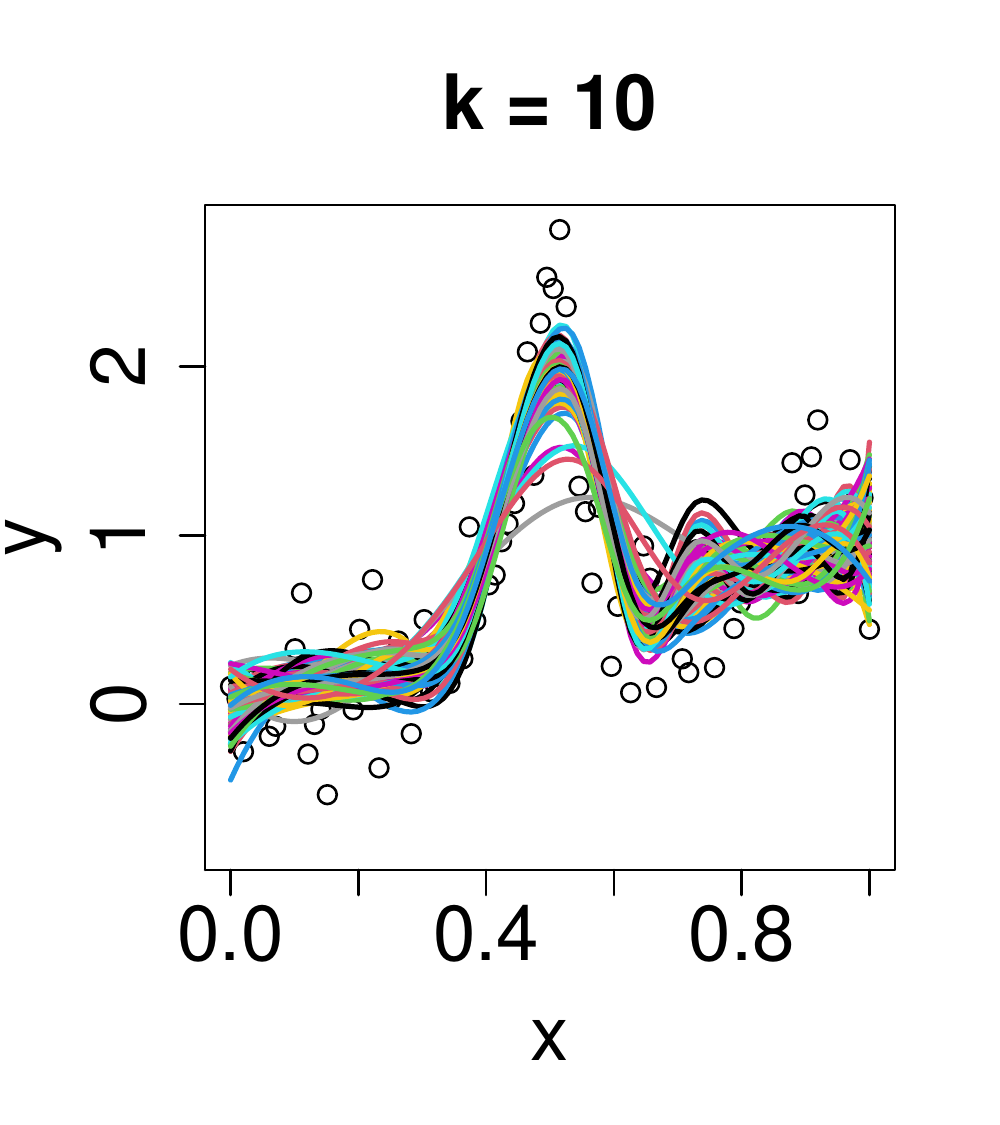}}&
{\includegraphics[scale=0.50]{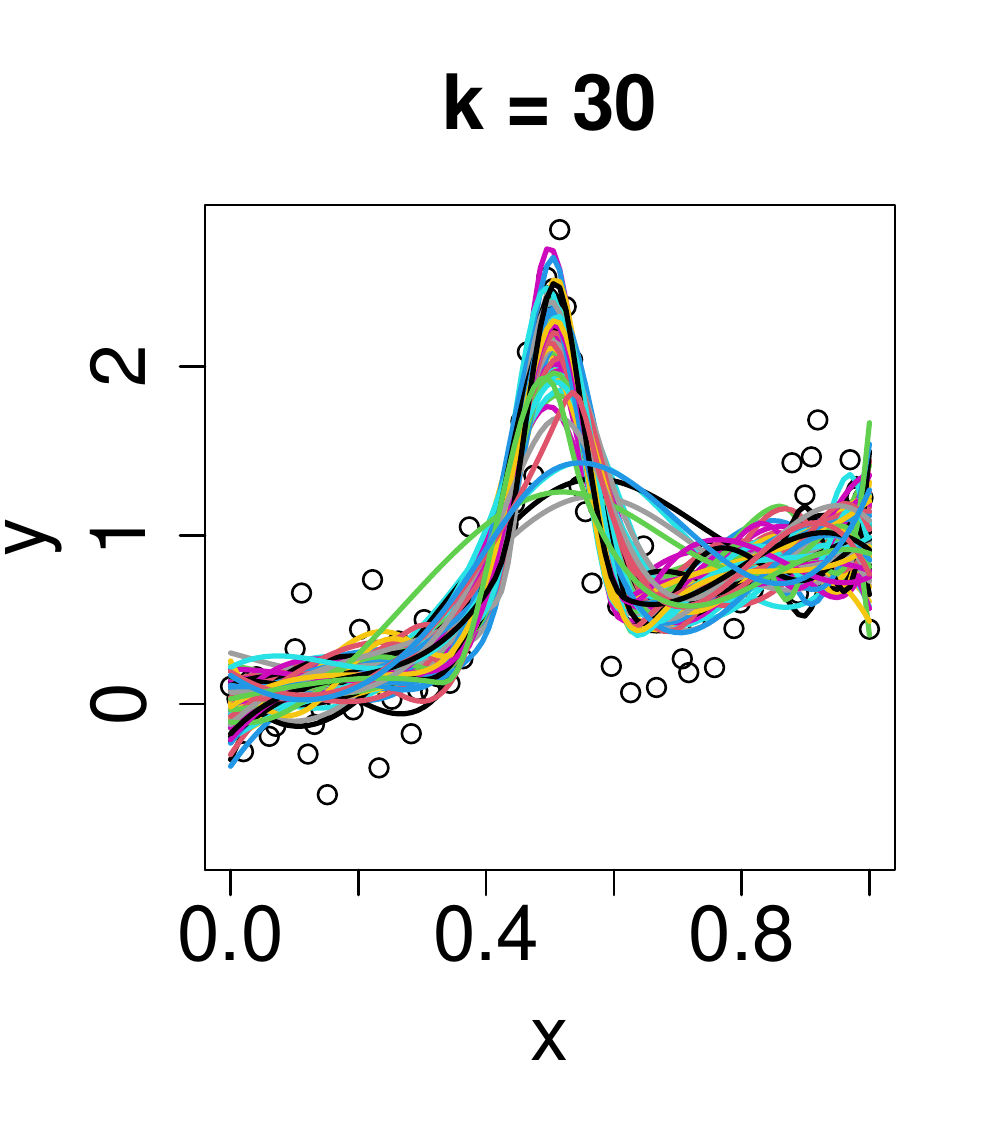}}&
{\includegraphics[scale=0.50]{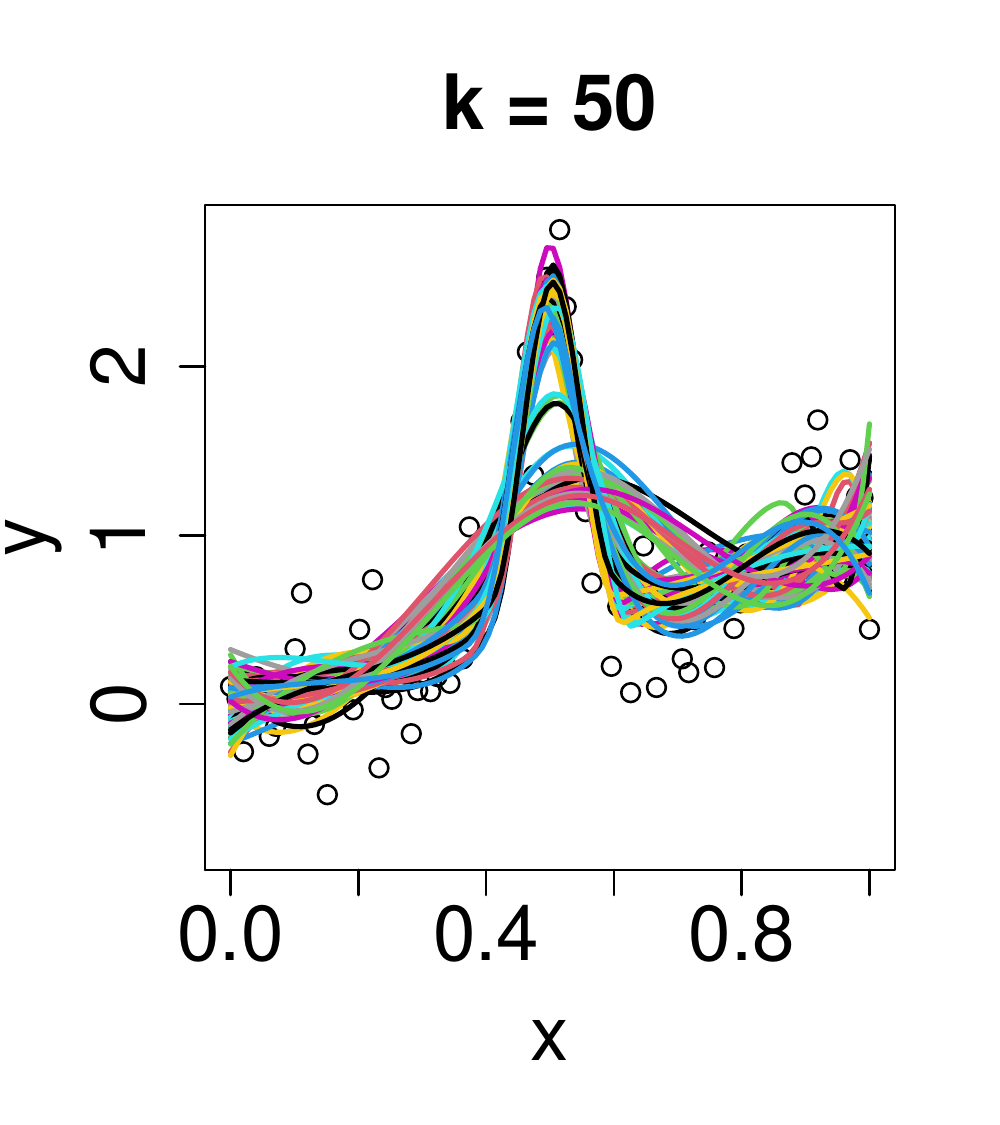}}
\\
\end{tabular}
\end{center}\vspace{-0.5cm}
\caption{Fittings for each of the 100 replicates, for different numbers of knots and according to the selection criteria BF.}\label{fig:100ajuste_1bump}
\end{figure}

Figure \ref{fig:100ajuste_1bump} exhibits the 100 fitted models for each of the replicates considering $ K = 10 $, $ 30 $ and $ 50 $, and the BF selection criterion. One can see that the variability increases as the value of $ K $ increases. The same occurs when considering the other selection criteria.

\begin{figure}[h!]
\begin{center}
\begin{tabular}{cc}
{\includegraphics[scale=0.55]{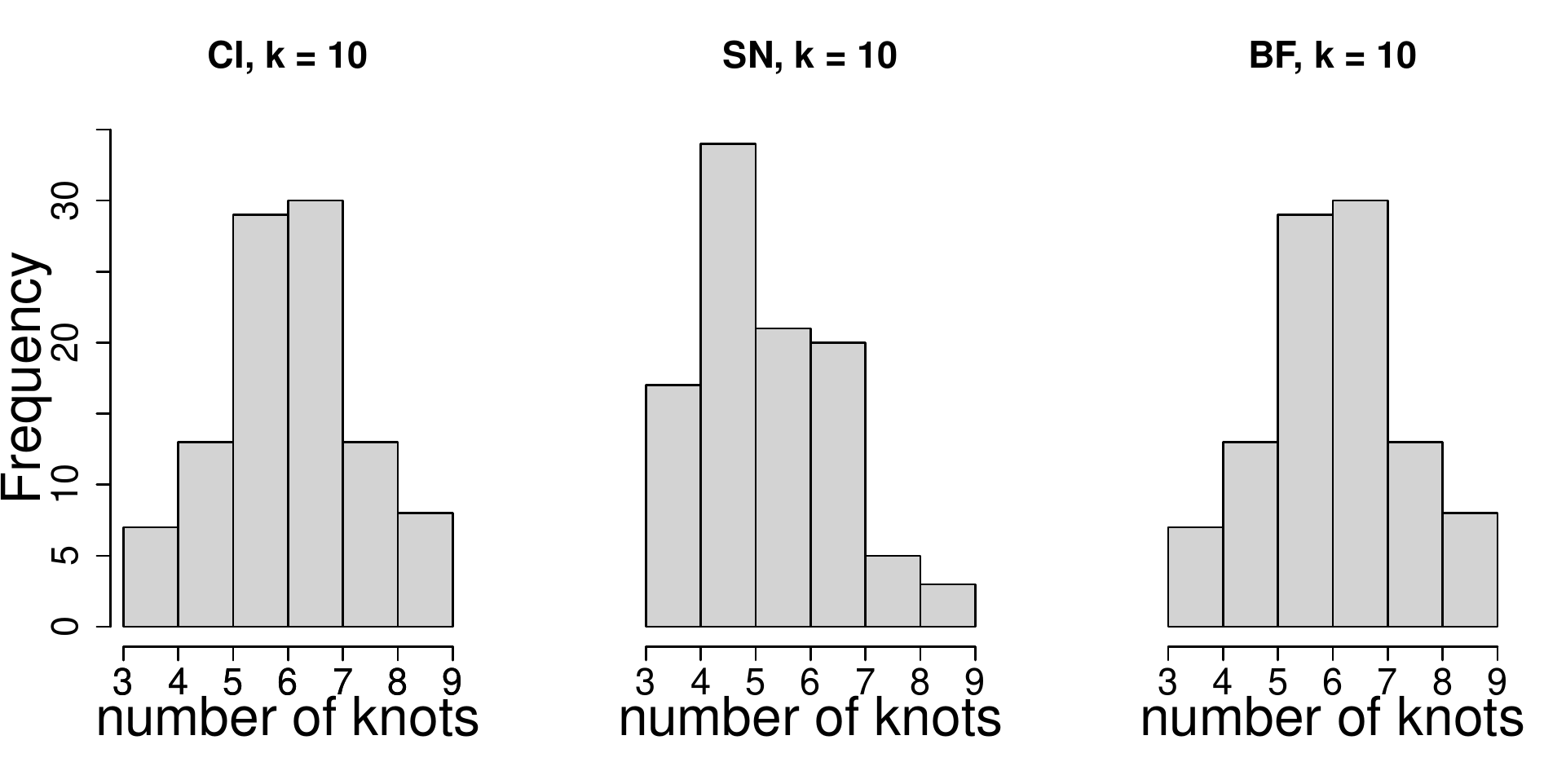}}\\
{\includegraphics[scale=0.55]{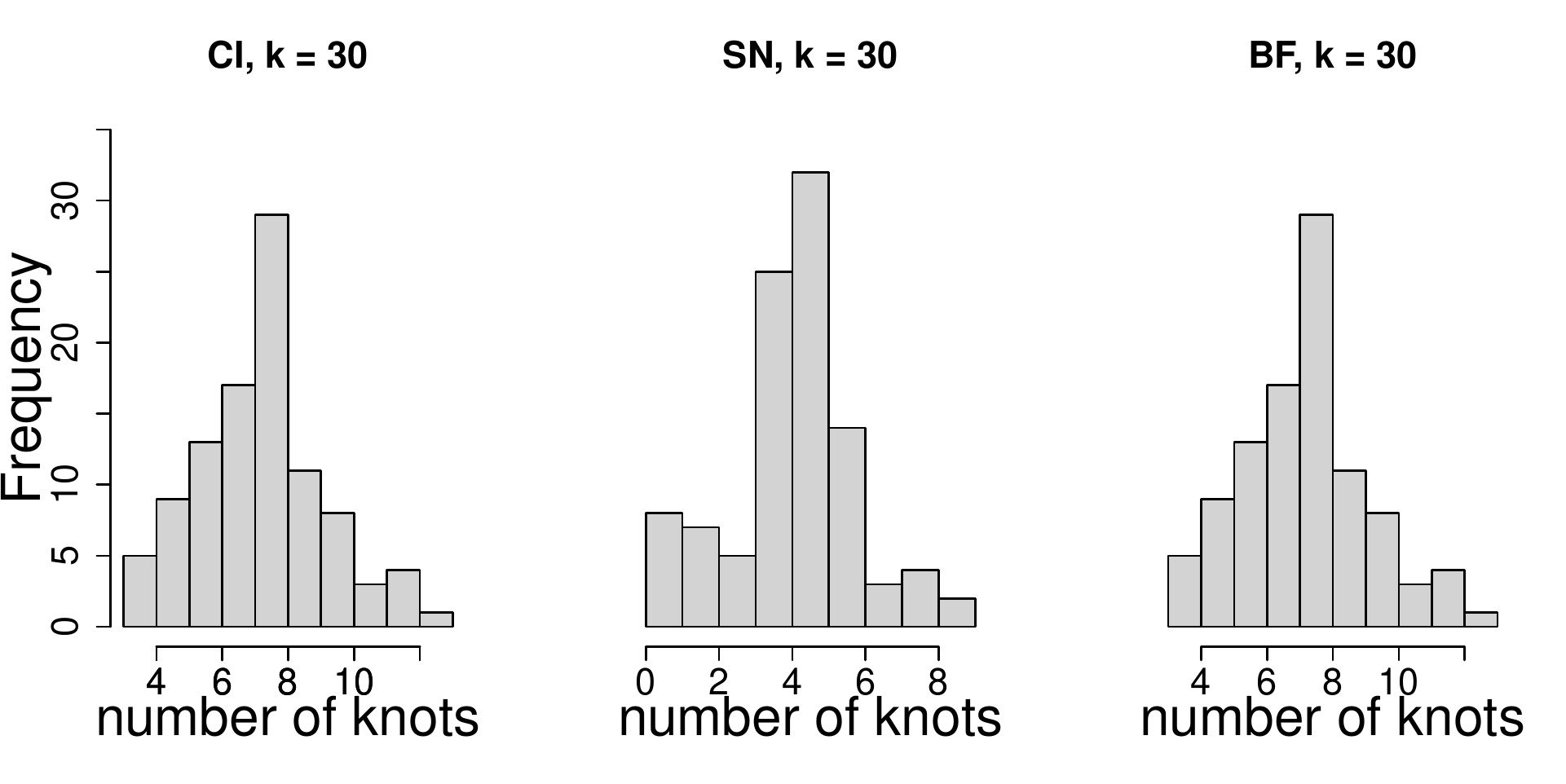}}\\
{\includegraphics[scale=0.55]{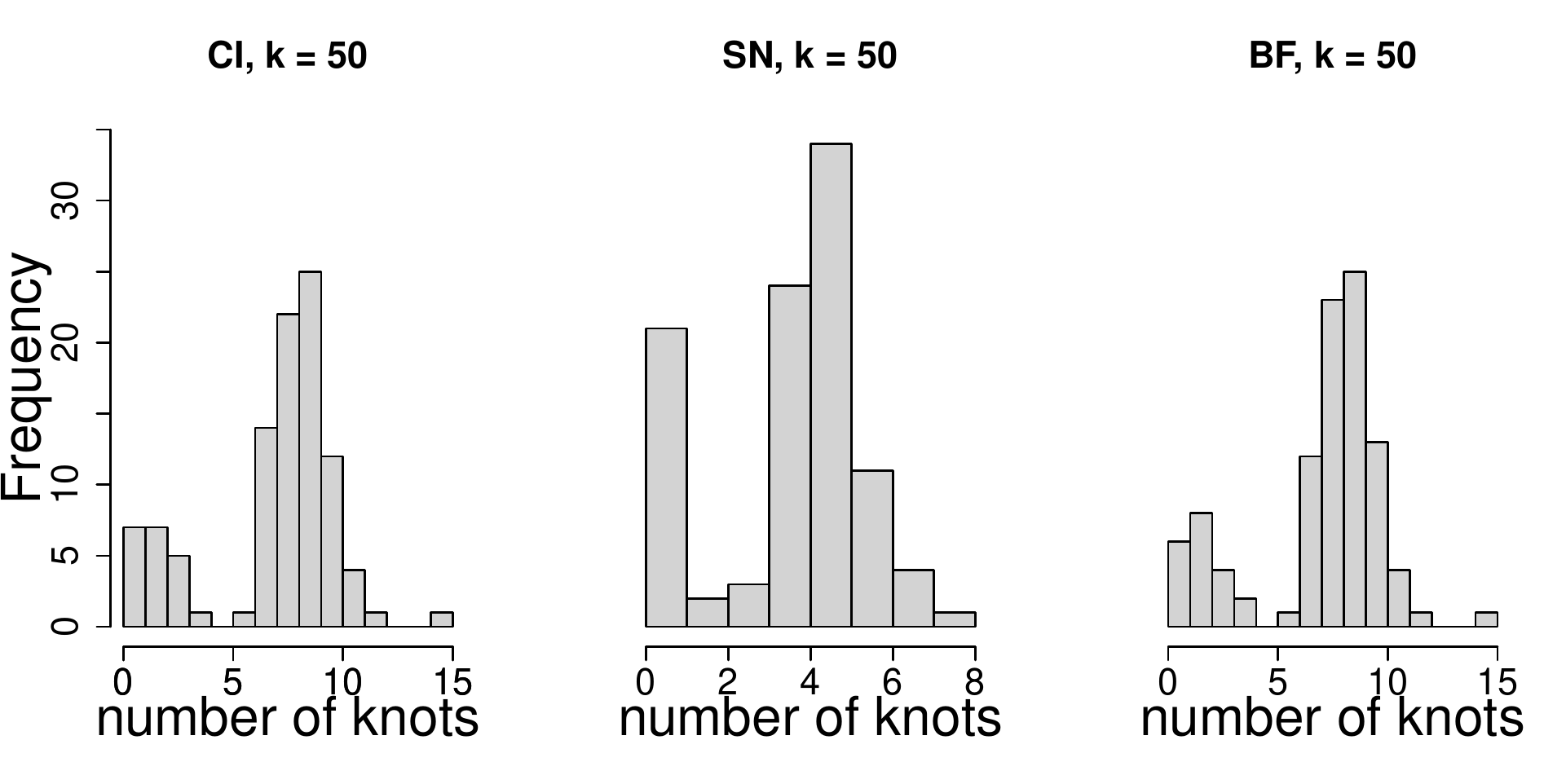}}\\
\end{tabular}
\end{center}\vspace{-0.5cm}
\caption{Frequency of the number of  selected knots in the 100 replicates.}\label{fig:freq_nos_1bump}
\end{figure}

Figure \ref{fig:freq_nos_1bump} shows the frequency of the number of knots selected in the 100 replicates for each of the selection criteria (CI, SN and BF) and maximum number of knots ($ K = 10, 30 $ and $ 50 $). Note that CI and BF criteria,  when $K = 10$, select between 5 and 6 knots as the most frequent ones. On the other hand, when $ K = 30 $ knots  7 knots among them are selected more frequently. In the case where the maximum number of knots is $ K = 50 $ we have a bimodal behavior for the frequency of the number of selected knots and one can see  that the penalty is more severe as $ K $ grows.

ELBO as a model comparison measure can be used to propose the maximum number of knots. It is worth mentioning that the larger the ELBO the more that model is preferable. From the Table \ref{tab:elbo_medio_1bump}, which presents the average ELBO for each value of $ K $, we have that the model proposed with $ K = 30 $ knots  which selects more frequently 7 of these knots as significant, is preferable.
\begin{table}[h!]
\caption{Average ELBO } 
\begin{center}
{\footnotesize
\begin{tabular}{c|c|c|c|c|c}
  \hline
Criterion & k = 10 & k = 20 & k = 30 & k = 40 & k = 50 \\
\hline FB & -32.24 & -16.10 & {\bf{-2.55}}  & -19.59  & -22.23 \\
\hline IC & -32.24 & -16.37 & {\bf{-2.55}} & -20.91 & -46.14\\
\hline SN & {\bf{-37.15}} & -39.14 & -69.16 & -244.80 & -390.80\\
\hline 
\end{tabular}}
\end{center}\label{tab:elbo_medio_1bump}
\end{table}

\clearpage

\subsubsection{Exercise 5: Double structure/Two bumps}


Similar to exercise 4, we proposed exercise 5, however, now considering a curve with 2 bumps. The curve $f$ is a  mixing two normal distributions:
$$f(x) = 0.3 N(x|0.4,0.01) + 0.7 N(x|0.8,0.01),$$
where $N(x|a,b)$ is a normal distribution with mean $a$ and variance $b$. In this exercise we assume $n=300$ and the other conditions of simulation 4 were kept. That is, 100 replicates, $\phi= 1/0.3$, $x_i$ equally space in $[0,1]$, $p=3$, knots placed at  the $x_i$ quantiles and maximum number of knots $K= 10, 20, 30, 40, 50$. We omitted the graphics for $K=20$ and $K=40$, as they are similar.

\begin{figure}[h!]
\begin{center}
\begin{tabular}{ccc}
{\includegraphics[scale=0.50]{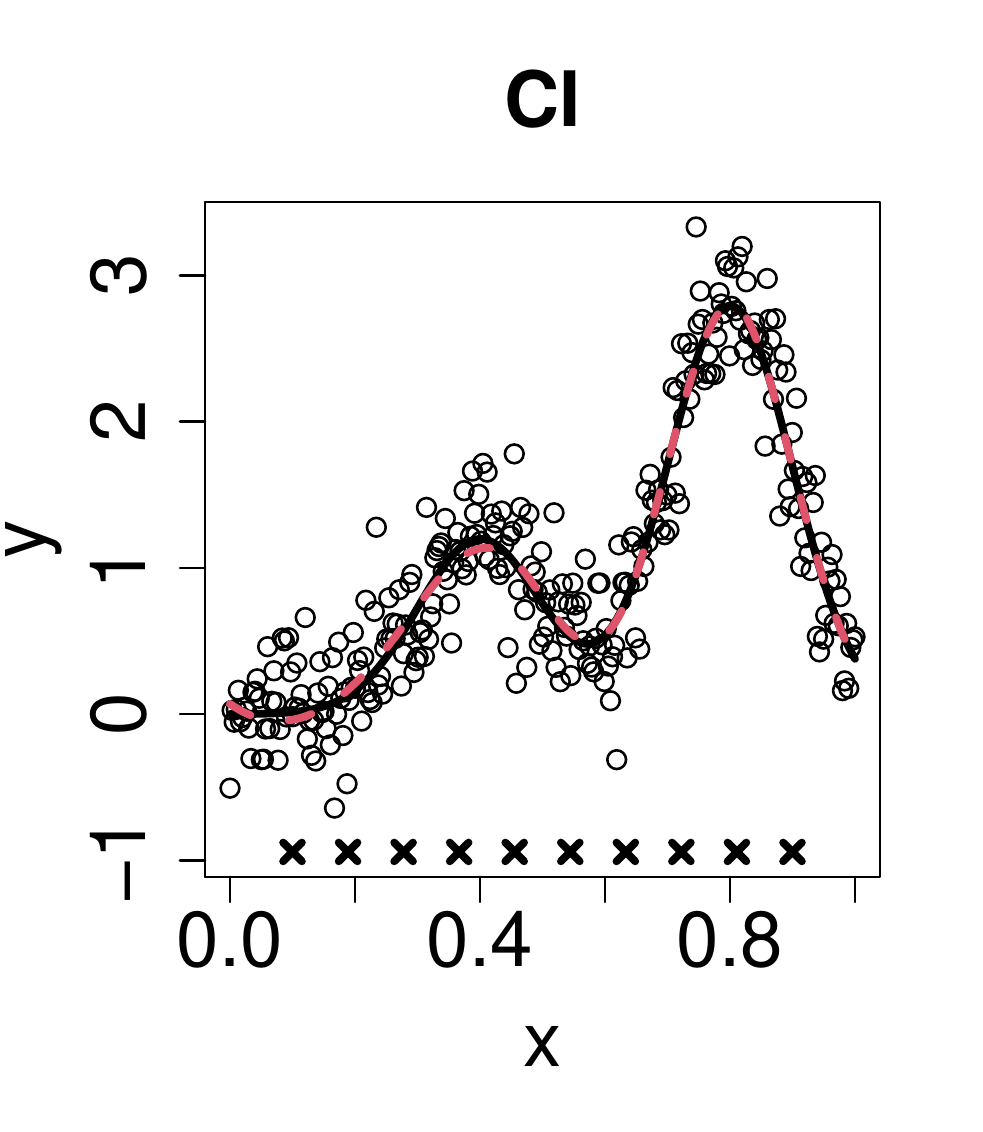}}&
{\includegraphics[scale=0.50]{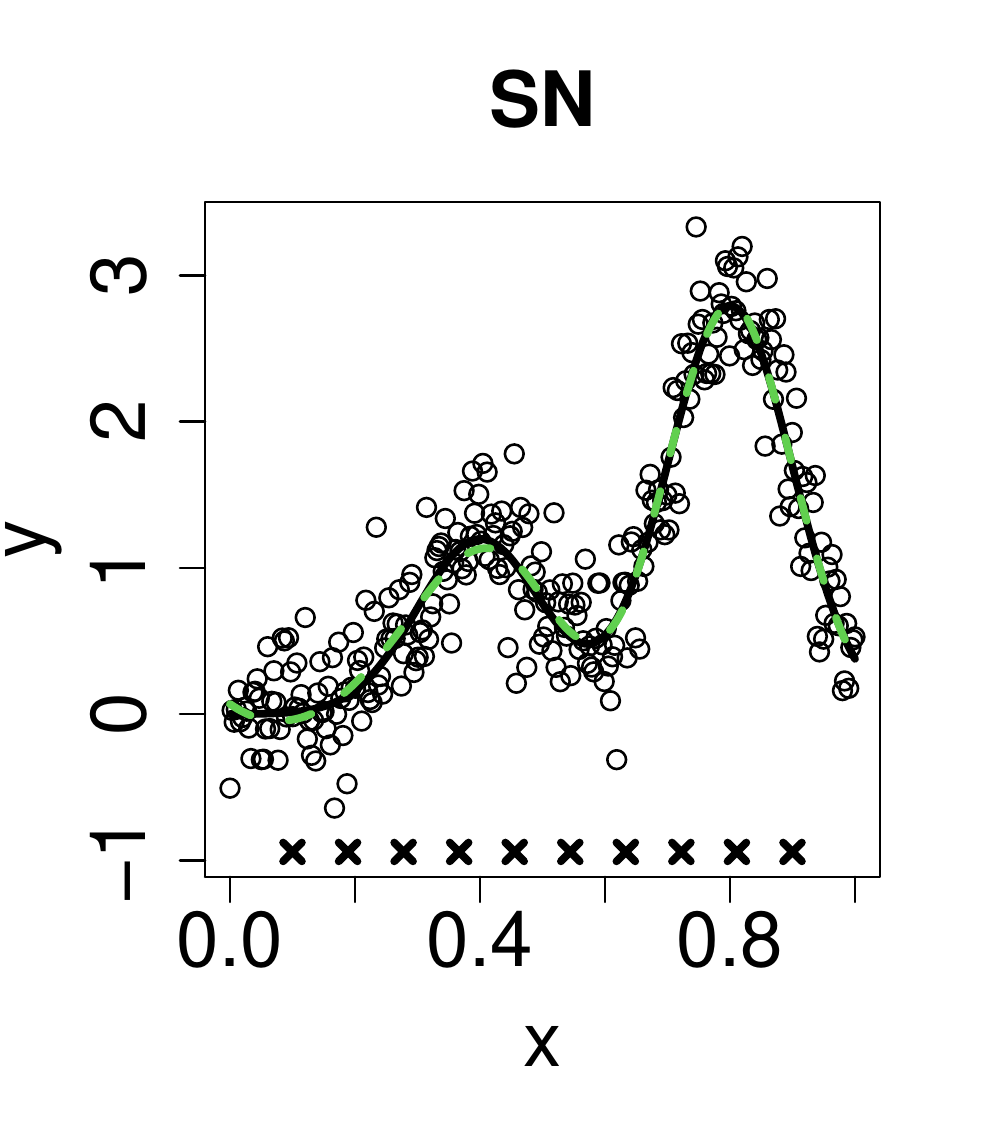}}&
{\includegraphics[scale=0.50]{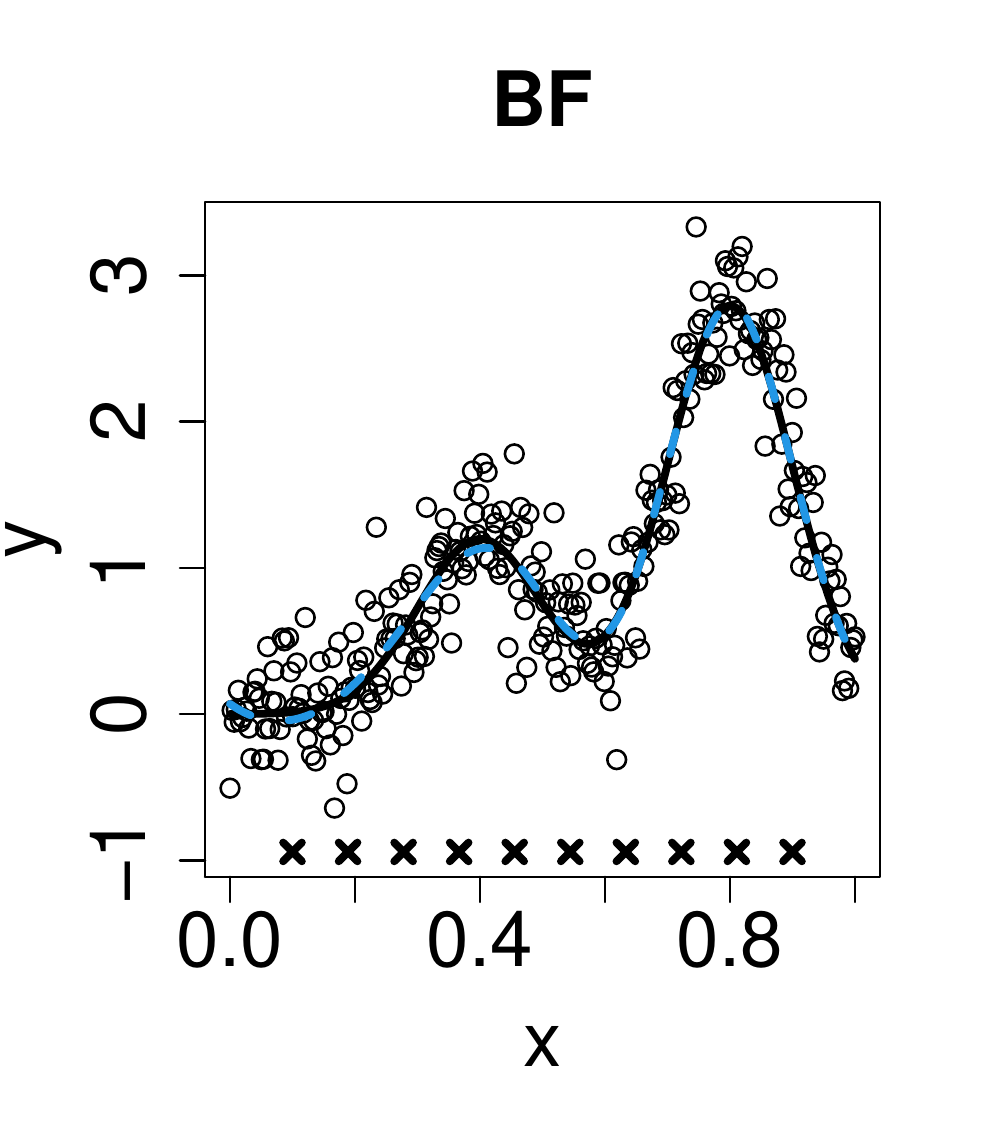}}\\
{\includegraphics[scale=0.50]{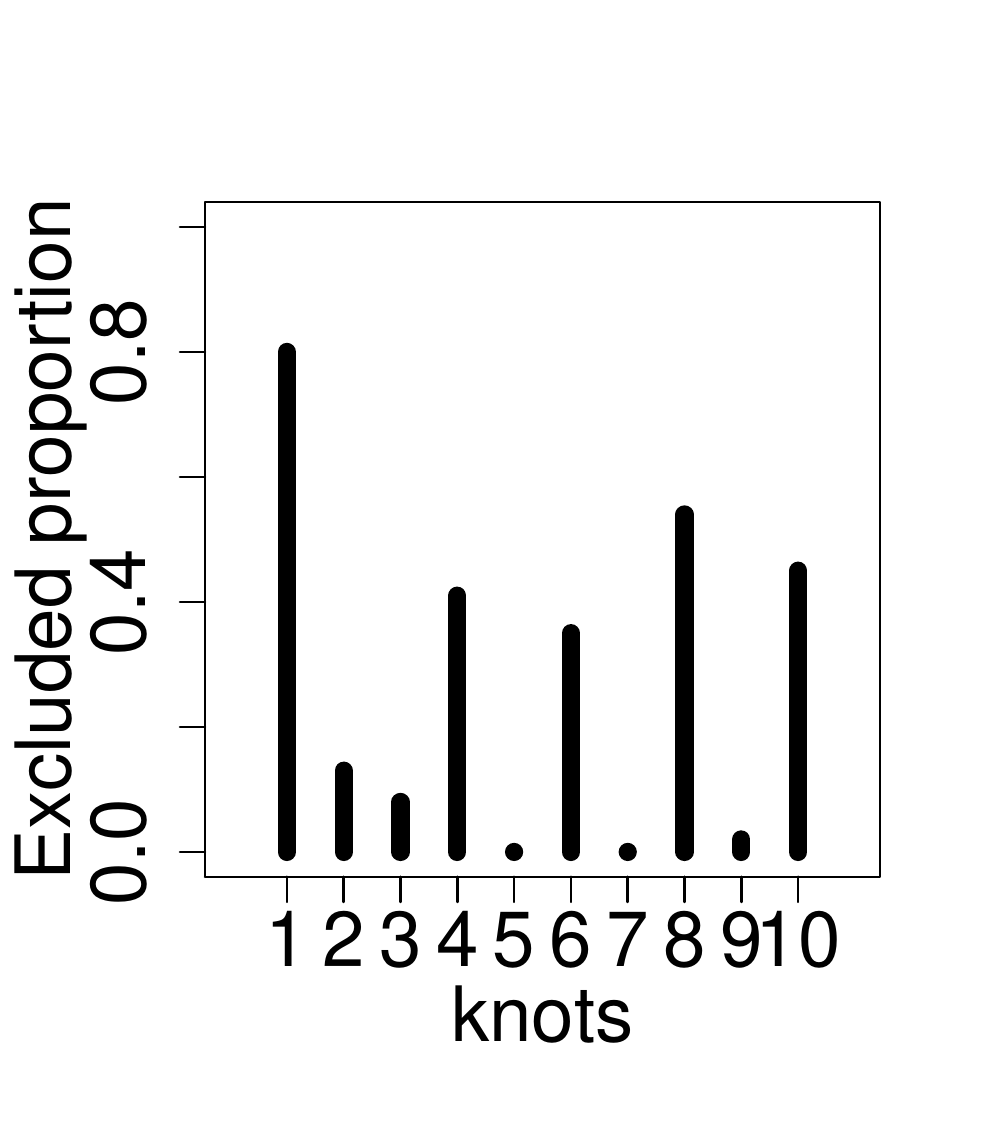}}&
{\includegraphics[scale=0.50]{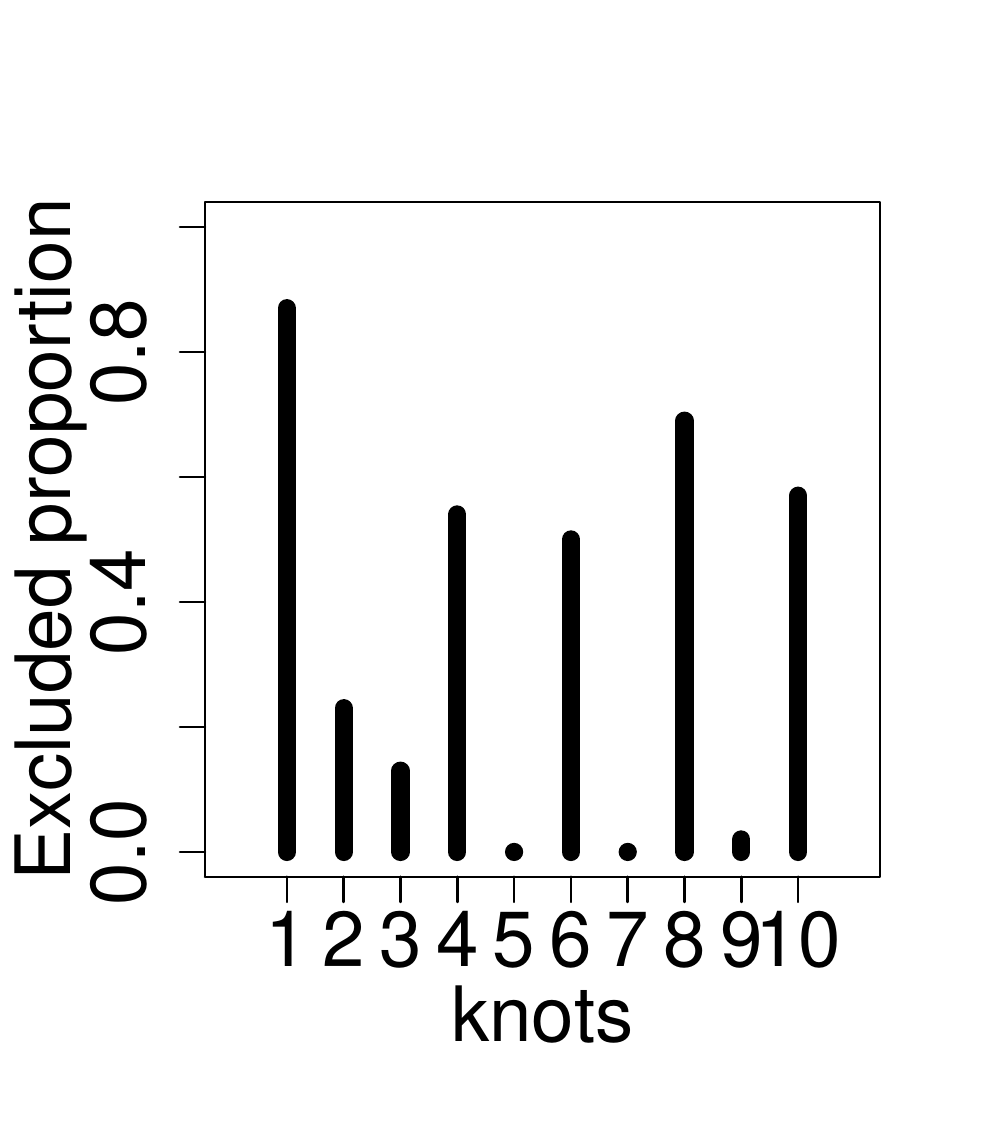}}&
{\includegraphics[scale=0.50]{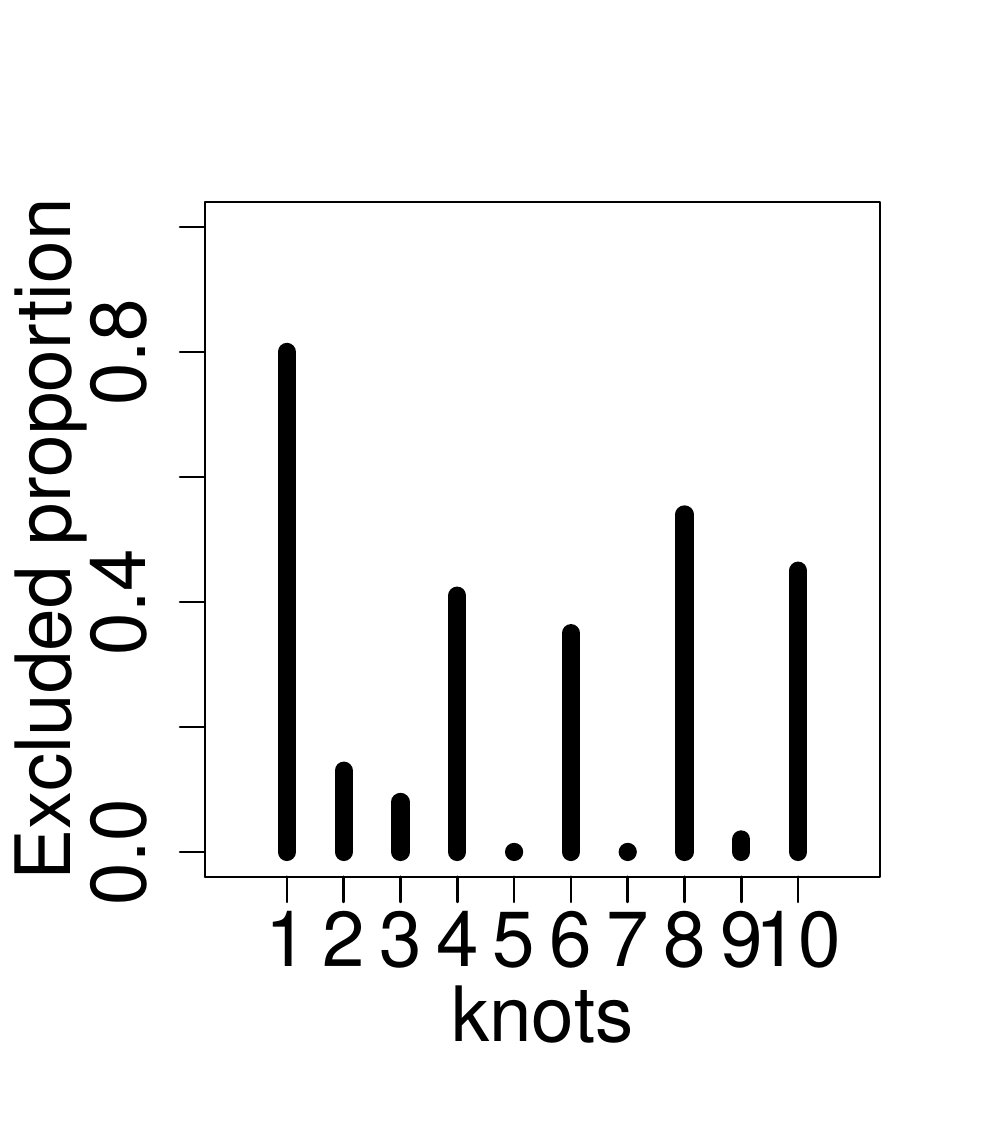}}\\
\end{tabular}
\end{center}\vspace{-0.5cm}
\caption{Proportion of excluded knots and the average of the fittings -  K=10.}\label{fig:ajuste_2bump_k10}
\end{figure}

\begin{figure}[h!]
\begin{center}
\begin{tabular}{ccc}
{\includegraphics[scale=0.50]{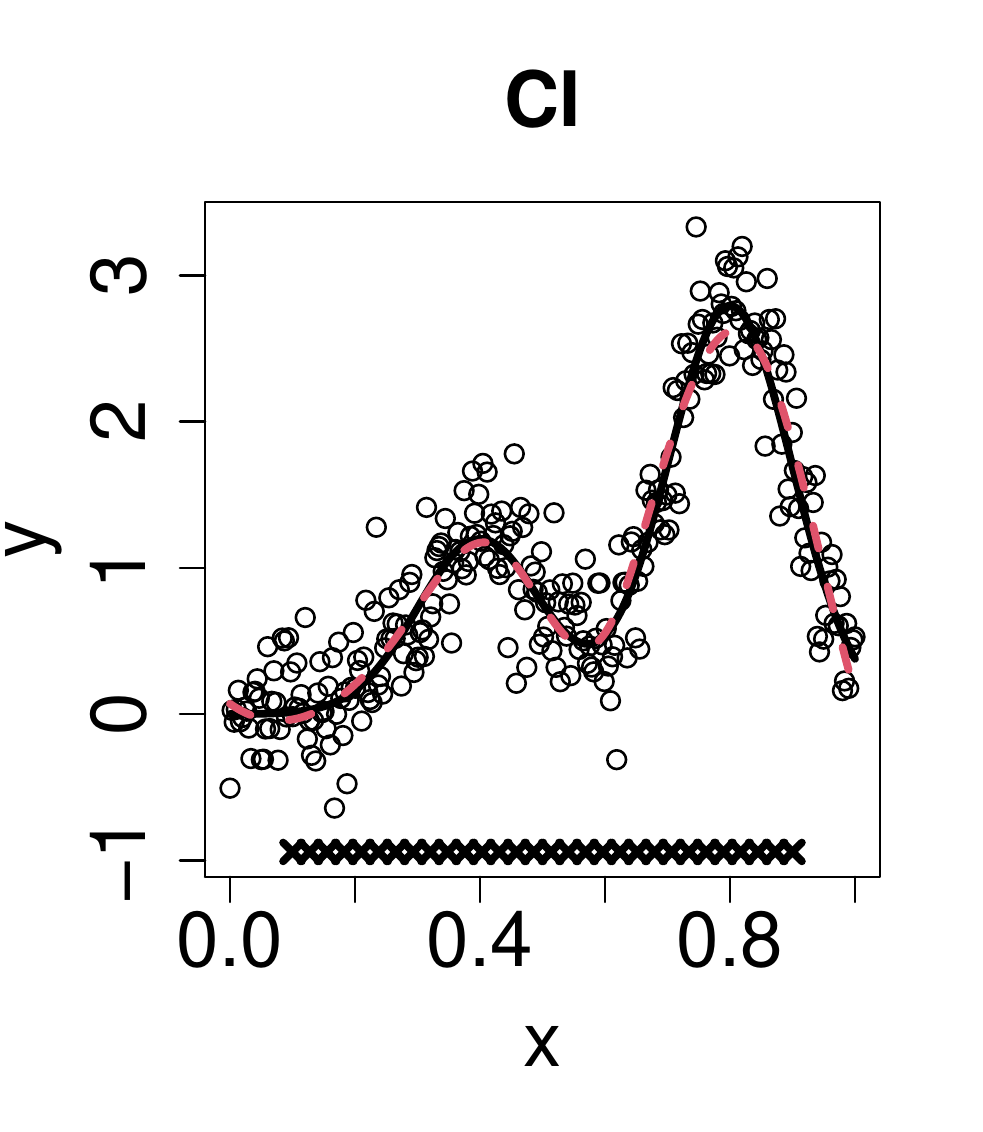}}&
{\includegraphics[scale=0.50]{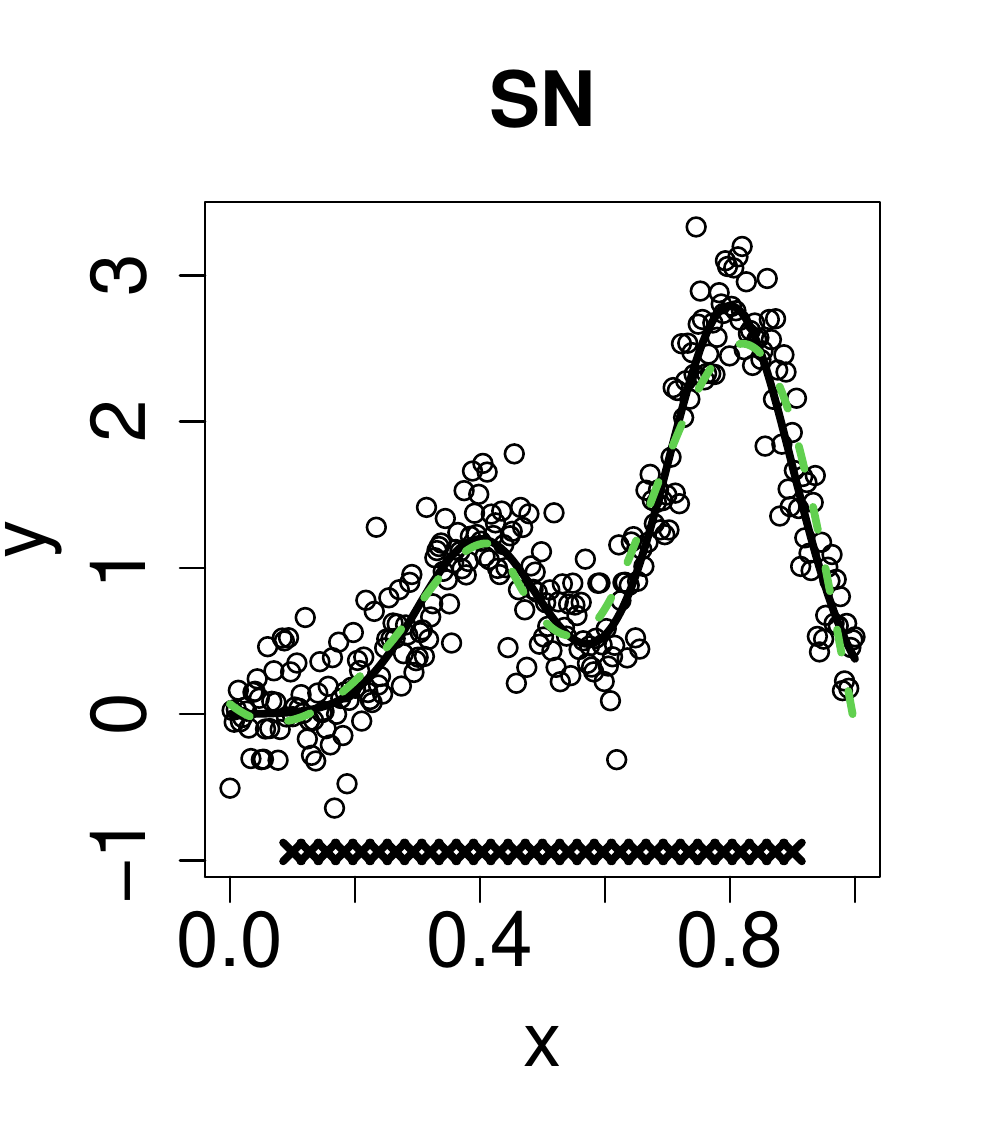}}&
{\includegraphics[scale=0.50]{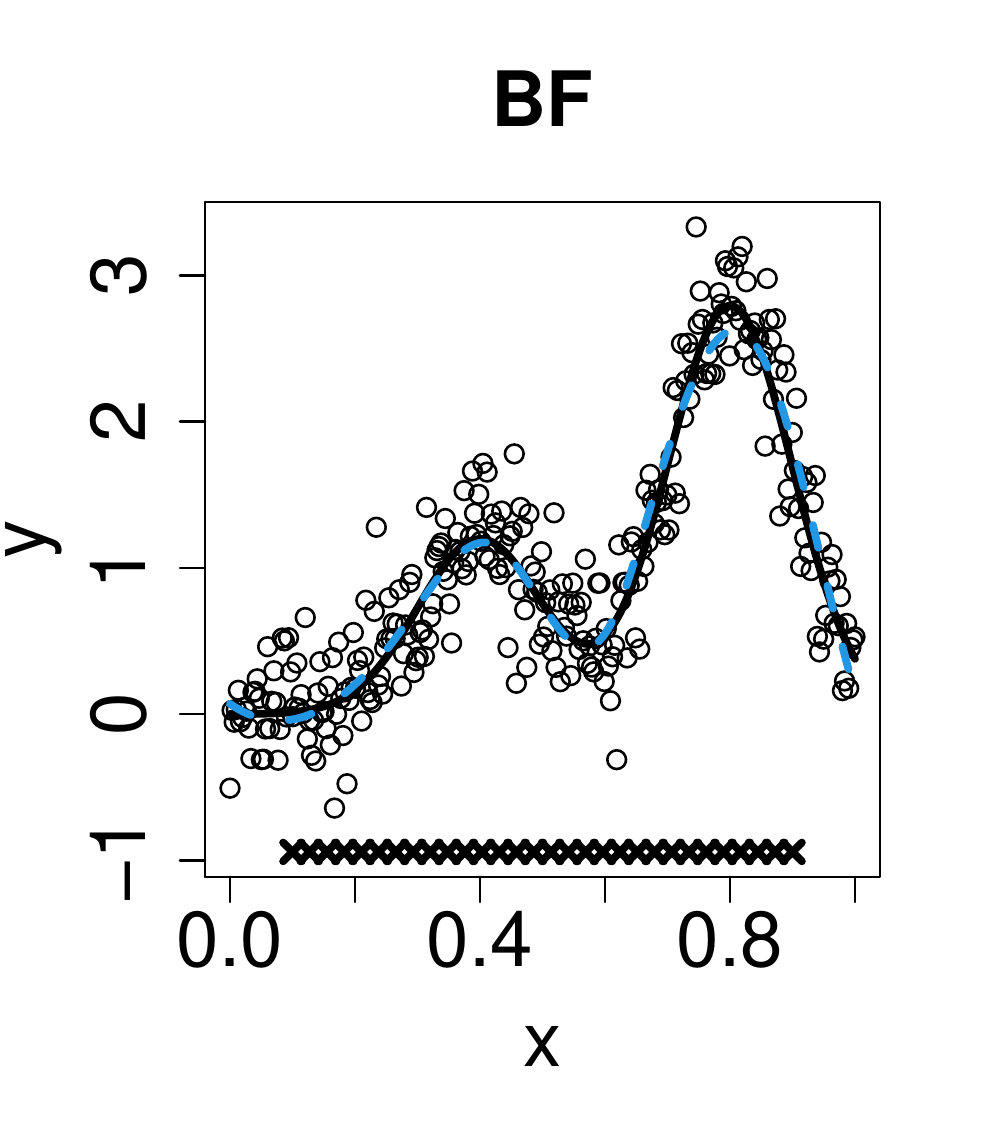}}\\
{\includegraphics[scale=0.50]{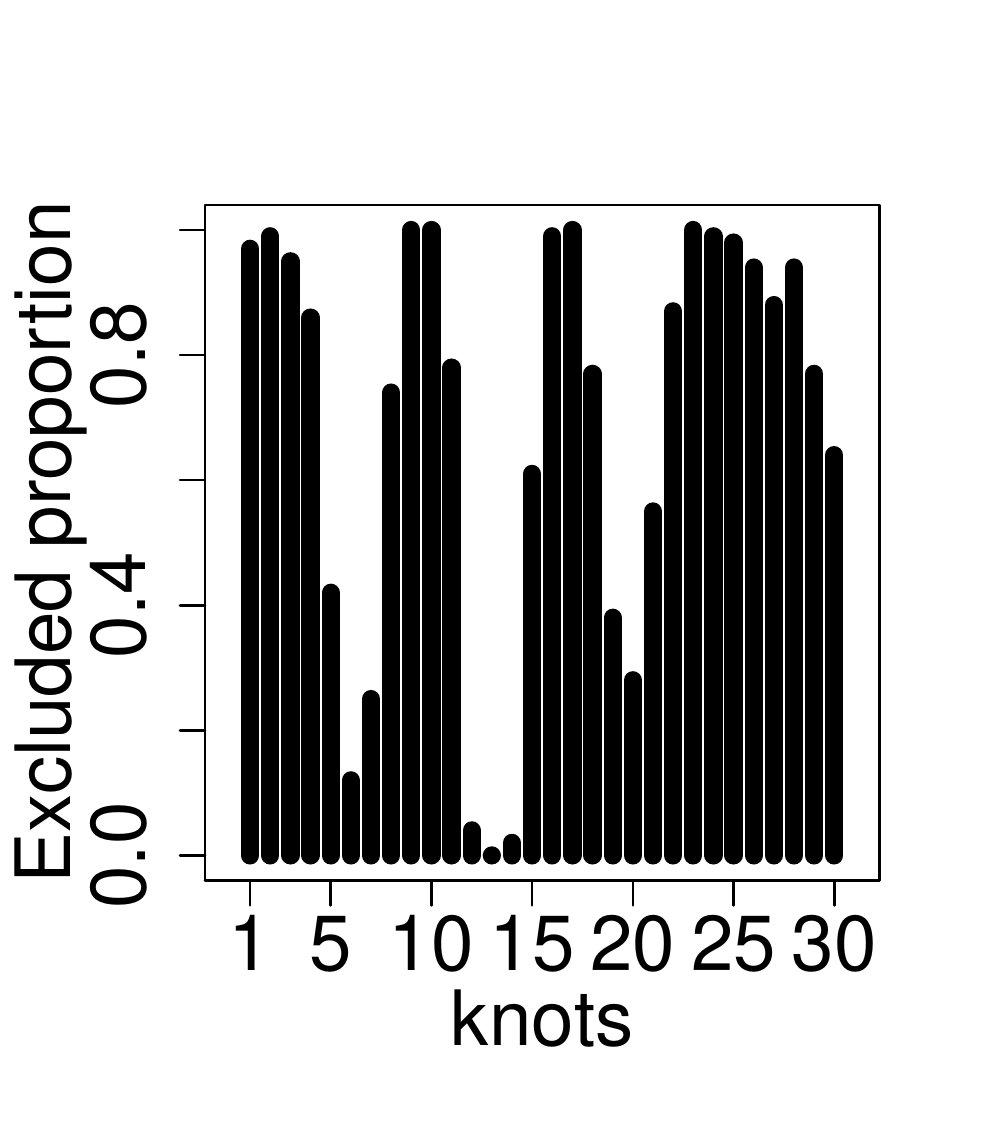}}&
{\includegraphics[scale=0.50]{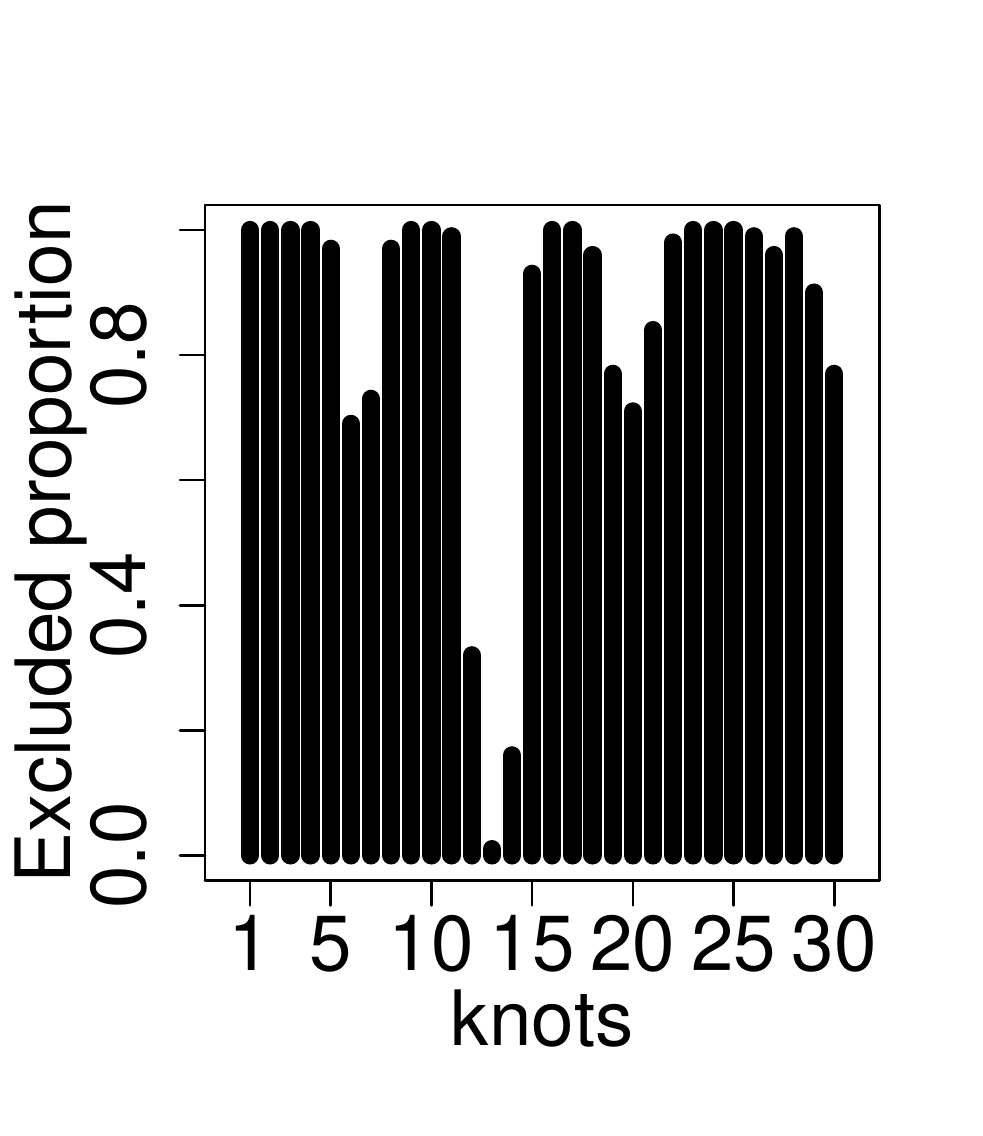}}&
{\includegraphics[scale=0.50]{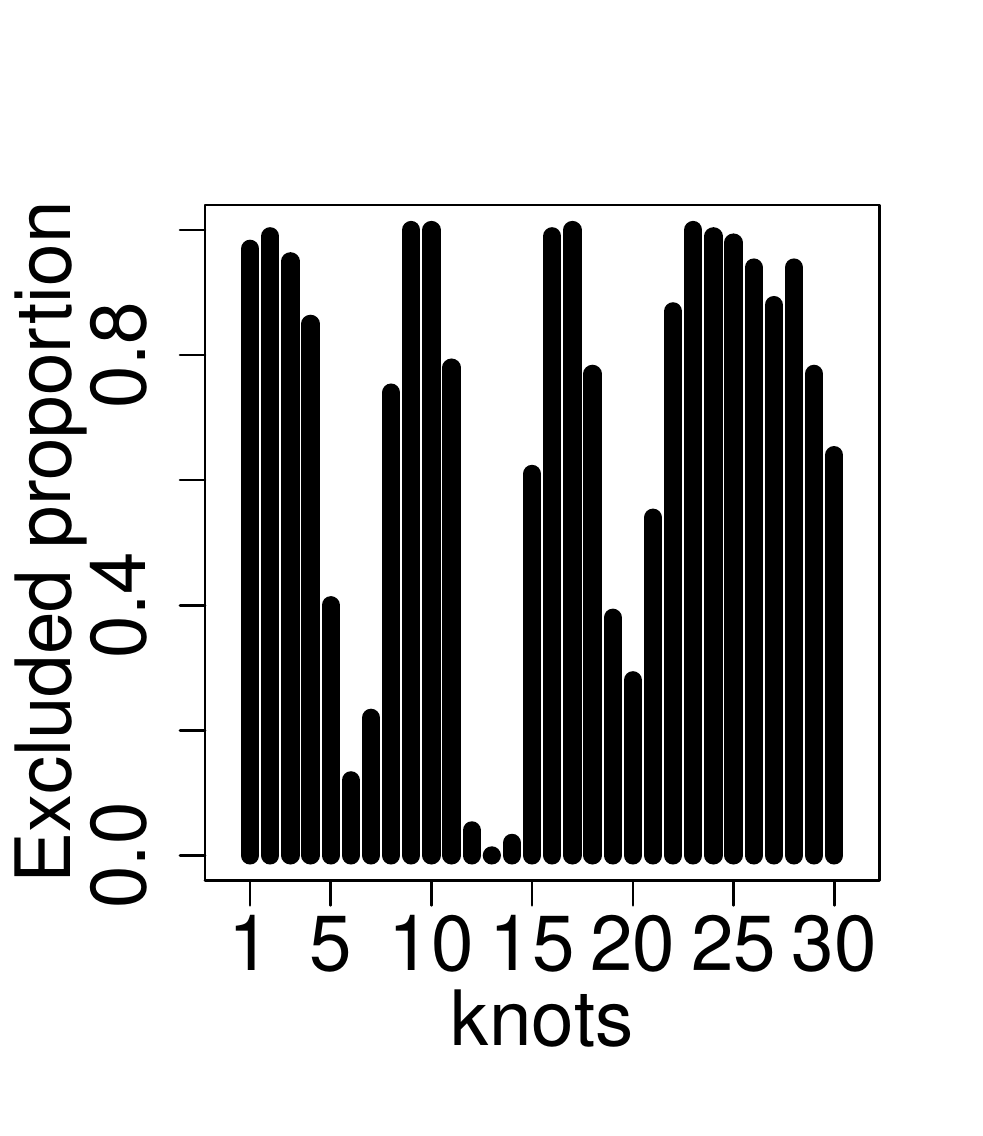}}\\
\end{tabular}
\end{center}\vspace{-0.5cm}
\caption{Proportion of excluded knots and the average of the fittings -  K=30.}\label{fig:ajuste_2bump_k30}
\end{figure}

\begin{figure}[h!]
\begin{center}
\begin{tabular}{ccc}
{\includegraphics[scale=0.50]{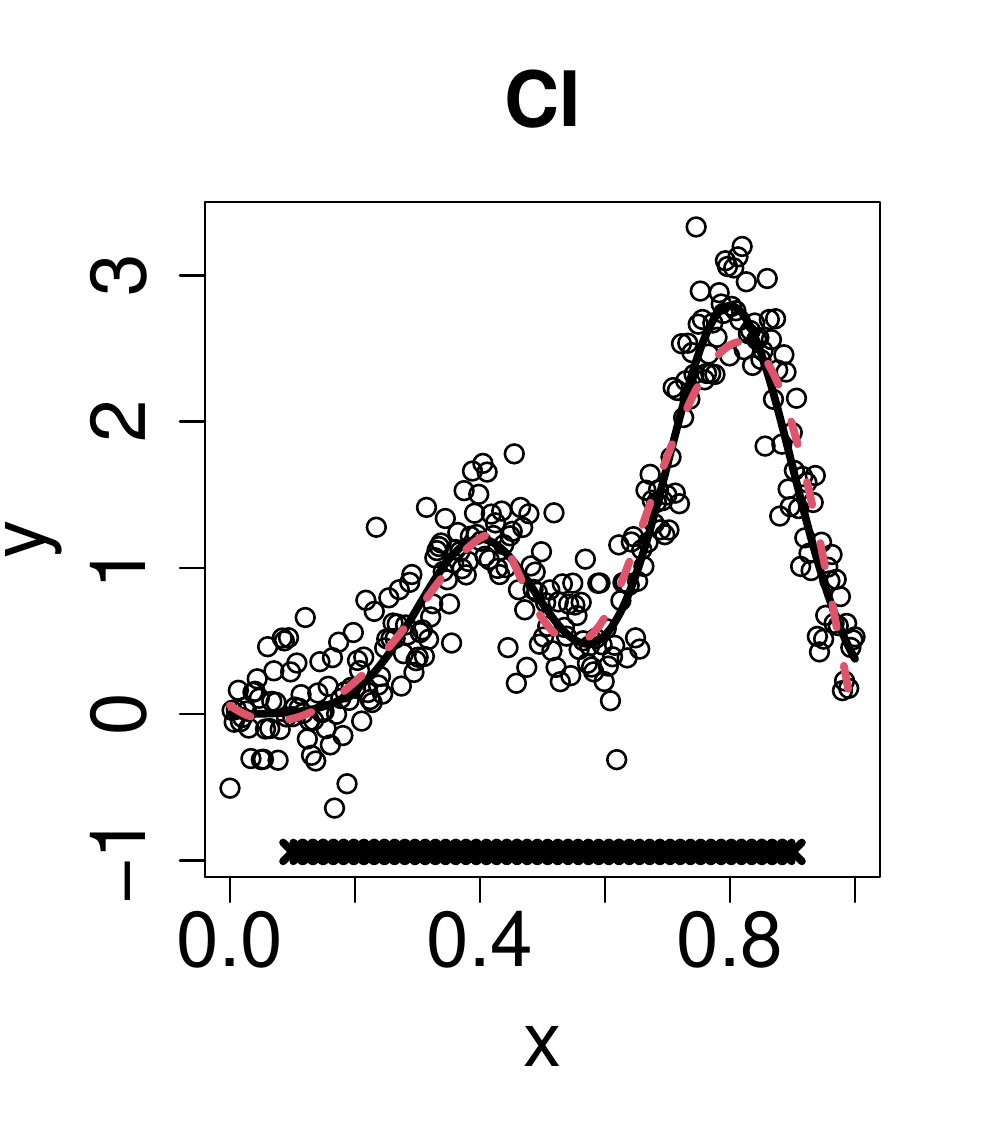}}&
{\includegraphics[scale=0.50]{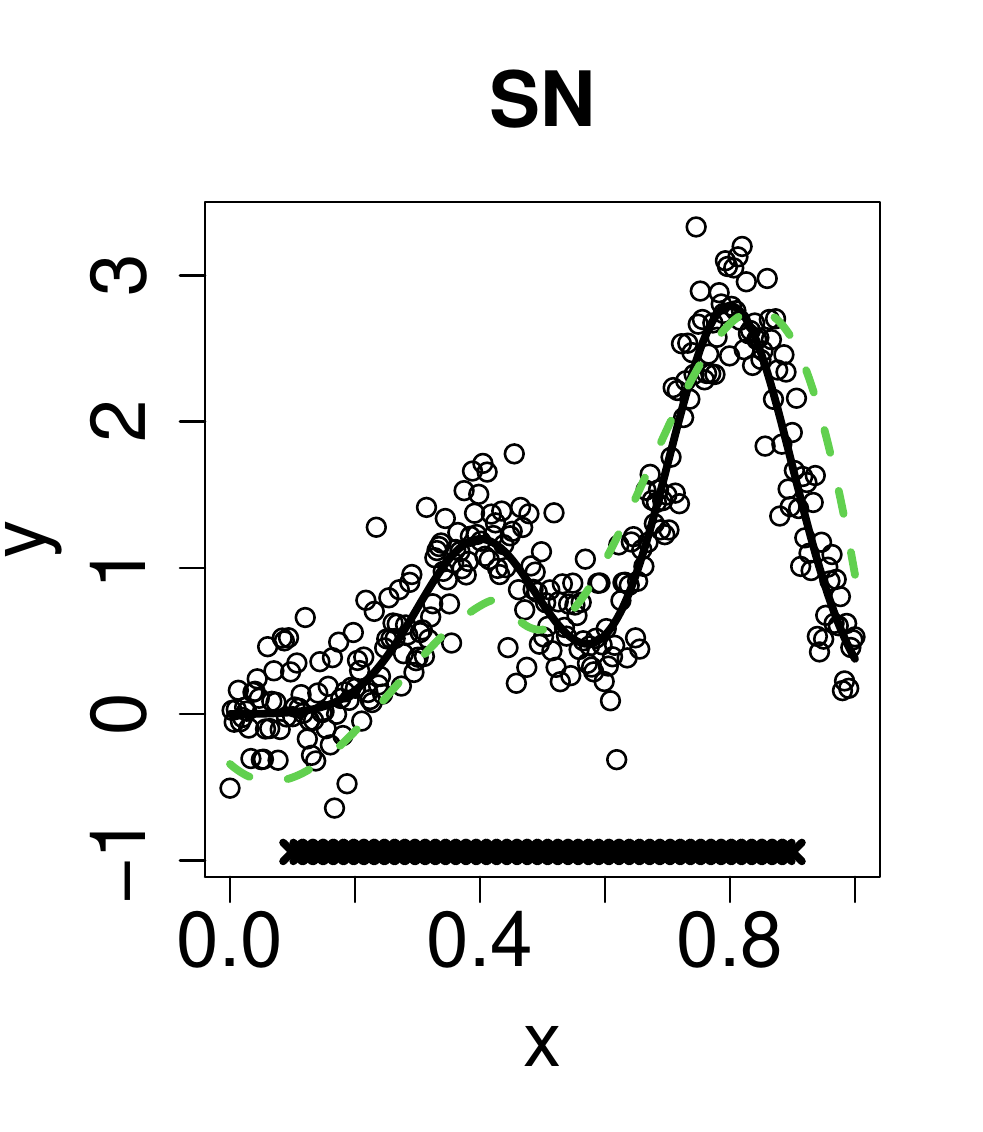}}&
{\includegraphics[scale=0.50]{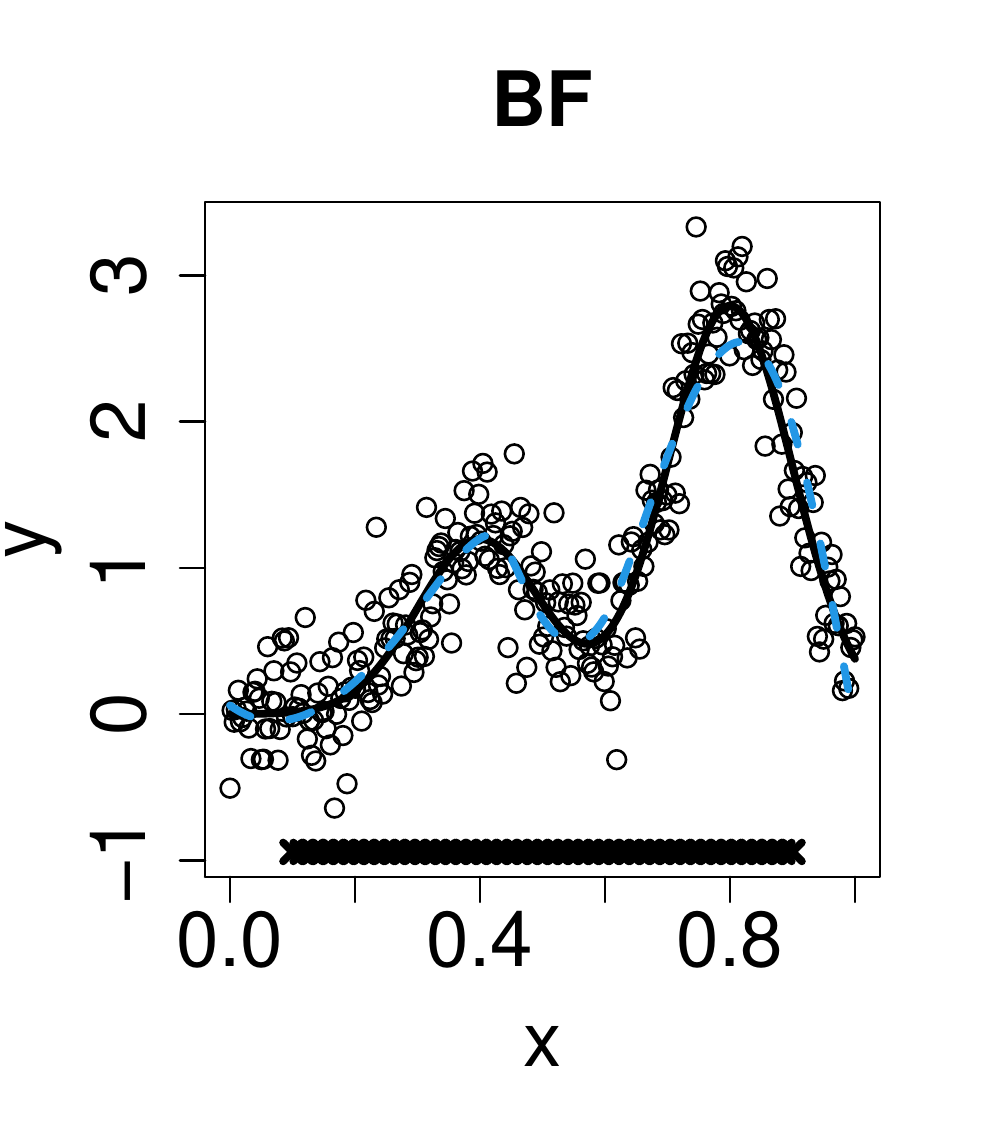}}\\
{\includegraphics[scale=0.50]{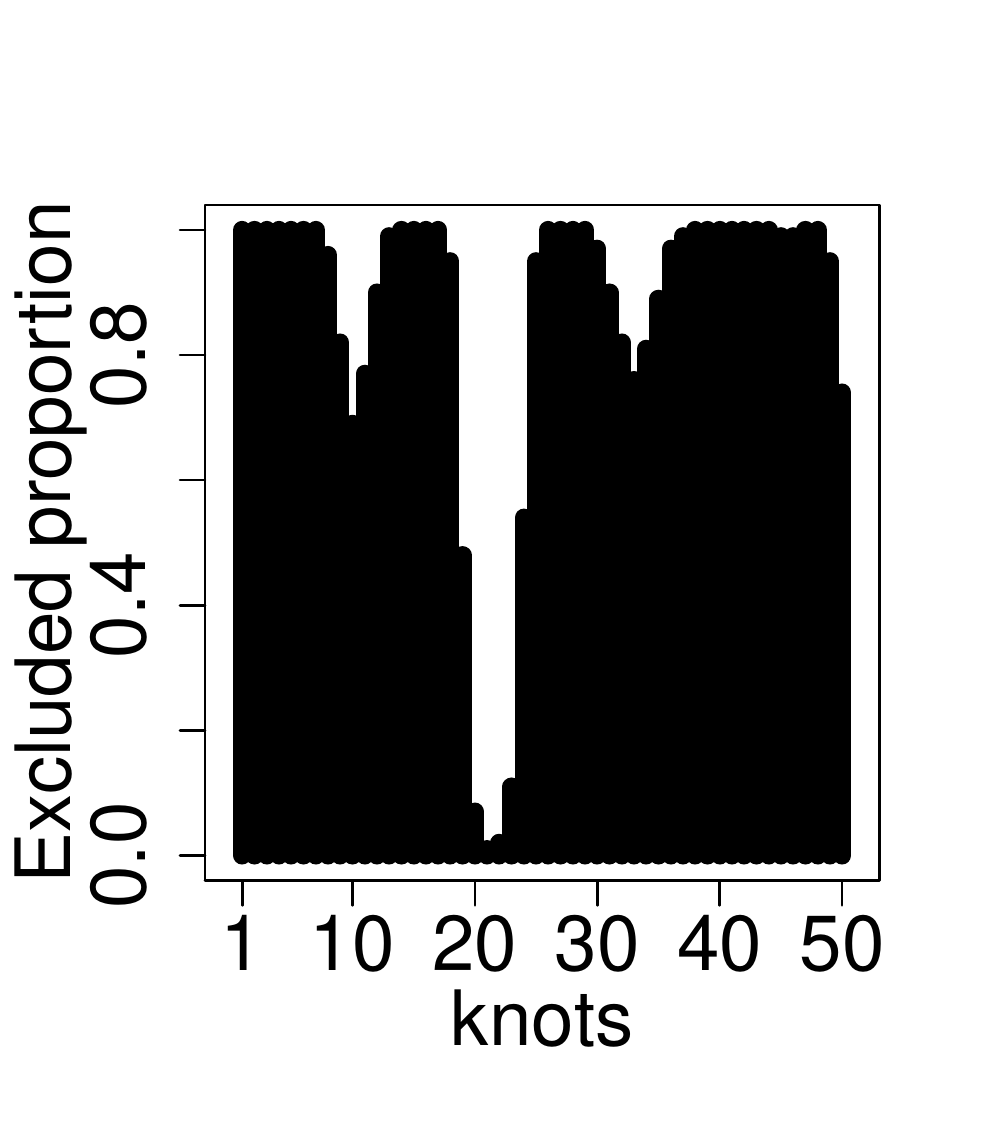}}&
{\includegraphics[scale=0.50]{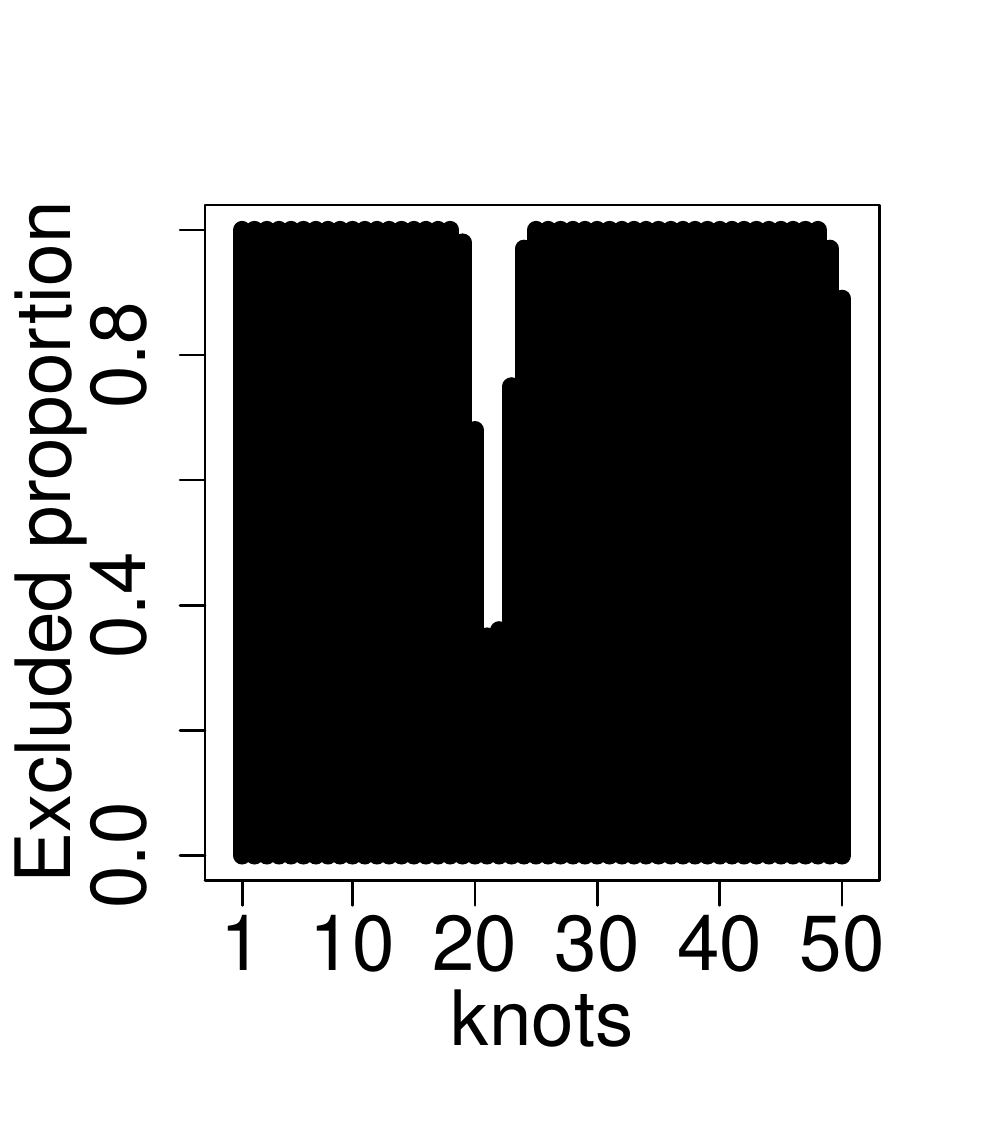}}&
{\includegraphics[scale=0.50]{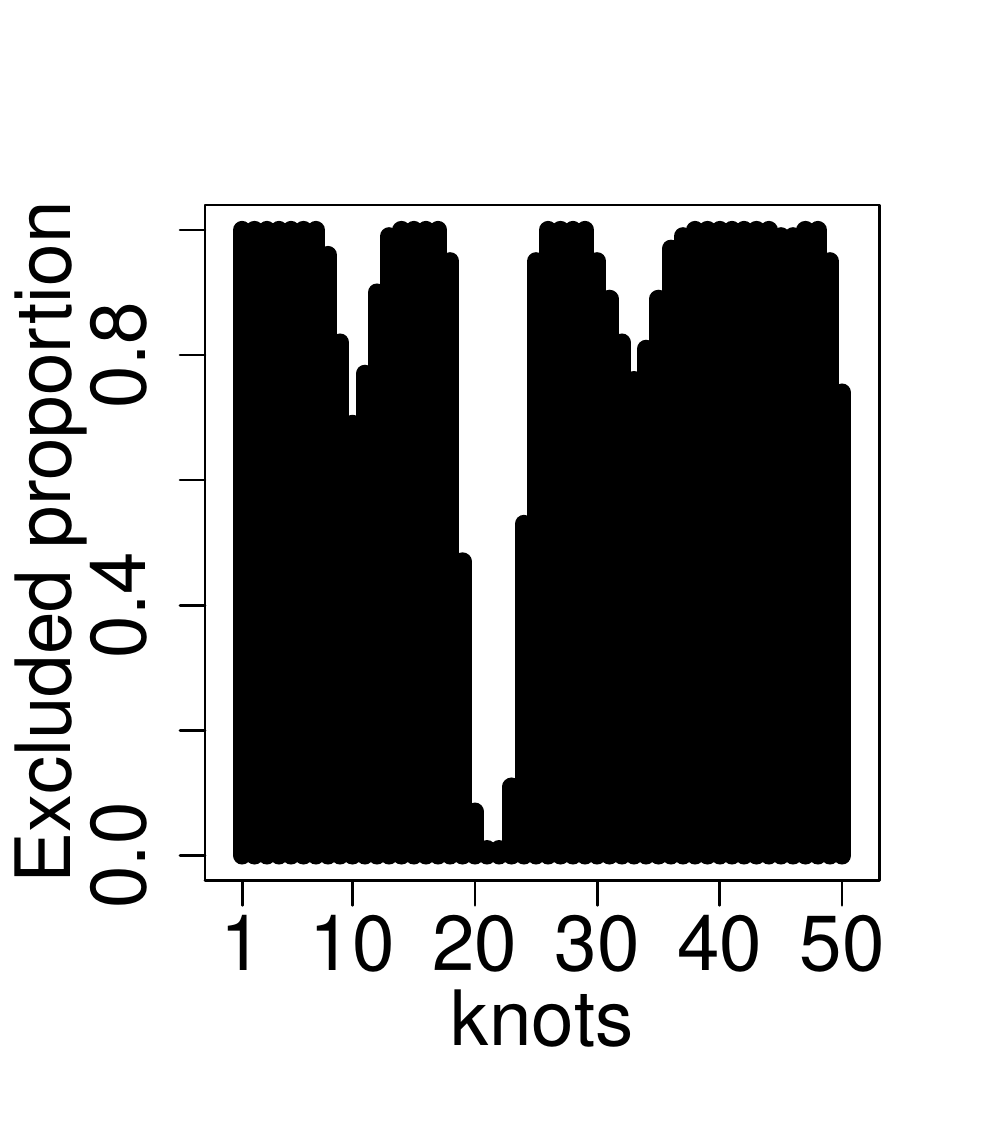}}\\
\end{tabular}
\end{center}\vspace{-0.5cm}
\caption{Proportion of excluded knots and the average of the fitted models  - K=50 nós.}\label{fig:ajuste_2bump_k50}
\end{figure}

The results obtained for the case of 1 bump are similar to those obtained here. Once again, it is possible to see in the Figures \ref{fig:ajuste_2bump_k10}, \ref{fig:ajuste_2bump_k30} and \ref{fig:ajuste_2bump_k50} the average of the fittings of the 100 replicates for the three selection criteria in the dashed lines (first row of plots) and the proportion of excluded knots (second row of plots) for different values of $K$. In the three figures one may notice that the knots that are most selected as significant are positioned in the ups and downs of the bumps. For the case of 2 bumps, the  higher the value of $K$, the more severe the exclusion of knots is. The SN criterion does not show good results as the maximum number of knots increases and the performance of the CI and BF criteria are similar in all cases.

\begin{figure}[h!]
\begin{center}
\begin{tabular}{ccc}
{\includegraphics[scale=0.50]{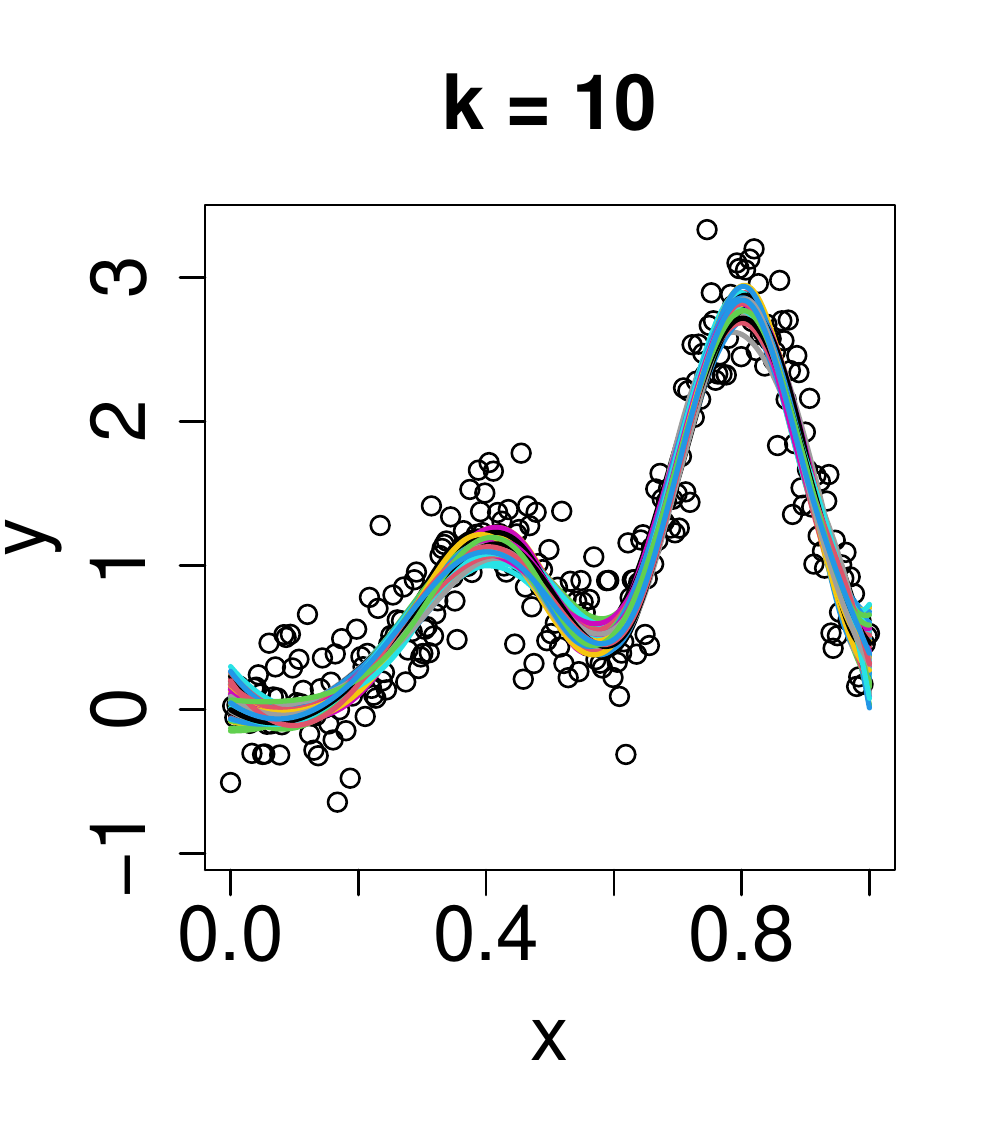}}&
{\includegraphics[scale=0.50]{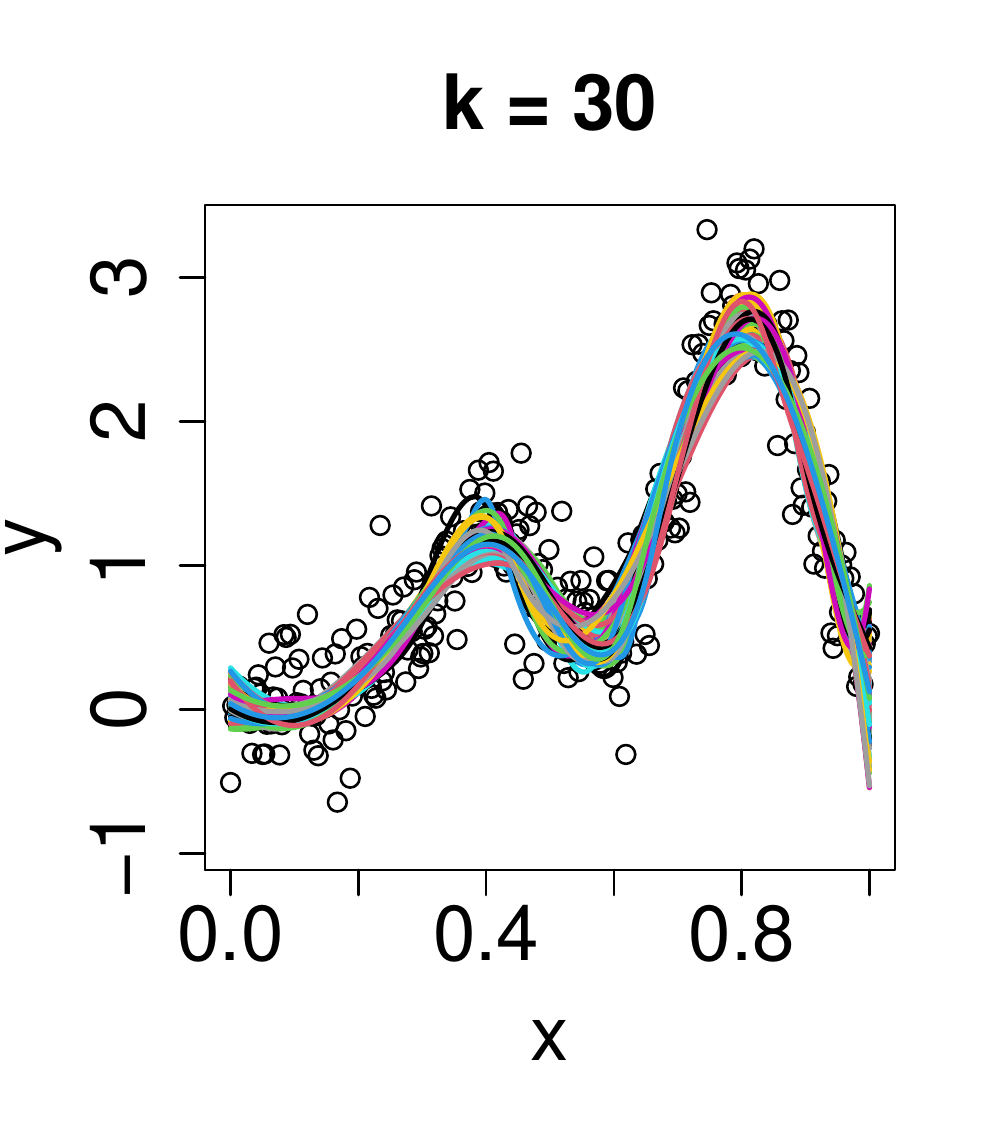}}&
{\includegraphics[scale=0.50]{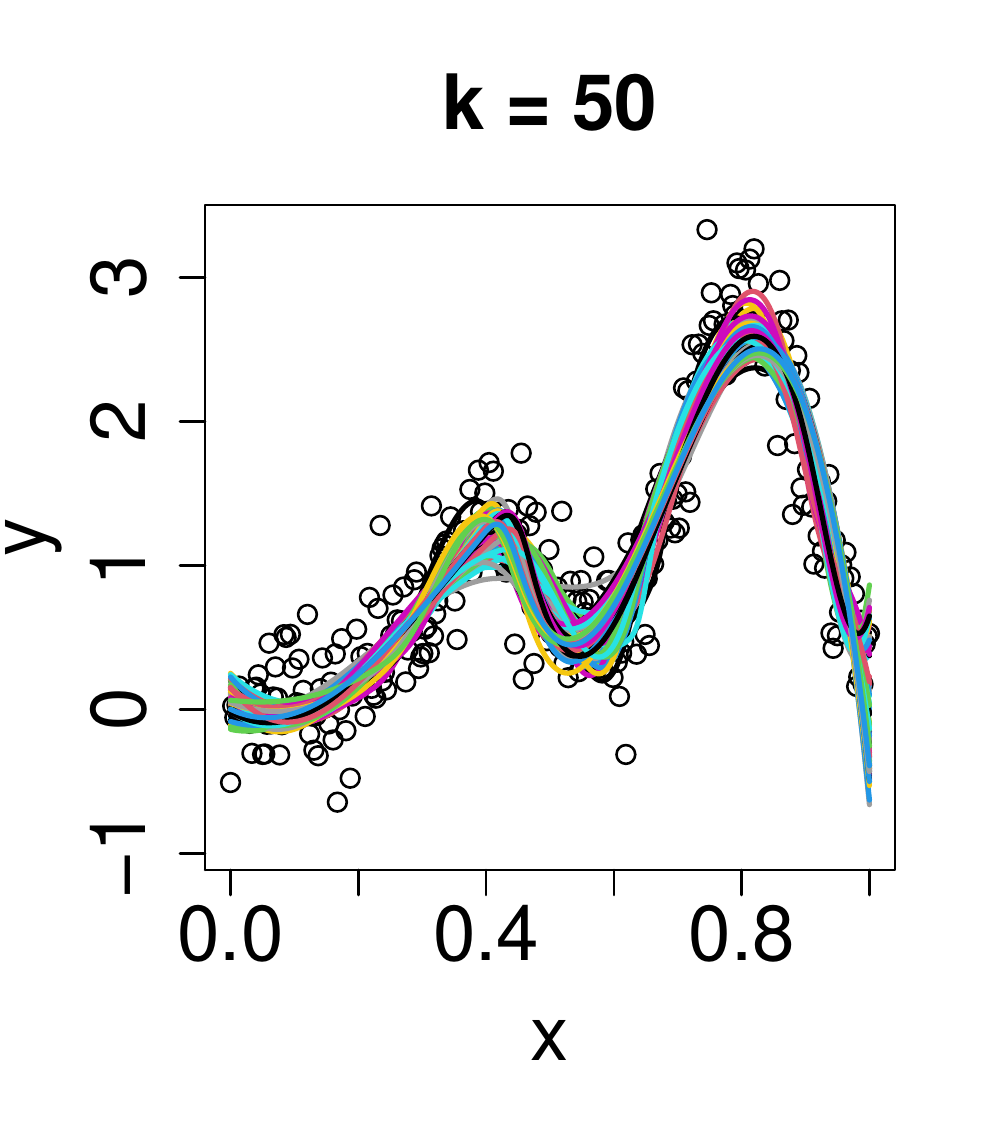}}\\
\end{tabular}
\end{center}\vspace{-0.5cm}
\caption{Fit of 100 replicates  for different values of knots.}\label{fig:100ajuste_2bump}
\end{figure}

In Figure \ref{fig:100ajuste_2bump} shows the fittings of 100 replicates according to the BF criterion for different numbers of knots. It can be seen that  there is less variability than in the case of 1 bump, possibly due to the increase in the sample size to $ n = 300 $ . Nonetheless, it is  possible to see  that the when the  maximum number of knots increases ,  the variability between the fitted models also increases.

\begin{figure}[h!]
\begin{center}
\begin{tabular}{cc}
{\includegraphics[scale=0.55]{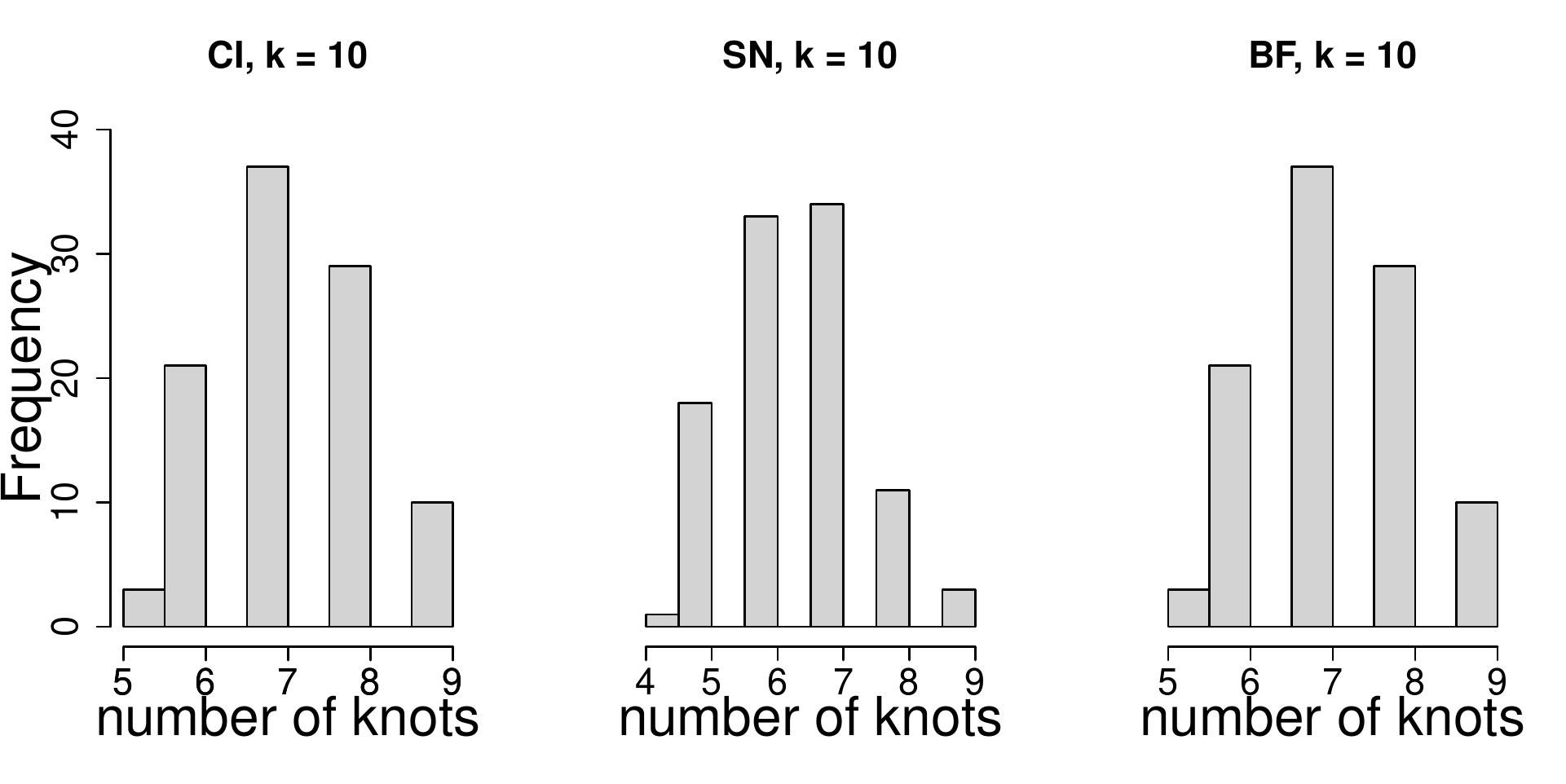}}\\
{\includegraphics[scale=0.55]{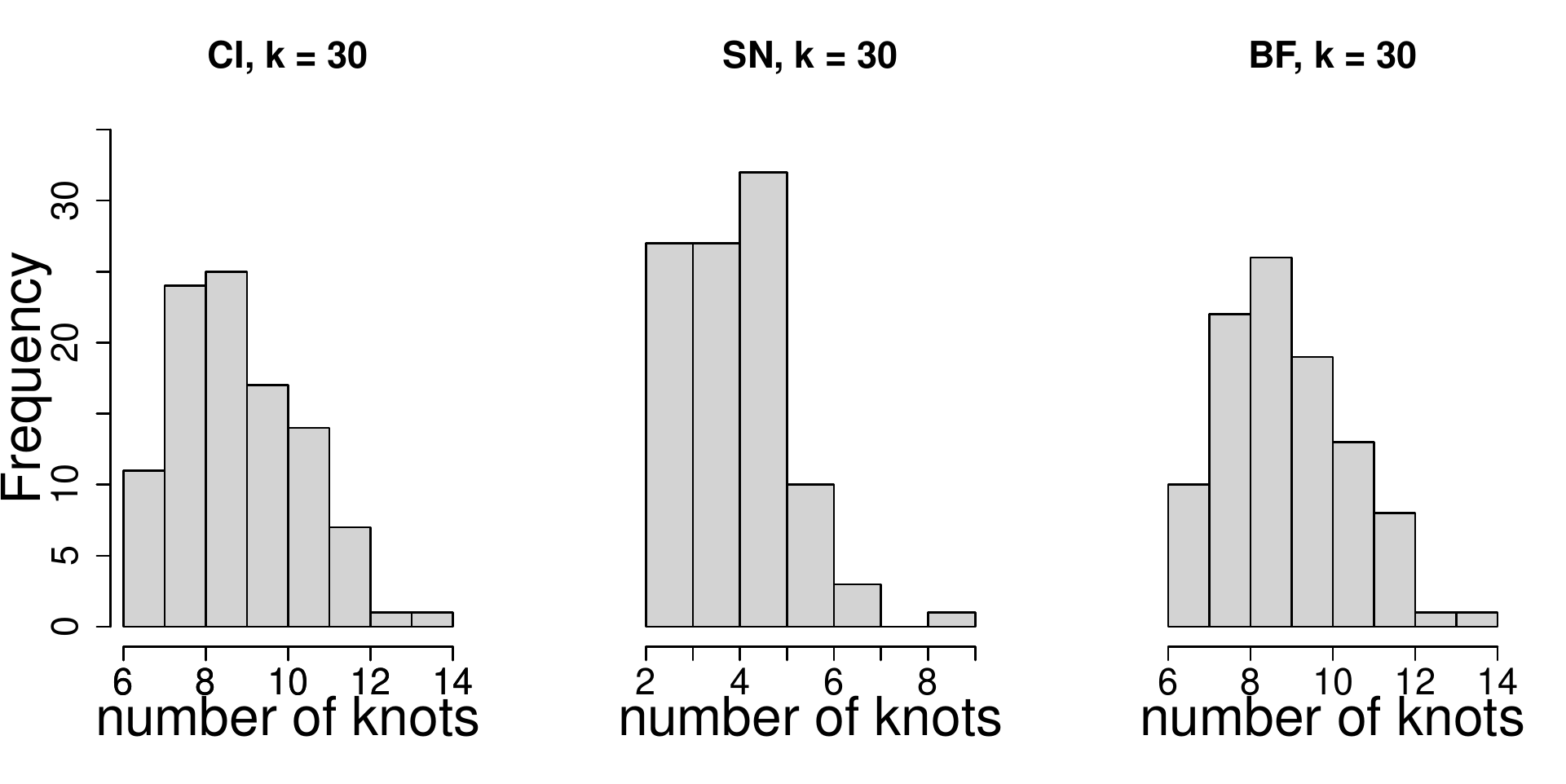}}\\
{\includegraphics[scale=0.55]{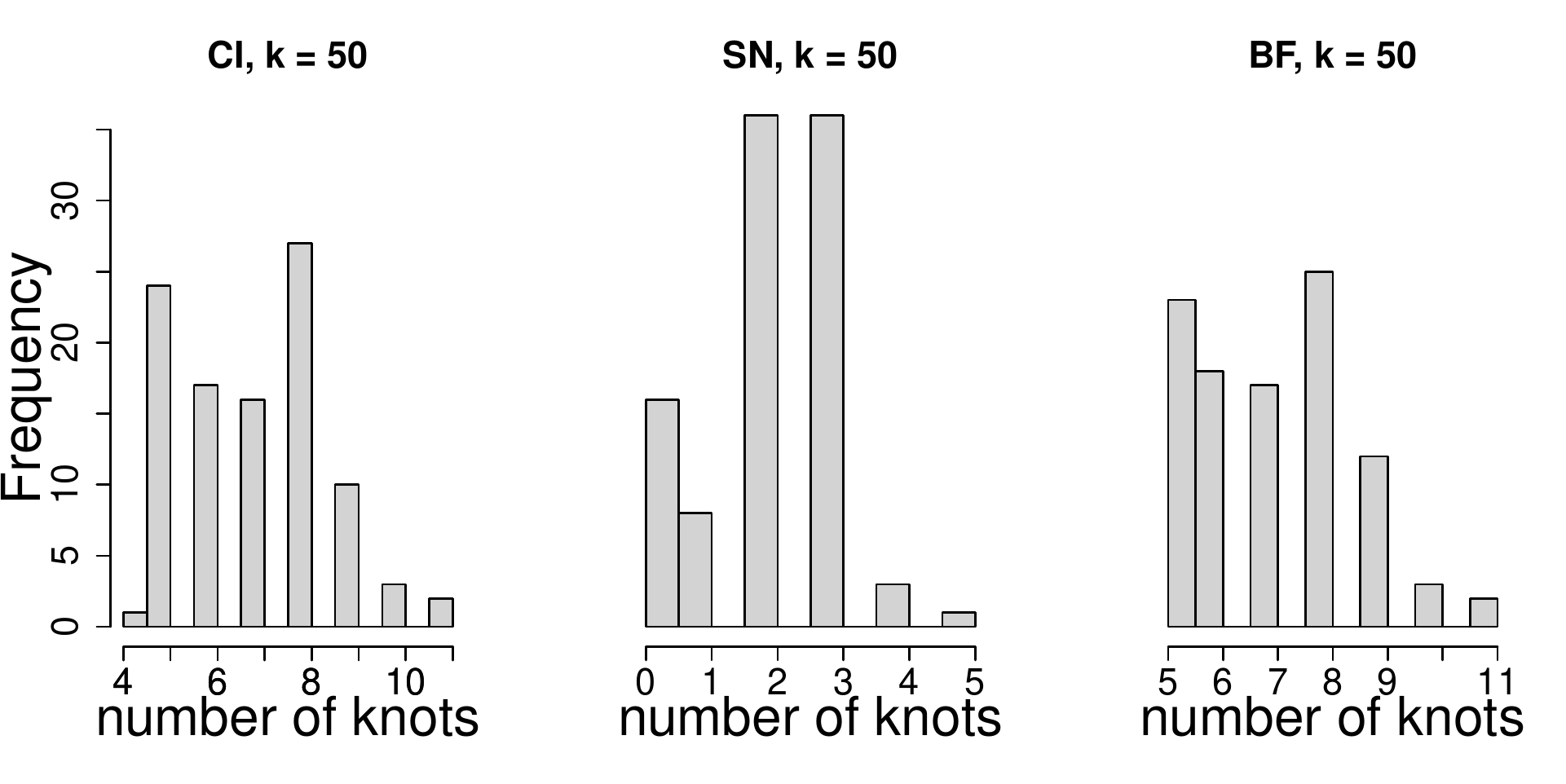}}\\
\end{tabular}
\end{center}\vspace{-0.5cm}
\caption{Frequency of selected knots in 100 replicates.}\label{fig:freq_nos_2bump}
\end{figure}

Our analysis shows that the SN criterion does not provide a good fit for $K = 50$ and in  Figure \ref{fig:freq_nos_2bump} we see that this criterion tends to underestimate the number of significant knots as $K$ increases. We will analyze in more detail the frequency of the number of knots selected in the 100 replicates using the CI and BF criteria. Note that both criteria present similar results. In Figure \ref{fig:freq_nos_2bump} we observed that when the maximum number of knots is 10, both the CI and BF criteria indicate more frequently that 7 out of 10 knots are significant. When $K = 30$ or $ K= 50$, the criteria CI and BF most frequently indicate 8 knots as significant.

In order to define the maximum number of knot, the average ELBO was calculated for each value of $K$ in a fixed grid. For the FB criterion we have that the average ELBO is -74.36 when $K = 10$. This value increases (-68.67 when $K =  20$) until it reaches  the maximum value of -67.15 when $K = 30$. The average ELBO value for $K = 40$ is -70.44. Thus, for BF criterion $K = 30$ is the initial guess for the maximum number of knots.The same can be seen in the CI criterion. This result coincides with that obtained in exercise 4.

\clearpage

\section{Applications to real data}


In this section, two applications with real data will be analyzed. The first is considered more usual in the literature in the area and the second addresses a current issue related to the world pandemic of Covid-19. In both cases, the Penalized Spline Regression model  with polynomials of degrees 2 and 3 is fiited to the data. In addition, the maximum number varies between 10, 20 and 30. It is noteworthy that in the first example the knots are equally spaced positioned. Lasso, through variational inference (VB), is used in both applications together with the Bayes Factor (BF) criterion for the selection of the most significant knots. ELBO will be the measure considered for comparing models

\subsection{Age and Income data}

\begin{figure}[h!]
\begin{center}
\begin{tabular}{cc}
{\includegraphics[scale=0.49]{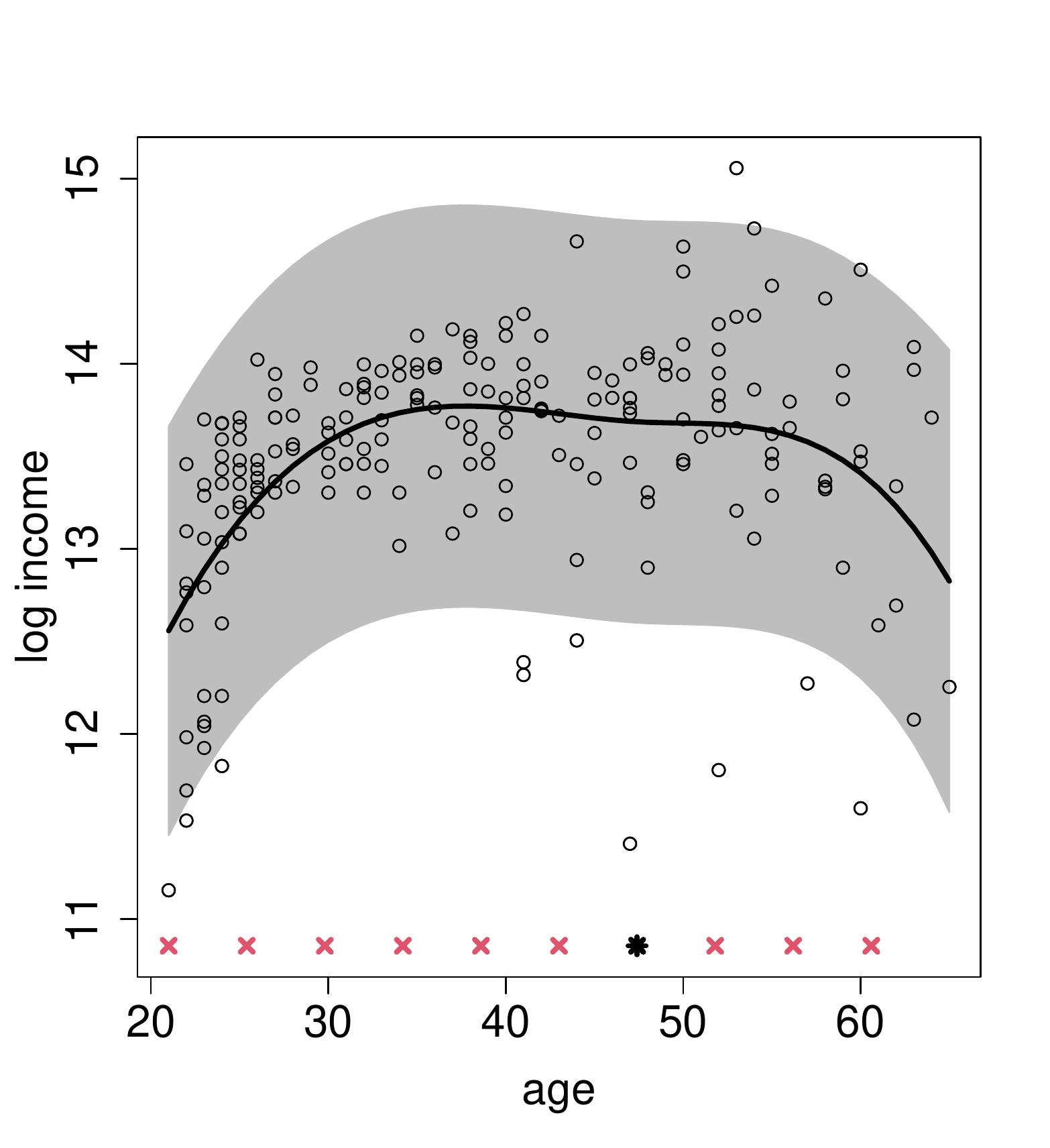}}&
{\includegraphics[scale=0.49]{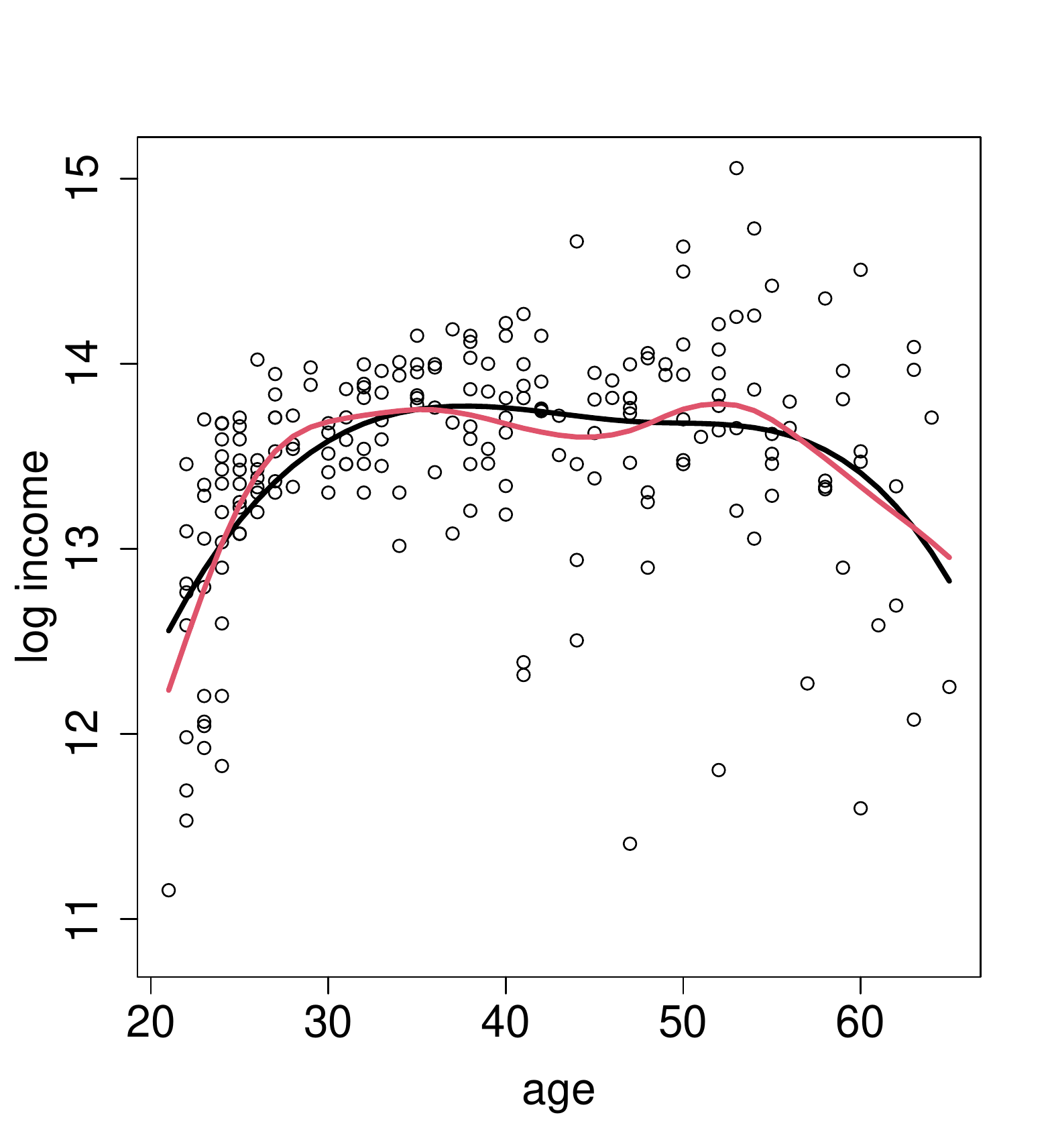}}\\
\end{tabular}
\end{center}\vspace{-0.5cm}
\caption{Left: The fit of the spline regression model (black solid line)  with $ p = 3 $ and one significant knot among $K=10$ knots and the log-income and age data (black dots) with the  credible interval of $95\%$  (shaded area). Right: The fit of the proposed model (solid black line) and a fitted model using the R function ”smooth.spline" (solid red line) to the log-income and age data (dots).} \label{fig:ajuste_ageincome}
\end{figure}

The first data set considers the income and age of 205 Canadians (\cite{Ullah85}. These data have been widely used in applications of non-parametric regression models. See for example \cite{Ruppert02}. A logarithmic transformation was applied to income, as can be seen in the data represented by black dots in the two plot  in Figure \ref{fig:ajuste_ageincome}.
Table \ref{tab:elbo_ageincome} shows the ELBO measure computed for each fitted model by varying the degree of the polynomial $(p)$ and the maximum number of knots $(k)$. For both $ p = 2 $ and $ p = 3 $, ELBO achieves its maximum value at $ K = 10 $. Comparing these two quantities, the largest ELBO occurs for $ p = 3 $ and $ K = 10 $. The graph on the left of Figure \ref{fig:ajuste_ageincome} shows the fitting of the penalized spline regression model (solid black line) and its credible interval of $95\%$ (gray shaded area). Note that the credible interval covers a large part of the observed points. At the bottom of the graph we have the symbol "x" representing the 10 knots  positioned and in black the only knot considered significant among the 10, according to the FB criterion. At the level of comparison, the graph on the right of Figure \ref{fig:ajuste_ageincome} shows the fitted model obtained with the R function "smooth.spline" (solid line in red) and the fitting of the proposed model (black solid line). In this application, the results of these two fitted models are similar.
\begin{table}[h!]
\caption{ELBO - Age and income data} 
\begin{center}
{\footnotesize
\begin{tabular}{c|c|c|c}
  \hline
& k = 10 & k = 20 & k = 30 \\
\hline $p = 2$ & {\bf{-221.49}} & -225.45 & -237.22 \\
\hline $p =3$ & {\bf{-220.77}} & -227.03 & -223.93 \\
\hline 
\end{tabular}}
\end{center}\label{tab:elbo_ageincome}
\end{table}

\subsubsection{Covid-19 data}
\label{subsubsection:Covid-19_data}


In order to observe the trend of the daily cases of Covid-19 in the USA and Brazil, the penalized spline regression model was fitted to the data (logarithmic scale). United States data ranges from March 1, 2020 to November 30, 2020, while data from Brazil ranges from March 10, 2020 to November 30, 2020.

\begin{figure}[h!]
\begin{center}
\begin{tabular}{cc}
{\includegraphics[scale=0.49]{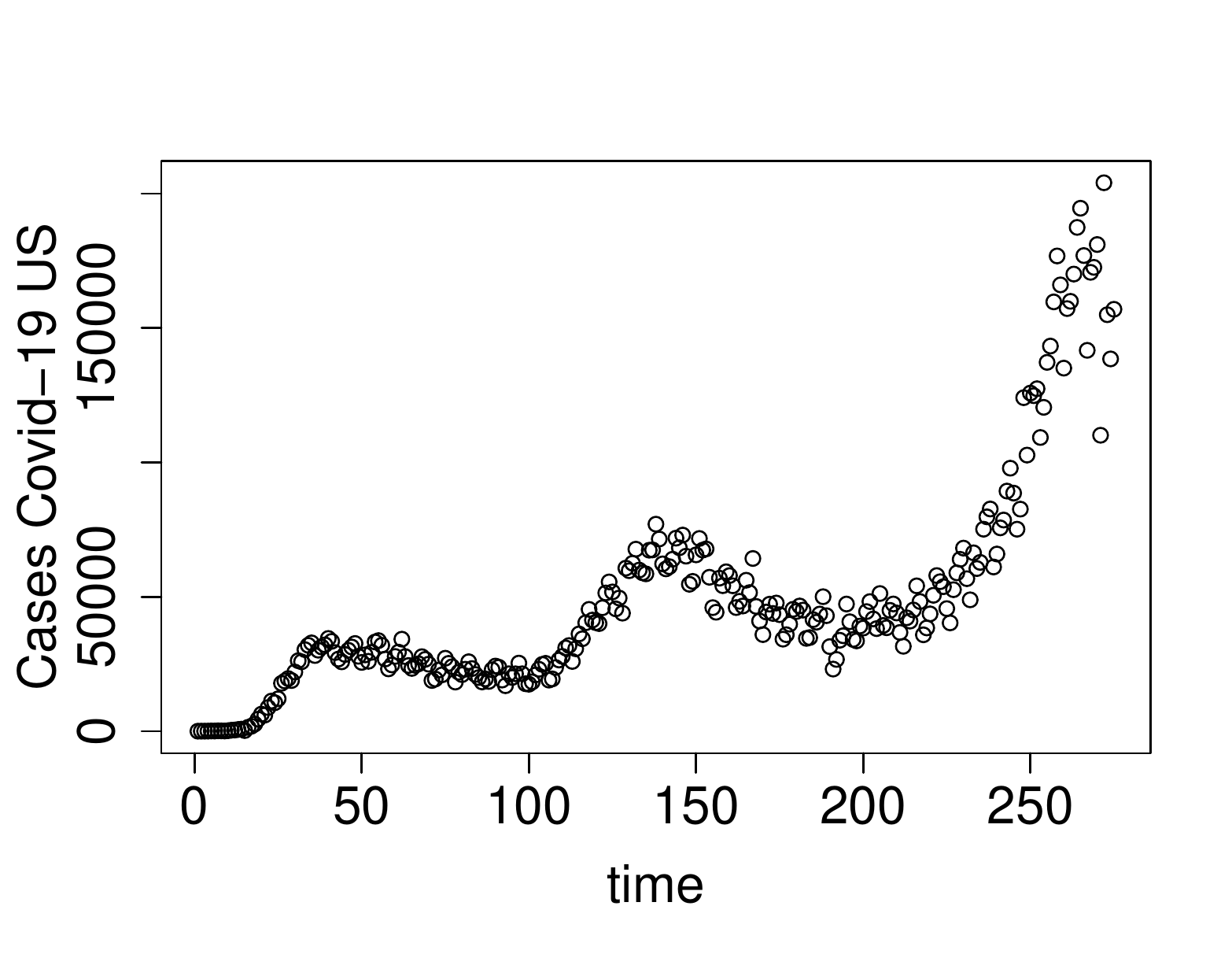}}& {\includegraphics[scale=0.49]{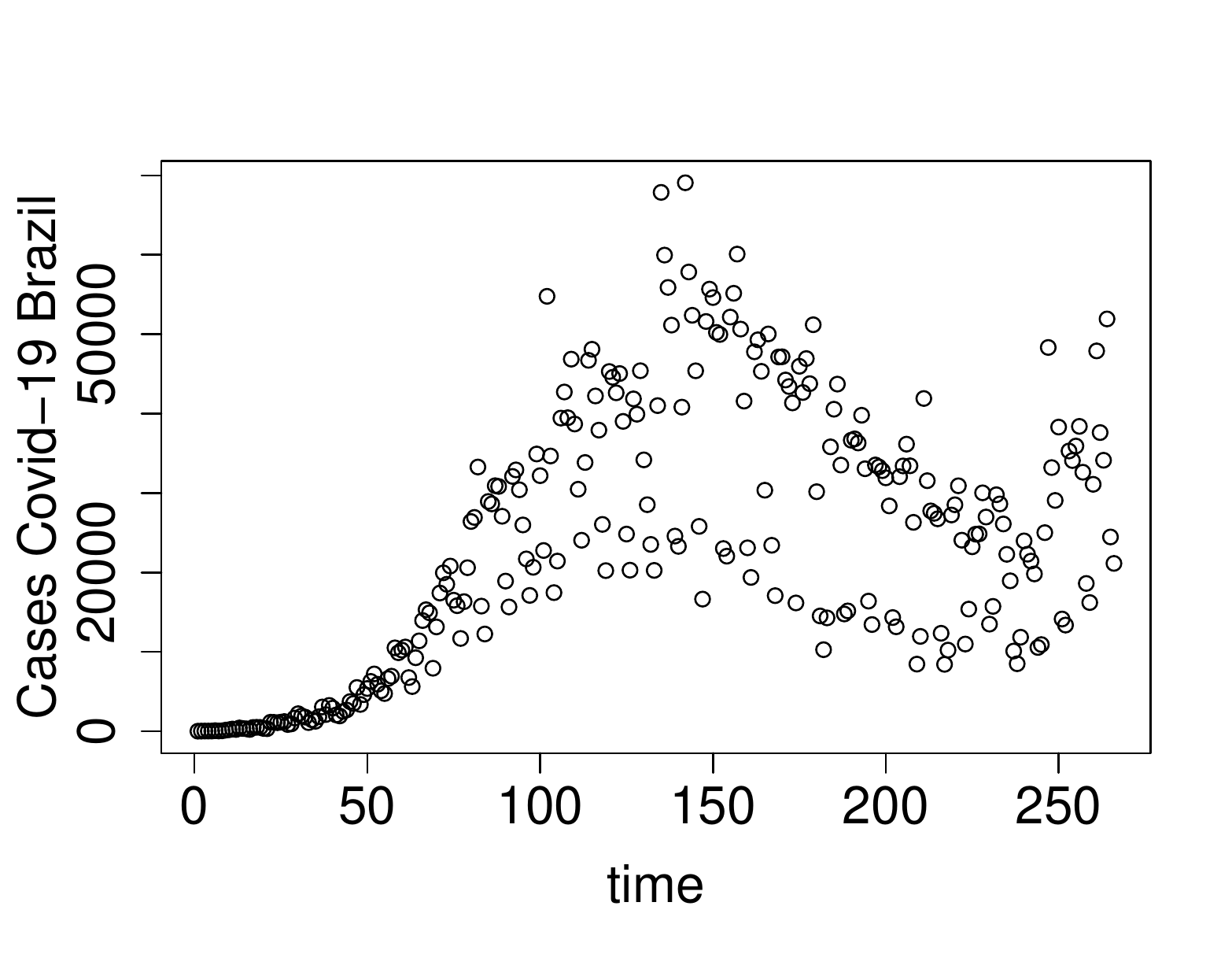}}
\end{tabular}
\caption{Daily cases of Covid-19 in US (left) and in Brazil (right).}\label{fig:casos_covid}
\end{center}
\end{figure}

\begin{figure}[h!]
\begin{center}
\begin{tabular}{cc}
{\includegraphics[scale=0.49]{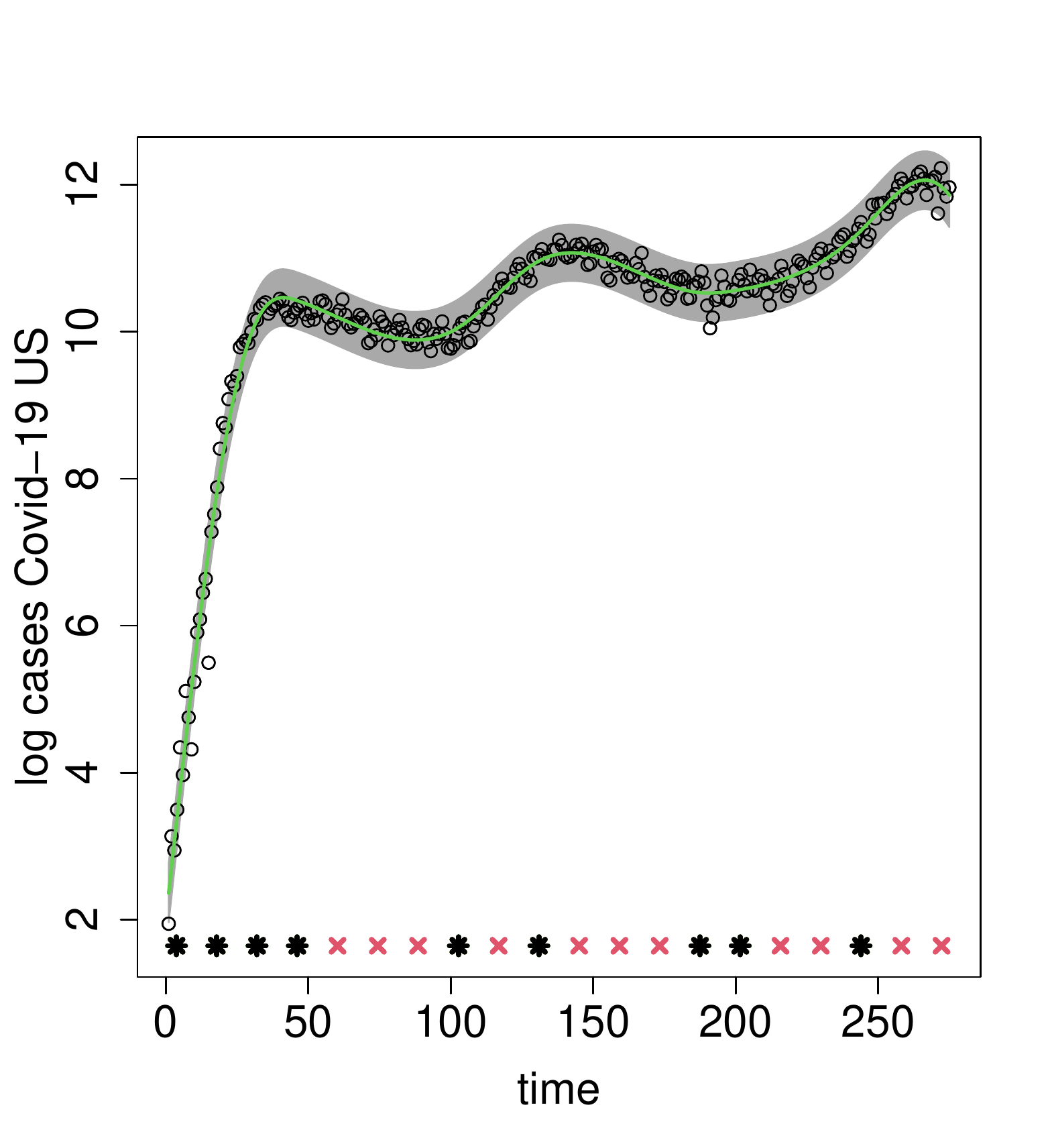}} & {\includegraphics[scale=0.49]{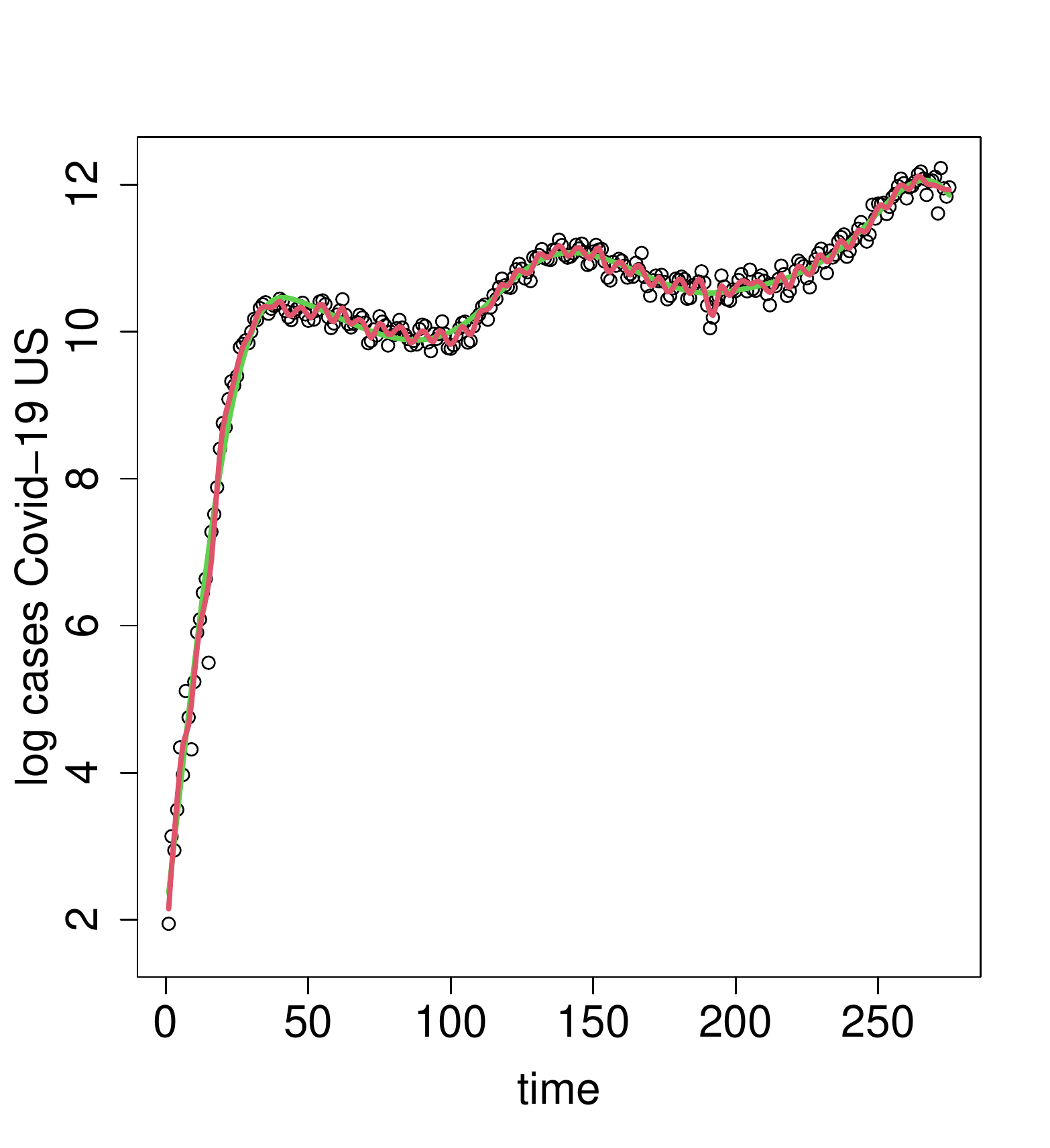}}\\
\end{tabular}
\vspace{-0.5cm}
\caption{Left: The fit of the penalized spline regression model (solid green line) with $p=3$ and 9 significant knots (black asterisks) out of $K=20$ knots (x red) with the $95\%$ credible interval (shaded area) and the fit for the logarithm of the number of daily cases of Covid-19 in the USA (black dots). Right: The fit of the proposed model (solid green line) and the fit of a model using the R function  "smooth.spline" (solid red line) to the log data of the number of daily cases of Covid-19 in the USA (dots).}\label{fig:ajuste_covidUS}
\end{center}
\end{figure}

\begin{figure}[h!]
\begin{center}
\begin{tabular}{cc}
{\includegraphics[scale=0.49]{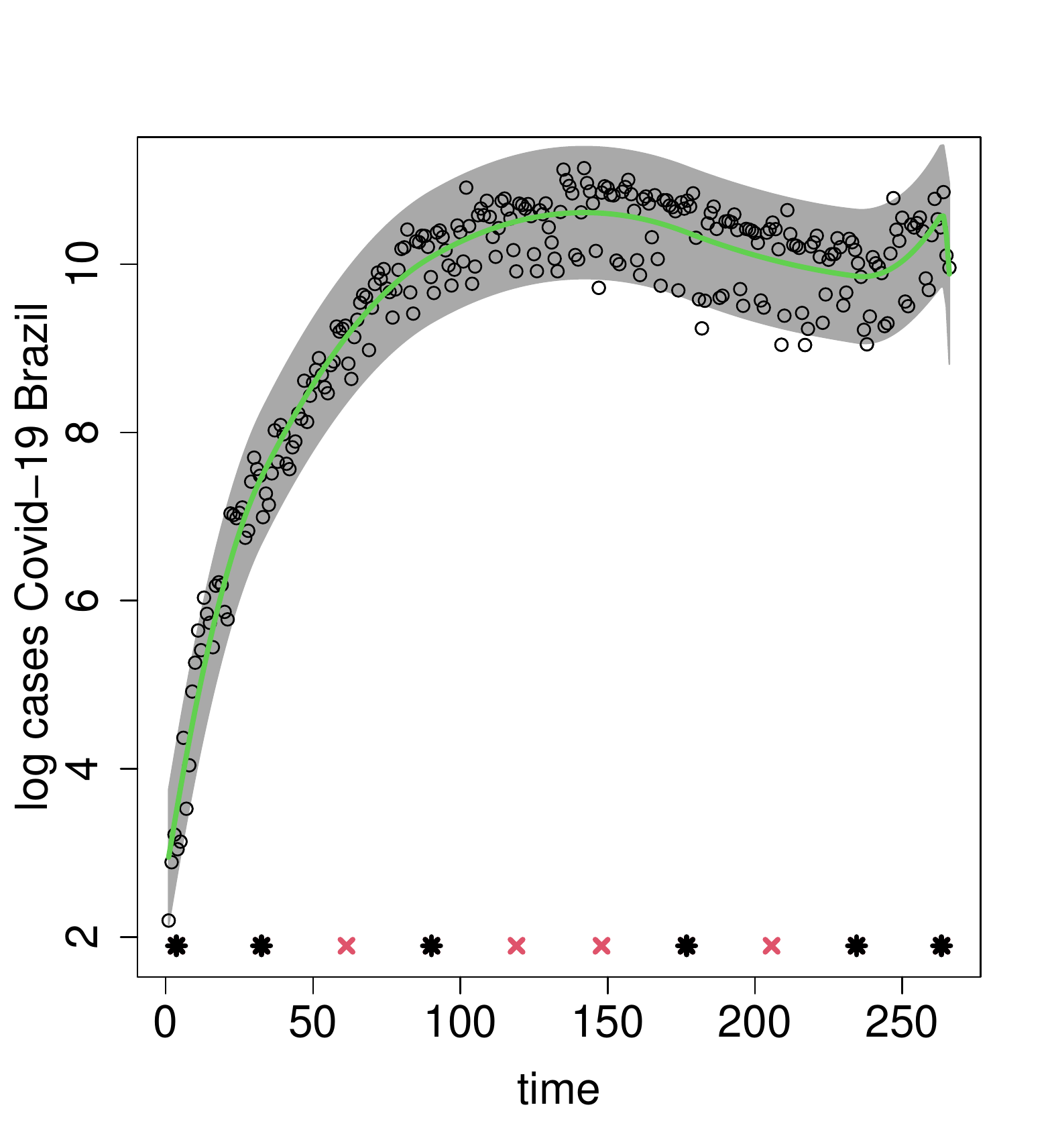}} & {\includegraphics[scale=0.49]{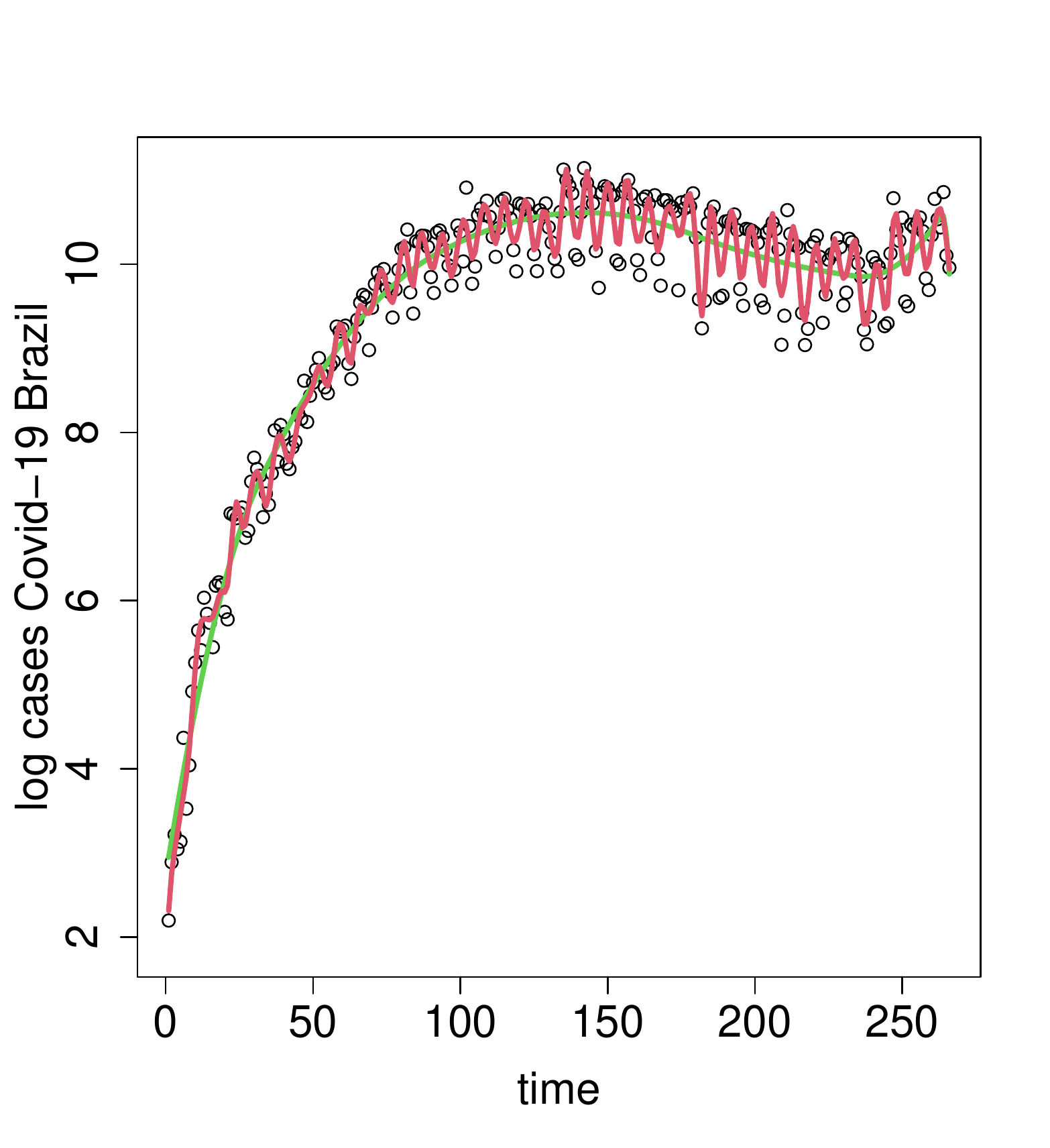}}\\
\end{tabular}
\end{center}\vspace{-0.5cm}
\caption{Left: the fit of the penalized regression spline model (solid green line) with $p =2$ and 6 significant knots (black asterisks) out of $ K=10$ knots (x in red) with the $95\%$ credible interval (shaded area) and the fit of the logarithm of the number of daily cases of Covid-19 in Brazil (black dots). Right: The curve fitting of the proposed model (solid green line) and the fit using R function "smooth.spline" (solid red line) to the logarithm  of the number of daily cases of Covid-19 in Brazil (dot).}\label{fig:ajuste_covidBR}
\end{figure}

Figure \ref{fig:casos_covid} shows the number of daily cases of Covid-19 in the USA (left) and Brazil (right) on the original scale. The black dots in the Figures \ref{fig:ajuste_covidUS} and \ref{fig:ajuste_covidBR}
show the same data set in logarithmic scale.



In this example, analyzing real data, we consider models with splines  of degrees 2 and 3. The maximum number of knots  varies every 10 knots. To find the maximum number of knots, we use the ELBO measure by starting the grid with $K = 10 $ knots. After obtaining the optimal $ K$ value  BF criterion is applied to select which knots are the most  significant and  their positions.


The inference procedure was performed from the Bayesian point of view through variational inference. The  prior distribution remains the same as for studies with artificial data.

Table \ref{tab:elbo_covid} exhibits the results of ELBO for different values of $p$ and $K$, for data from the USA and Brazil.

Marked in bold are the cases in which ELBO achieves the highest values for $ p = 2 $ and $ p = 3 $. In the North American case, ELBO is maximum when $ p = 3 $ and $ K = 20 $. It is worth mentioning that among the 20 knots, 9 were significant according to the BF criterion. Considering the Brazil data, one can see that the largest ELBO occurs when $ p = 2 $ and $ K = 10 $, and only 6 of these 10 knots are significant, according to the BF. The following results are presented only for models with the largest ELBO.
\pagebreak 

\begin{table}[h!]
\caption{ELBO - US and Brazil Covid-19 data.} 
\begin{center}
{\footnotesize
\begin{tabular}{c|c|c|c||c|c|c}
  \hline
& \multicolumn{3}{c||}{US} &  \multicolumn{3}{c}{Brazil} \\
\hline & k = 10 & k = 20 & k = 30 & k = 10 & k = 20 & k = 30 \\
\hline $p = 2$ & 57.25 & {\bf{108.96}} & 95.25 & {\bf{-145.32}} & -168.23 & -164.55 \\
\hline $p =3$ & 18.89 & {\bf{116.10}} & 24.53 & -181.52 & {\bf{-179.88}} & -197.66\\
\hline 
\end{tabular}}
\end{center}\label{tab:elbo_covid}
\end{table}

The plot to the left of the Figure \ref{fig:ajuste_covidUS} shows the fit (solid green line) of the penalized spline regression model with  3 polynomial of degree 3 and 9 significant knots(out of a total of 20 knots) for the US Covid-19 data. The shaded area represents the $ 95\% $ credible interval and it contains most of the observed data (black dots). Note that significant knots (black asterisks) are located such that they cover the bumps  of the curve. The "x" in red are the excluded knots that were not considered in the fit. The plot on the right compares the fits of the proposed model (solid green line) with the  R function "smooth .spline" (red solid line). Note  that the second fit captures, in addition to the signal, the noise contained in the data.

Figure \ref{fig:ajuste_covidBR} shows the fit (solid green line) of the proposed model with $ p = 2 $ and 6 significant knots (black asterisks at the bottom of the graph). The excluded knots are represented by "x" in red. The $ 95\% $ credible interval (gray shadow) covers much of the observed data points. The plot on the right makes a comparison between the fit of the proposed model and the R function "smooth.spline" (red solid line). As in the case of the U.S. data, we observed that the fit by using "smooth.spline" does not smooth the data as the proposed model and follows  the series random noise more closely.

\pagebreak 


\pagebreak 


\section{Conclusions}
This article proposes a new scalable procedure for selecting the number of knots in regression splines: A fully automatic Bayesian Lasso through variational inference. Simulation studies have shown effectiveness of this procedure in modeling different types of data sets. In addition, the numerical exercises show that this approach is much faster than the traditional one that is based on MCMC type algorithms. In real data sets the procedure was able to capture the trend existing in them. Thus providing a better understanding of the data dynamic.

\paragraph{Acknowledgments.} This paper was partially supported by Fapesp Grants (RD) 2018/04654, (RD and HSM) 2019/10800-0, (RD) 2019/00787-7. 

\clearpage

\appendix{Appendix 1: The variational distributions for Lasso}

The variational posterior for $\bfbeta$ and $\phi$ while holding $q_2(\bftau|\lambda)$ and $q_3(\lambda)$ fixed, is given by 

\begin{eqnarray*}
\log q_1^\ast(\bfbeta,\phi) &=& \log(p(\bfy|\bfbeta,\phi)) + E_{\tau}[\log(p(\bfbeta,\phi|\bftau))] + const\\
&=& \log\left[(2\pi)^{-n/2}|\phi^{-1}I_n|^{-1/2} \exp\left\{-\frac{1}{2}(\bfy-X\bfbeta)^T (\phi I_n) (\bfy-X\bfbeta)\right\}\right] + \\
&& + E_\tau \left\{ \log \left[(2\pi)^{-p/2} |\phi^{-1} \bfD_\tau|^{-1/2} \exp\left\{-\frac{1}{2} \bfbeta^T \phi \bfD_\tau^{-1} \bfbeta \right\}\right]\right\} + \\ 
&& + E_\tau \{\log [\phi^{a_0-1} \exp\{-\phi b_0\}]\} + const\\
&=& \left(\frac{n}{2}+\frac{p}{2}+a_0-1\right) \log \phi - \frac{\phi}{2} \{\bfbeta^T[E_\tau(\bfD_\tau^{-1}) + X^TX]\bfbeta + \bfy^T\bfy - 2\bfy^TX\bfbeta + 2b_0\} + const \\
&=& \log N(\bfbeta|m_\beta, \phi^{-1} C_\beta) \times Ga(\phi|a_\phi,b_\phi)
\end{eqnarray*}

It is easy to see that this is a normal-gamma distribution with parameters:

\begin{eqnarray*}
C_\beta^{-1} = E_{\tau} (\bfD_\tau^{-1}) + X^TX, \, \, \,  \, \, \,   &\mbox{and}&  \, \, \,  \, \, \, m_\beta = C_\beta X^T\bfy, \\
a_\phi = a_0 + n/2, \, \, \, \, \,  \,  &\mbox{and}&  \, \, \,  \, \, \, b_\phi = b_0 + \frac{1}{2} (\bfy^T\bfy - m_\beta^T C_\beta^{-1}m_\beta).
\end{eqnarray*}

The variational  distribution of   $\bftau$ while holding $q_3(\lambda)$ fixed is given by 

\begin{eqnarray*}
\log q_2^\ast(\tau_j) &=& E_{\lambda}[\log (p(\tau_j|\lambda))] + E_{\beta,\phi}[\log (p(\beta_j,\phi|\tau_j))] + const\\ 
&=& E_\lambda \{\log [\exp\{-\lambda \tau_j\}]\} + E_{\beta,\phi}\left\{\log \left[ (\phi^{-1} \tau_j)^{-1/2} \exp\left\{-\frac{\phi}{2\tau_j} \beta_j^2 \right\} \right]\right\} + const \\
&=& - \frac{1}{2} \log \tau_j - \frac{1}{2}\left(2 E_\lambda[\lambda] \tau_j + \frac{1}{\tau_j} E_{\beta,\phi}[\phi \beta_j^2] \right) + const \\
&=& \log GIG(\tau_j|c_\tau,d_\tau,f_{\tau_j})
\end{eqnarray*}
with GIG being generalized inverse Gaussian distribution, where
$$c_\tau = \frac{1}{2}\;;\; d_\tau = 2 E_\lambda[\lambda] \; ; \; f_{\tau_j} = E_{\beta,\phi}[\phi \beta_j^2].$$ 
Therefore, 
$$\log q_2^\ast(\bftau) = \log \prod_{j=1}^p GIG(\tau_j|c_\tau,d_\tau,f_{\tau_j}).$$
 
The variational distribution of $\lambda$ is:

\begin{eqnarray*}
\log q_3^\ast(\lambda) &=& \log (p(\lambda)) + E_\tau[\log (p(\bftau|\lambda))] + const\\
&=& \log [\lambda^{g_0-1} \exp\{-h_0 \lambda\}] + E_\tau\left(\log \left[\prod_{j=1}^p \lambda \exp\{-\tau_j \lambda\}\right] \right)+ const\\
&=& (g_0 + p - 1) \log \lambda - \lambda [h_0 + \sum_{j=1}^p E_\tau(\tau_j)]+ const\\
&=& \log Ga(\lambda|g_\lambda, h_\lambda)
\end{eqnarray*}

which is a gamma distribution with parameters $$g_\lambda = g_0 + p  \; ;\; h_\lambda = h_0 + \sum_{j=1}^p E_\tau(\tau_j).$$

The expected values can be computed as follows:

It is worth  pointing out that if $X \sim GIG(p,a,b)$, then its density is 
$$
f(x|p,a,b) = \frac{ (\frac{a}{b})^{\frac{p}{2} } } {2 \kappa_p(\sqrt{a b}) } \, x_{p-1} \, \exp\{- (ax + b/x)/2\}, \ \ x>0,
$$
where $\kappa_p(\cdot)$ is a modified Bessel function of the second kind, with 
\begin{eqnarray*}
 E[X]&=&\sqrt{\frac{b}{a} } \frac{ \kappa_{p+1}( \sqrt{ab} )}{\kappa_{p}( \sqrt{ab} )} \ \  \mbox{and}  \ \ Var[X] = \left(\frac{b}{a}\right) \left[\frac{ \kappa_{p+2}( \sqrt{ab} )}{\kappa_{p}(\sqrt{ab})} - \left(\frac{ \kappa_{p+1}(\sqrt{ab} )}{\kappa_{p}(\sqrt{ab} )} \right)^2\right]\\
 E[X^{-1}]&=&\sqrt{\frac{a}{b} } \frac{ \kappa_{p+1}(\sqrt{ab} )}{\kappa_{p}(\sqrt{ab} )}  - \frac{2 p}{b}
\end{eqnarray*}

Therefore,

$$E_\tau[\bfD_\tau^{-1}] = diag(E_\tau(\tau_1^{-1}), \ldots, E_\tau(\tau_p^{-1})),$$ 

$$E_\tau(\tau_j^{-1}) = \frac{\sqrt{d_\tau} \kappa_{c_\tau+1}(\sqrt{d_\tau f_{\tau_j}})}{\sqrt{f_{\tau_j}} \kappa_{c_\tau}(\sqrt{d_\tau f_{\tau_j}})} - \frac{2 c_\tau}{f_{\tau_j}},$$ 

$$E_\tau(\tau_j) = \frac{\sqrt{f_{\tau_j}} \kappa_{c_\tau+1}(\sqrt{d_\tau f_{\tau_j}})}{\sqrt{d_\tau} \kappa_{c_\tau}(\sqrt{d_\tau f_{\tau_j}})},$$

$$E_\lambda(\lambda) = \frac{g_\lambda}{h_\lambda}.$$ 

For the calculus of $E_{\beta,\phi}[\phi \beta_j^2]$, let $x|y \sim N(\mu_x,y^{-1}\sigma_x)$ and $y \sim Ga(a,b)$, so

$$E[X^2 Y] = E[E(X^2 Y|Y)] = E[YE(X^2|Y)] = E[Y(E^2(X|Y)+ Var(X|Y))] = \mu_x^2 E(Y) + \sigma_x = \frac{a \mu_x^2}{b} + \sigma_x.$$

Thus,

$$E_{\beta,\phi}[\phi \beta_j^2] = m_{\beta_j}^2 a_\phi/b_\phi + (C_\beta)_{jj}.$$ 

\clearpage

\appendix{Appendix 2: The variational distributions for Regression Spline}

The variational posterior for $\bfbeta^{(2)}$ and $\phi$ while holding $q_2(\bftau|\lambda)$, $q_3(\lambda)$ and $q_4(\bfbeta^{(1)})$ fixed, is given by 

\begin{eqnarray*}
\log q_1^\ast(\bfbeta^{(2)},\phi) &=& E_{\beta^{(1)}}[\log(p(\bfy|\bfbeta,\phi))] + E_{\tau}[\log(p(\bfbeta^{(2)},\phi|\bftau))] + const\\
&=& E_{\beta^{(1)}} \left\{\log\left[(2\pi)^{-n/2}|\phi^{-1}I_n|^{-1/2} \right. \right. \times \\ 
&& \times \left.  \left. \exp\left\{-\frac{1}{2}(\bfy-X_1\bfbeta^{(1)} - X_2\bfbeta^{(2)})^T (\phi I_n) (\bfy-X_1\bfbeta^{(1)} - X_2\bfbeta^{(2)})\right\}\right] \right\} + \\
&& + E_\tau \left\{ \log \left[(2\pi)^{-K/2} |\phi^{-1} \bfD_\tau|^{-1/2} \exp\left\{-\frac{1}{2} {\bfbeta^{(2)}}^T \phi \bfD_\tau^{-1} \bfbeta^{(2)} \right\}\right]\right\} + \\ 
&& + \log [\phi^{a_0-1} \exp\{-\phi b_0\}] + const\\
&=& \left(\frac{n}{2}+\frac{K}{2}+a_0-1\right) \log \phi +\\
&& - \frac{\phi}{2} \left\{{\bfbeta^{(2)}}^T[E_\tau(\bfD_\tau^{-1}) + X_2^TX_2]\bfbeta^{(2)} - 2 {\bfbeta^{(2)}}^T[X_2^T\bfy - X_2^TX_1 E_\beta^{(1)}(\bfbeta^{(1)})]\right\} + \\ 
&& - \phi \left\{b_0 +\frac{1}{2} [\bfy^T \bfy - 2 E_{\beta^{(1)}}({\bfbeta^{(1)}}^T) X_1^T \bfy + E_{\beta^{(1)}}({\bfbeta^{(1)}}^T X_1^T X_1 \bfbeta^{(1)}) ] \right\} + const \\
&=& \log N(\bfbeta^{(2)}|m_{\beta^{(2)}}, \phi^{-1} C_{\beta^{(2)}}) \times Ga(\phi|a_\phi,b_\phi)
\end{eqnarray*}

It is easy to see that this is a normal-gamma distribution with parameters:

\begin{eqnarray*}
C_{\beta^{(2)}}^{-1} = E_{\tau} (\bfD_\tau^{-1}) + X_2^TX_2, \, \, \,  \, \, \,   &\mbox{and}&  \, \, \,  \, \, \, m_{\beta^{(2)}} = C_{\beta^{(2)}} [X_2^T\bfy - X_2^TX_1 E_\beta^{(1)}(\bfbeta^{(1)})],
\end{eqnarray*}
$$a_\phi = a_0 + n/2 \;\;\;\; \mbox{and}$$
$$b_\phi =  \left\{b_0 +\frac{1}{2} [\bfy^T \bfy - 2 E_{\beta^{(1)}}({\bfbeta^{(1)}}^T) X_1^T \bfy + E_{\beta^{(1)}}({\bfbeta^{(1)}}^T X_1^T X_1 \bfbeta^{(1)}) ] - m_{\beta^{(2)}}^T C_{\beta^{(2)}}^{-1} m_{\beta^{(2)}} \right\}.$$

The variational  distribution of   $\bftau$ while holding $q_3(\lambda)$ fixed is given by 

\begin{eqnarray*}
\log q_2^\ast(\tau_j) &=& E_{\lambda}[\log (p(\tau_j|\lambda))] + E_{\beta,\phi}[\log (p(\beta_j,\phi|\tau_j))] + const\\ 
&=& E_\lambda \{\log [\exp\{-\lambda \tau_j\}]\} + E_{\beta,\phi}\left\{\log \left[ (\phi^{-1} \tau_j)^{-1/2} \exp\left\{-\frac{\phi}{2\tau_j} \beta_j^2 \right\} \right]\right\} + const \\
&=& - \frac{1}{2} \log \tau_j - \frac{1}{2}\left(2 E_\lambda[\lambda] \tau_j + \frac{1}{\tau_j} E_{\beta,\phi}[\phi \beta_j^2] \right) + const \\
&=& \log GIG(\tau_j|c_\tau,d_\tau,f_{\tau_j})
\end{eqnarray*}
with GIG being generalized inverse Gaussian distribution, where
$$c_\tau = \frac{1}{2}\;;\; d_\tau = 2 E_\lambda[\lambda] \; ; \; f_{\tau_j} = E_{\beta,\phi}[\phi \beta_j^2].$$ 
Therefore, 
$$\log q_2^\ast(\bftau) = \log \prod_{j=1}^p GIG(\tau_j|c_\tau,d_\tau,f_{\tau_j}).$$
 
The variational distribution of $\lambda$ is:

\begin{eqnarray*}
\log q_3^\ast(\lambda) &=& \log (p(\lambda)) + E_\tau[\log (p(\bftau|\lambda))] + const\\
&=& \log [\lambda^{g_0-1} \exp\{-h_0 \lambda\}] + E_\tau\left(\log \left[\prod_{j=1}^p \lambda \exp\{-\tau_j \lambda\}\right] \right)+ const\\
&=& (g_0 + p - 1) \log \lambda - \lambda [h_0 + \sum_{j=1}^p E_\tau(\tau_j)]+ const\\
&=& \log Ga(\lambda|g_\lambda, h_\lambda)
\end{eqnarray*}

which is a gamma distribution with parameters $$g_\lambda = g_0 + p  \; ;\; h_\lambda = h_0 + \sum_{j=1}^p E_\tau(\tau_j).$$

The expected values can be computed as follow:

It is worth  pointing out that if $X \sim GIG(p,a,b)$, then its density is 
$$
f(x|p,a,b) = \frac{ (\frac{a}{b})^{\frac{p}{2} } } {2 \kappa_p(\sqrt{a b}) } \, x_{p-1} \, \exp\{- (ax + b/x)/2\}, \ \ x>0,
$$
where $\kappa_p(\cdot)$ is a modified Bessel function of the second kind, with 
\begin{eqnarray*}
 E[X]&=&\sqrt{\frac{b}{a} } \frac{ \kappa_{p+1}( \sqrt{ab} )}{\kappa_{p}( \sqrt{ab} )} \ \  \mbox{and}  \ \ Var[X] = \left(\frac{b}{a}\right) \left[\frac{ \kappa_{p+2}( \sqrt{ab} )}{\kappa_{p}(\sqrt{ab})} - \left(\frac{ \kappa_{p+1}(\sqrt{ab} )}{\kappa_{p}(\sqrt{ab} )} \right)^2\right]\\
 E[X^{-1}]&=&\sqrt{\frac{a}{b} } \frac{ \kappa_{p+1}(\sqrt{ab} )}{\kappa_{p}(\sqrt{ab} )}  - \frac{2 p}{b}
\end{eqnarray*}

Therefore,

$$E_\tau[\bfD_\tau^{-1}] = diag(E_\tau(\tau_1^{-1}), \ldots, E_\tau(\tau_p^{-1})),$$ 

$$E_\tau(\tau_j^{-1}) = \frac{\sqrt{d_\tau} \kappa_{c_\tau+1}(\sqrt{d_\tau f_{\tau_j}})}{\sqrt{f_{\tau_j}} \kappa_{c_\tau}(\sqrt{d_\tau f_{\tau_j}})} - \frac{2 c_\tau}{f_{\tau_j}},$$ 

$$E_\tau(\tau_j) = \frac{\sqrt{f_{\tau_j}} \kappa_{c_\tau+1}(\sqrt{d_\tau f_{\tau_j}})}{\sqrt{d_\tau} \kappa_{c_\tau}(\sqrt{d_\tau f_{\tau_j}})},$$

$$E_\lambda(\lambda) = \frac{g_\lambda}{h_\lambda}.$$ 

For the calculus of $E_{\beta,\phi}[\phi \beta_j^2]$, let $x|y \sim N(\mu_x,y^{-1}\sigma_x)$ and $y \sim Ga(a,b)$, so

$$E[X^2 Y] = E[E(X^2 Y|Y)] = E[YE(X^2|Y)] = E[Y(E^2(X|Y)+ Var(X|Y))] = \mu_x^2 E(Y) + \sigma_x = \frac{a \mu_x^2}{b} + \sigma_x.$$

Thus,

$$E_{\beta,\phi}[\phi \beta_j^2] = m_{\beta_j}^2 a_\phi/b_\phi + (C_\beta)_{jj}.$$ 

\clearpage

\bibliography{myref}
 
\end{document}